%% file: TPC.tex
\documentclass[a4paper,11pt]{article}
\pdfoutput=1 

\usepackage{jheppub} 
                     
\usepackage[utf8]{inputenc}
\usepackage{graphicx} 
\usepackage{grffile}
\usepackage{subcaption}
\usepackage{dcolumn}  
\usepackage{makecell}
\usepackage{colordvi}
\usepackage{color}
\usepackage{epstopdf}
\usepackage{amssymb}
\usepackage{amsmath}
\usepackage{bm}
\usepackage{slashed}
\usepackage{enumerate}
\usepackage{url}
\usepackage{morefloats}

\usepackage{natbib}

\usepackage[colorlinks=true,
urlcolor=blue,
linkcolor=blue,
citecolor=blue,
linktocpage=true,
pdfproducer=medialab,
pdfa=true,
anchorcolor=blue]{hyperref}

\usepackage{tabularx,array}

\newcolumntype{C}{>{\centering\arraybackslash}X}
\usepackage{multirow}

\usepackage{lineno}

\usepackage{float}
\usepackage{xcolor}

\usepackage[title]{appendix}

\renewcommand{\arraystretch}{1.1}

\renewcommand{\thesection}{\arabic{section}}

\usepackage{changepage}
\usepackage{booktabs}

\input belle2sym.tex
\input utility.tex

\preprint{}

\begin{document}

\title{Analysis note: two-particle correlation in \boldmath $e^+e^-$ collisions at \boldmath $91$--$209$~GeV with archived ALEPH data}

\author[b]{Yu-Chen Chen,}
\author[b]{Yen-Jie Lee,}
\author[b]{Yi Chen,}
\author[a]{Paoti Chang,}
\author[b]{Chris McGinn,}
\author[b]{Tzu-An Sheng,}
\author[b]{Gian Michele Innocenti,}
\author[c]{Marcello Maggi}

\affiliation[a]{National Taiwan University, Taipei, Taiwan}
\affiliation[b]{Massachusetts Institute of Technology, Cambridge, USA}
\affiliation[c]{INFN Sezione di Bari, Bari, Italy}%

\emailAdd{janice\_c@mit.edu}
\emailAdd{yenjie@mit.edu}
\emailAdd{chenyi@mit.edu}
\emailAdd{pchang@phys.ntu.edu.tw}

\date{\today}

\abstract{
The first measurement of two-particle angular correlations for charged particles produced in \(e^+e^-\) annihilation up to \(\sqrt{s} = 209\) GeV is presented. Hadronic \(e^+e^-\) data, archived at center-of-mass energies ranging from 91 to 209 GeV, were collected using the ALEPH detector at LEP between 1992 and 2000. The angular correlation functions have been measured across a wide range of pseudorapidities and the full azimuth in bins of charged particle multiplicity. This is the first such measurement using LEP-II data. With LEP-II data at 91 GeV, neither the beam coordinate analysis nor the thrust coordinate analysis reveals significant long-range correlations, consistent with the finding in the previous measurement with the LEP-I sample. Results for \(e^+e^-\) data at energies above 91 GeV, which allow for higher event multiplicities reaching approximately 50, are presented for the first time. A long-range near-side excess in the correlation function has been identified in the thrust axis analysis. Moreover, the two-particle correlation functions were decomposed using a Fourier series, and the resulting Fourier coefficients \(v_n\) were compared with event generator outputs. In events with high multiplicity, featuring more than 50 particles, the extracted \(v_2\) and \(v_3\) magnitudes from the data are higher than those from the Monte Carlo reference.
}

\keywords{Two-particle correlation, Quark-gluon plasma, electron-positron annihilation}


\maketitle
\flushbottom


\section{Introduction}
\label{sec:Introduction}
In heavy-ion collision experiments, two-particle angular correlations~\cite{STAR:2005ryu,STAR:2009ngv,PHOBOS:2009sau,Chatrchyan:2012wg,Aamodt:2011by,Adam:2019woz} are extracted for studying the Quark-Gluon Plasma (QGP)~\cite{Busza:2018rrf}. In these measurements, a long-range angular correlation, known as the ridge-like structure~\cite{STAR:2009ngv,PHOBOS:2009sau}, has been observed in various collision systems and at different collision energies. Since the beginning of LHC operations, this ridge structure has also been observed in high-multiplicity proton-proton collisions by the CMS collaboration~\cite{Khachatryan:2010gv} and confirmed by experiments at the LHC and RHIC using smaller collision systems than ion-ion collisions, such as proton-proton~\cite{Aad:2015gqa}, proton-ion~\cite{CMS:2012qk,ALICE:2012eyl,ALICE:2013snk,ATLAS:2012cix,Aaij:2015qcq}, and deuteron-ion~\cite{PHENIX:2013ktj} collisions. In heavy-ion collisions, the ridge structure is associated with the fluctuating initial state of the ions~\cite{Alver:2010gr}. However, the physical origin of the ridge structure in small systems remains under debate~\cite{Dumitru:2010iy,Dusling:2013qoz,Bozek:2011if,He:2015hfa,Nagle:2018nvi}. 

Recently, there has been growing interest in measuring two-particle correlations in even smaller collision systems, such as photonuclear PbPb~\cite{ATLAS:2021jhn,CMS:2022doq}, \(ep\)~\cite{ZEUS:2019jya}, and \(e^+e^-\)~\cite{Badea:2019vey,Belle:2022fvl,The:2022lun}. These measurements serve as complementary counterparts to results in larger collision systems and can be used to identify the minimal conditions for collective behavior~\cite{Nagle:2017sjv}. The use of electron beams eliminates complications such as multiple parton interactions and initial state correlations. So far, no significant ridge-like signal has been observed in the most elementary electron-positron annihilations, providing additional insights into the ridge signal~\cite{Bierlich:2019wld,Bierlich:2020naj,Castorina:2020iia,Agostini:2021xca,Larkoski:2021hee,Baty:2021ugw}.

In this analysis note, we report the first measurement of two-particle angular correlation functions as well as the extraction of flow coefficients in high multiplicity \(e^+e^-\) annihilation events at \(\sqrt{s}= 91\)–\(209\) GeV using archived ALEPH data from LEP-II. Due to the high collision energies above the \(Z^0\) pole, this dataset allows for studies of higher multiplicity events compared to those from LEP-I. Additionally, the analyzed dataset encompasses various underlying physics processes.

The analysis note is structured as follows: Section~\ref{sec:Sample} presents the data sample and its corresponding event thrust distributions. The thrust-axis two-particle correlation function and the applied corrections are detailed in sections~\ref{sec:TwoParticleCorrelationFunction} and~\ref{sec:Corrections}, respectively. Section~\ref{sec:Rst} presents the two-particle correlations and the projected azimuthal differential associated yields, and it also provides an estimate of the ridge signal upper limits when the associated yield signal is not significant. Section~\ref{sec:flow} presents the extracted flow coefficient. The note concludes with a brief summary in Section~\ref{sec:summary}.
\section{Experimental setup, data sample and the event thrust}
\label{sec:Sample}

The ALEPH detector is described in ~\cite{Decamp:1990jra}. The central part of the detector is designed for the efficient reconstruction of charged particles. Their trajectories are measured by a two-layer silicon strip vertex detector, a cylindrical drift chamber, and a large time projection chamber (TPC). These tracking detectors are situated inside a 1.5 T axial magnetic field generated by a superconducting solenoidal coil. The transverse momenta of the charged particles are reconstructed with a resolution of \(\delta p_t/p_t = 6 \times 10^{-4} p_t \oplus 0.005\) (GeV/c).

Electrons and photons are identified in the electromagnetic calorimeter (ECAL), which is located between the TPC and the superconducting coil. The ECAL is a sampling calorimeter, comprised of lead plates and proportional wire chambers segmented into \(0.9^\circ \times 0.9^\circ\) projective towers. These are read out in three depth sections and have a total thickness of approximately 22 radiation lengths. Isolated photons are reconstructed with a relative energy resolution of \(0.18/\sqrt{E} + 0.009\) GeV.

The iron return yoke, constructed with 23 layers of streamer tubes, also serves as the hadron calorimeter (HCAL) for the detection of charged and neutral hadrons. The relative energy resolution for hadrons is \(0.85/\sqrt{E}\). Muons are identified based on their patterns in the HCAL and by the muon chambers, which are made of two double layers of streamer tubes located outside the HCAL.

The information from the trackers and calorimeters is integrated using an energy-flow algorithm~\cite{ALEPH:1994ayc}. This algorithm generates a set of charged and neutral particles, called energy-flow objects. The charged particles reconstructed in the energy-flow algorithm are used in the correlation function analysis

The analysis is performed using the data sample taken by the ALEPH detector from 1992 to 2000, corresponding to collision energies of $\sqrt{s} = 91$--209~GeV and a total integrated luminosity of 745~pb$^{-1}$ of data. 
We have divide the data into two datasets according to their center-of-mass energies. One is the {\it $Z$-resonance dataset}, which provides results extending the previous measurement with only LEP-I data~\cite{Badea:2019vey}. The other one focuses exclusively on {\it high-energy events} at LEP-II, with collision energies greater than those of the $Z$-boson mass.

Unlike the sample taken at the 91.2 GeV~\cite{Badea:2019vey, Chen:2021uws}, where $Z$-decays dominate over other processes, a wide variety of processes beyond the $e^+e^- \to q\bar{q}$ fragmentation also make non-negligible contributions to the high-energy sample. The initial-state QED radiation is significant in high-energy events, an effect called ``radiative-return-to-$Z$.'' We follow the selection criteria studied by the ALEPH collaboration~\cite{ALEPH:2003obs}. To ensure that the analyzed events are produced from high center-of-mass energy collisions, we calculate the effective center-of-mass energy ($\sqrt{s'}$) by clustering the event into two jets, and obtain
\begin{equation}
s' = \frac{\sin \theta_1 + \sin \theta_2 - | \sin(\theta_1 + \theta_2) |}{ \sin \theta_1 + \sin \theta_2 + | \sin(\theta_1 + \theta_2) | } \times s,
\end{equation}
where $\theta_{1,2}$ are the angles of the two jets with respect to the beam (or the $z$) direction.
With the clustered two jets, the visible two-jet invariant-mass ($M_{\rm vis}$) can be calculated and also be used as a useful variable to suppress the QED radiation background.
In the analysis, the $\sqrt{s'}$ is required to be greater than $0.9 \sqrt{s}$, and $M_{\rm vis}$ should be larger than $0.7 \sqrt{s}$.

We follow the same hadronic event selection criteria as in the previous LEP-I work~\cite{Badea:2019vey}, requiring the polar angle of the event sphericity axis to be from $7\pi/36$ to $29\pi/36$. Events that have fewer than five tracks and a total reconstruction charged-particle energy smaller than $15$~GeV are rejected. The event selections are summarized in Table~\ref{tab:SelectionSummary}. The physics results and their implication are summarized in a separate letter for journal publication~\cite{Chen:2023njr}.

\subsection{Monte Carlo}
To study the reconstruction effects and correct the data, we utilize the Monte Carlo (MC) events simulated by the ALEPH collaboration. Archived $\textsc{pythia}$ 6.1~\cite{Sjostrand:2000wi} MC simulation samples, which were produced under the years 1994 and 1997-2000 run detector conditions by the ALEPH collaboration, were the available archived MC sample at the time of this analysis. The MC samples are used for the derivation of tracking efficiency and event selection corrections.

Different MC processes are weighted according to the cross-sections calculated by event generators. In Table~\ref{tab:physics_processes}, we show the number of simulated events and the corss-sections associated to different MC processes in the collision energy $\sqrt{s}=207$~GeV. Different collision-energy benchmarks: $\sqrt{s}=183,187,192,196,200,202,205,207$~GeV are used for the full LEP-II analysis.

\begin{table}[h]
\centering
\caption{List of physics processes with Monte Carlo event counts and cross-sections for LEP-II analysis. Shown in the table is the simulation with a collision energy $\sqrt{s}=207$~GeV. }
\begin{tabular}{|c|l|r|r|}
\hline
TAG & Process & \# Events & Cross-Section (picobarns) \\
\hline
\multicolumn{4}{|c|}{\(\gamma\gamma \rightarrow \text{hadrons}\)} \\
\hline
GGUD & \( \gamma\gamma  \rightarrow uu/dd \) & 2,000,000 & \( 1795.310 \pm 1.847 \) \\
GGSS & \( \gamma\gamma  \rightarrow ss \) & 200,000 & \( 88.486 \pm 0.287 \) \\
GGCC & \( \gamma\gamma  \rightarrow cc \) & 700,000 & \( 271.910 \pm 0.457 \) \\
GGBB & \( \gamma\gamma  \rightarrow bb \) & 25,000 & \( 1.410 \pm 0.012 \) \\
\hline
\multicolumn{4}{|c|}{\( e^+e^- \rightarrow \text{hadrons} \)} \\
\hline
KQQ  & \( e^+e^- \rightarrow qq \) & 2,000,000 & \( 81.007 \pm 0.022 \) \\
TT   & \( e^+e^- \rightarrow \tau^+\tau^- \) & 100,000 & \( 6.742 \pm 0.007 \) \\
\hline
\multicolumn{4}{|c|}{\( e^+e^- \rightarrow 4f \)} \\
\hline
KWW4F & \( W^+W^- \) & 2,350,000 & \( 18.958 \pm 0.005 \) \\
KWENU & \( W e \nu \) & 100,000 & \( 0.774 \pm 0.001 \) \\
PZEE  & \( Z e e \) & 200,000 & \( 8.470 \pm 0.019 \) \\
PZZ   & \( ZZ \) & 200,000 & \( 2.340 \pm 0.005 \) \\
ZNN   & \( Z \nu\nu \) & 50,000 & \( 0.017 \pm 0.000 \) \\
\hline
\end{tabular}
\label{tab:physics_processes}
\end{table}


\subsection{Charged particle selection}
\label{sec:chgdSel}
Charged particle candidates are required to have at least four TPC hits, and to be restricted from the interaction point within a loose allowed distance, where the radial displacement $d_0<2$~cm and the longitudinal displacement $z_0<10$~cm. 
The transverse momentum of a track should be greater than 0.2~GeV, and the absolute value of the cosine of the polar angle should be smaller than 0.94. The charged particle selections are summarized in Table~\ref{tab:SelectionSummary}.


\subsection{Event multiplicity distribution}

After applying all the selection criteria, the subsequent offline multiplicity, denoted as \({\rm N}_{\rm Trk}^{\rm Offline}\), is used for studying the correlation function's multiplicity dependence. It is important to note that \({\rm N}_{\rm Trk}^{\rm Offline}\) represents the count of charged particles that have passed the selection criteria outlined in the ``Charged particle selection'' subsection~\ref{sec:chgdSel}, without any additional efficiency corrections.

\begin{figure}[ht]
\centering
    \begin{subfigure}[b]{0.32\textwidth}
        \includegraphics[width=\textwidth,angle=0]{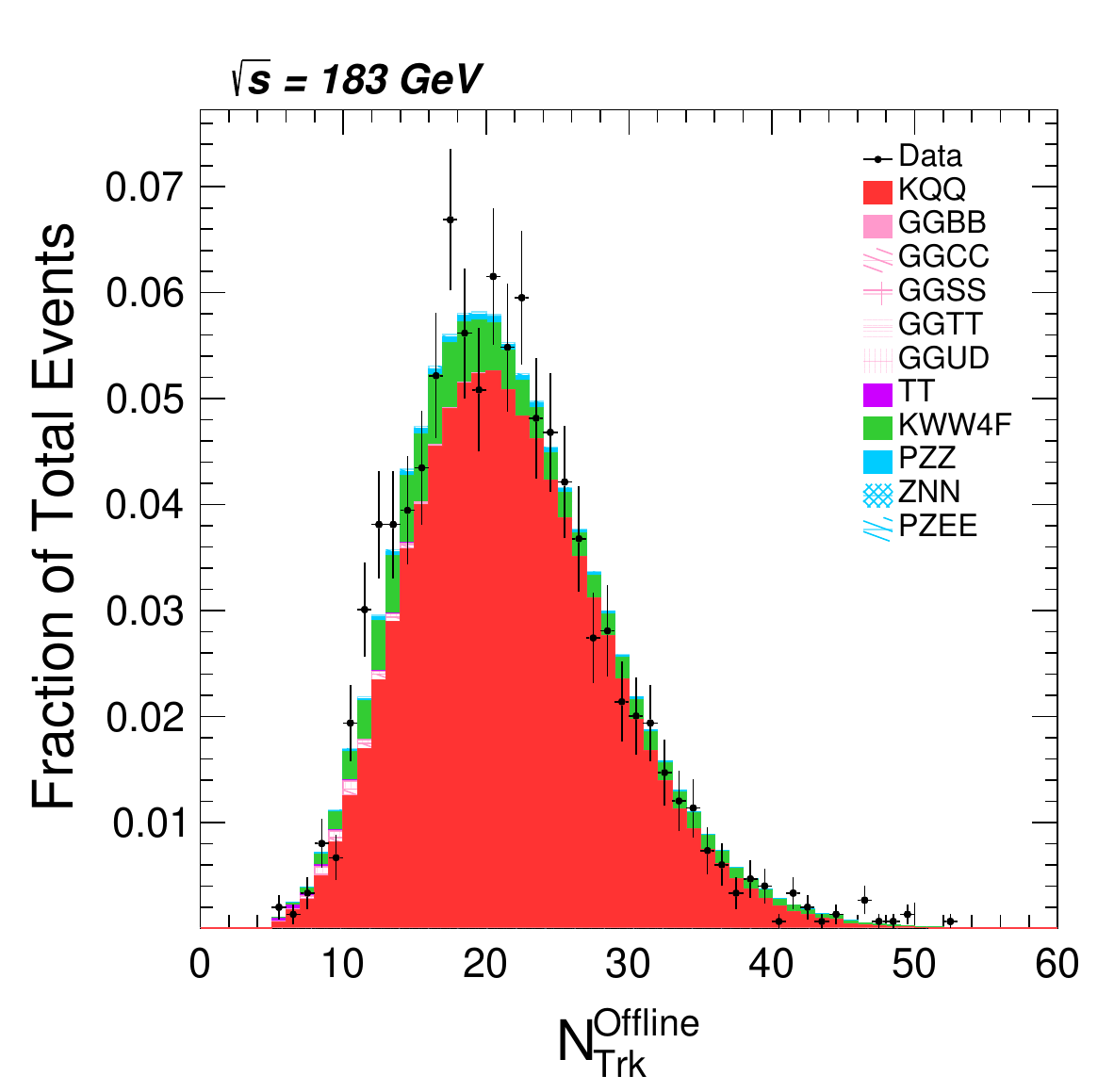}
        \caption{$\sqrt{s}=183$~GeV (Year 1997)}
    \end{subfigure}
    \begin{subfigure}[b]{0.32\textwidth}
        \includegraphics[width=\textwidth,angle=0]{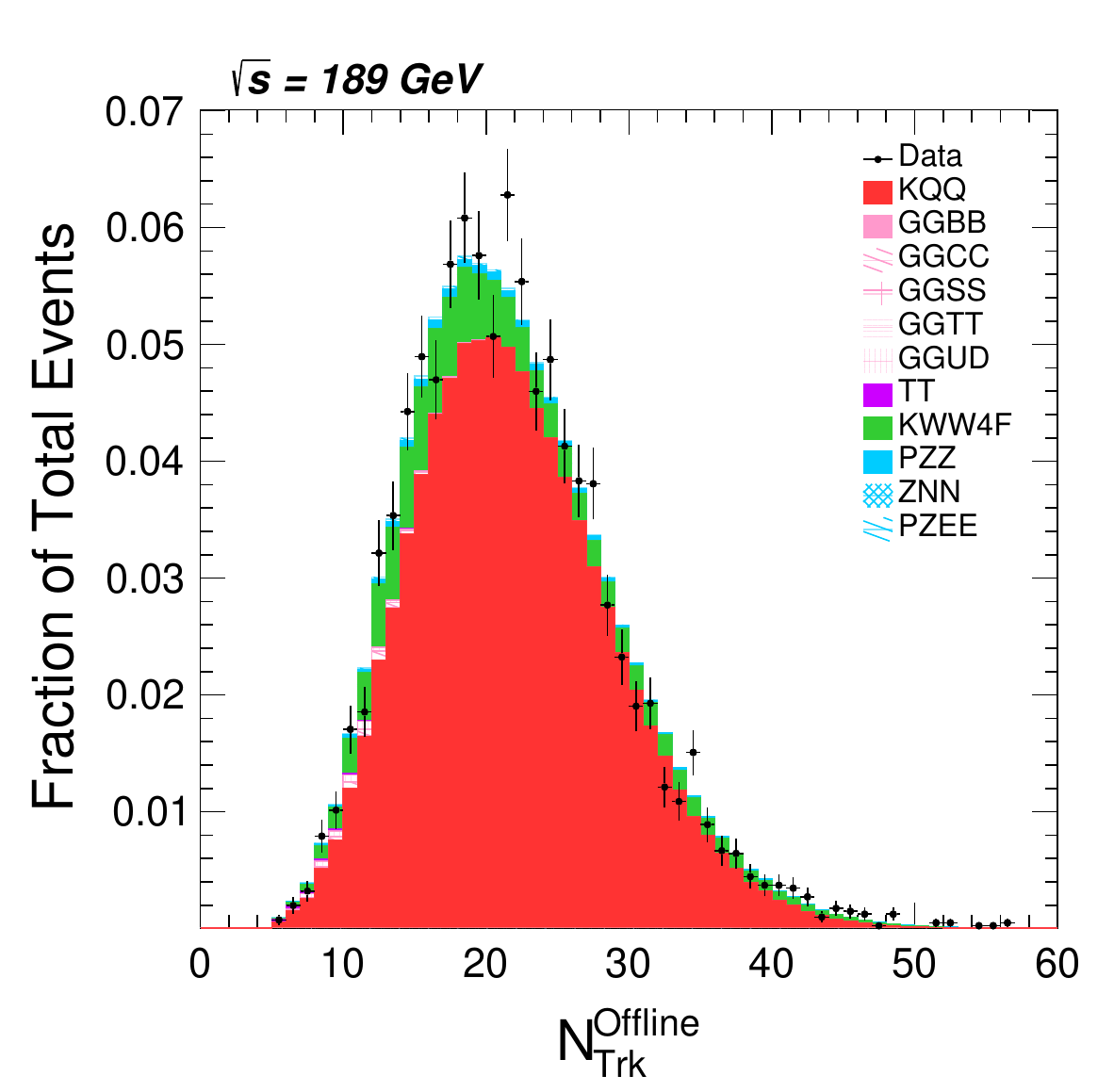}
        \caption{$\sqrt{s}=189$~GeV (Year 1998)}
    \end{subfigure}
    \begin{subfigure}[b]{0.32\textwidth}
        \includegraphics[width=\textwidth,angle=0]{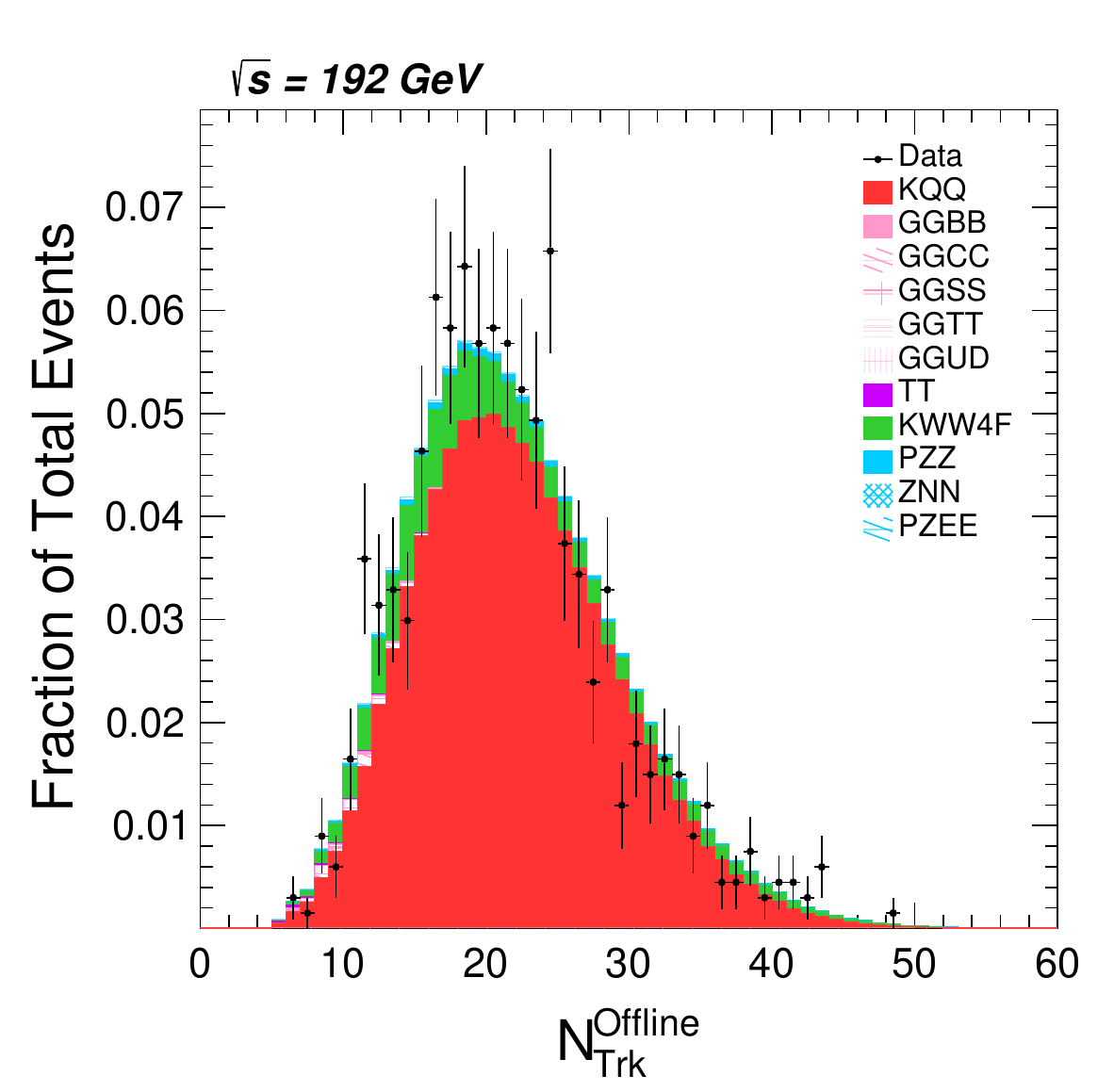}
        \caption{$\sqrt{s}=192$~GeV (Year 1999)}
    \end{subfigure}
    \begin{subfigure}[b]{0.32\textwidth}
        \includegraphics[width=\textwidth,angle=0]{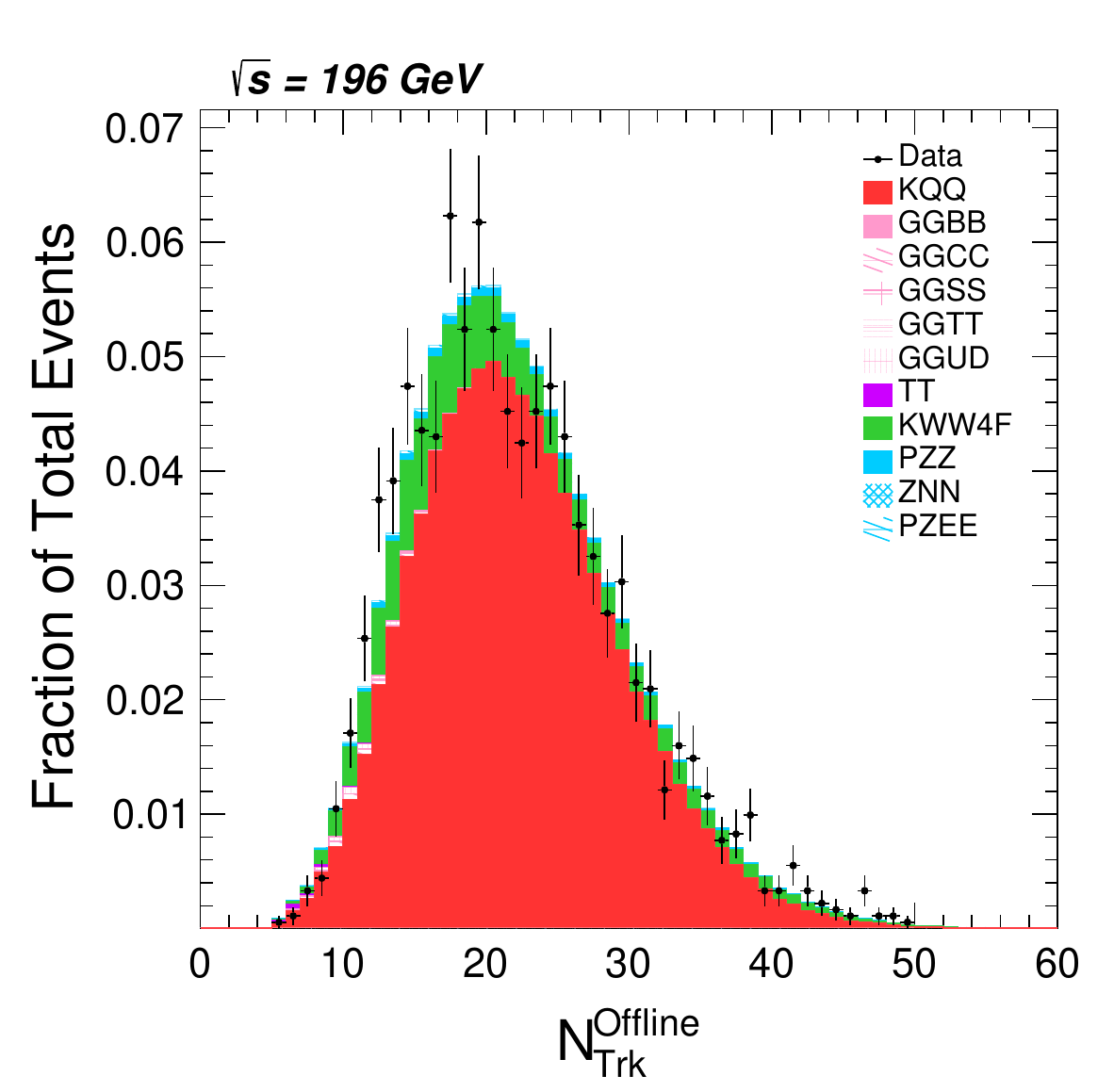}
        \caption{$\sqrt{s}=196$~GeV (Year 1999)}
    \end{subfigure}
    \begin{subfigure}[b]{0.32\textwidth}
        \includegraphics[width=\textwidth,angle=0]{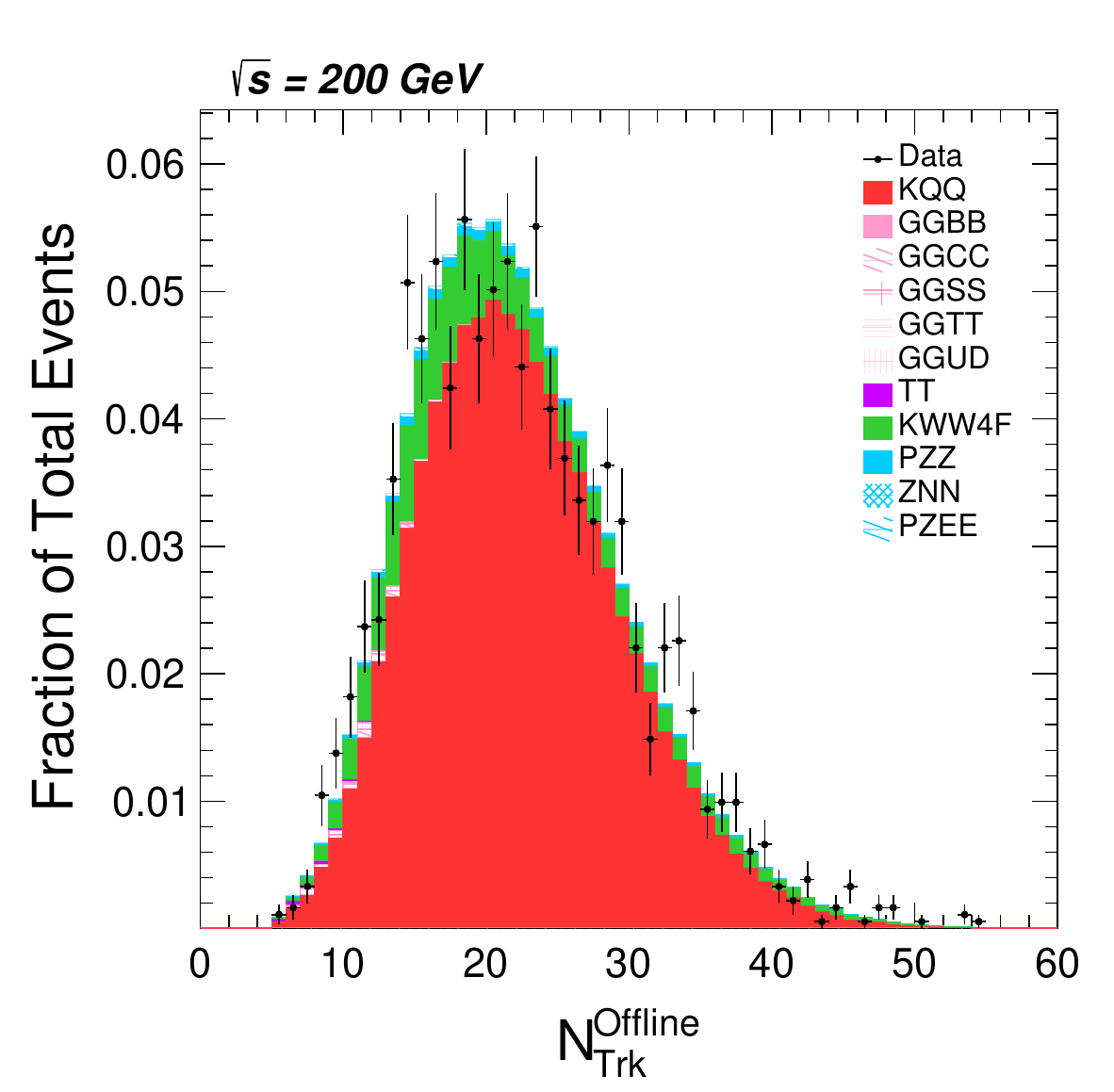}
        \caption{$\sqrt{s}=200$~GeV (Year 1999)}
    \end{subfigure}
    \begin{subfigure}[b]{0.32\textwidth}
        \includegraphics[width=\textwidth,angle=0]{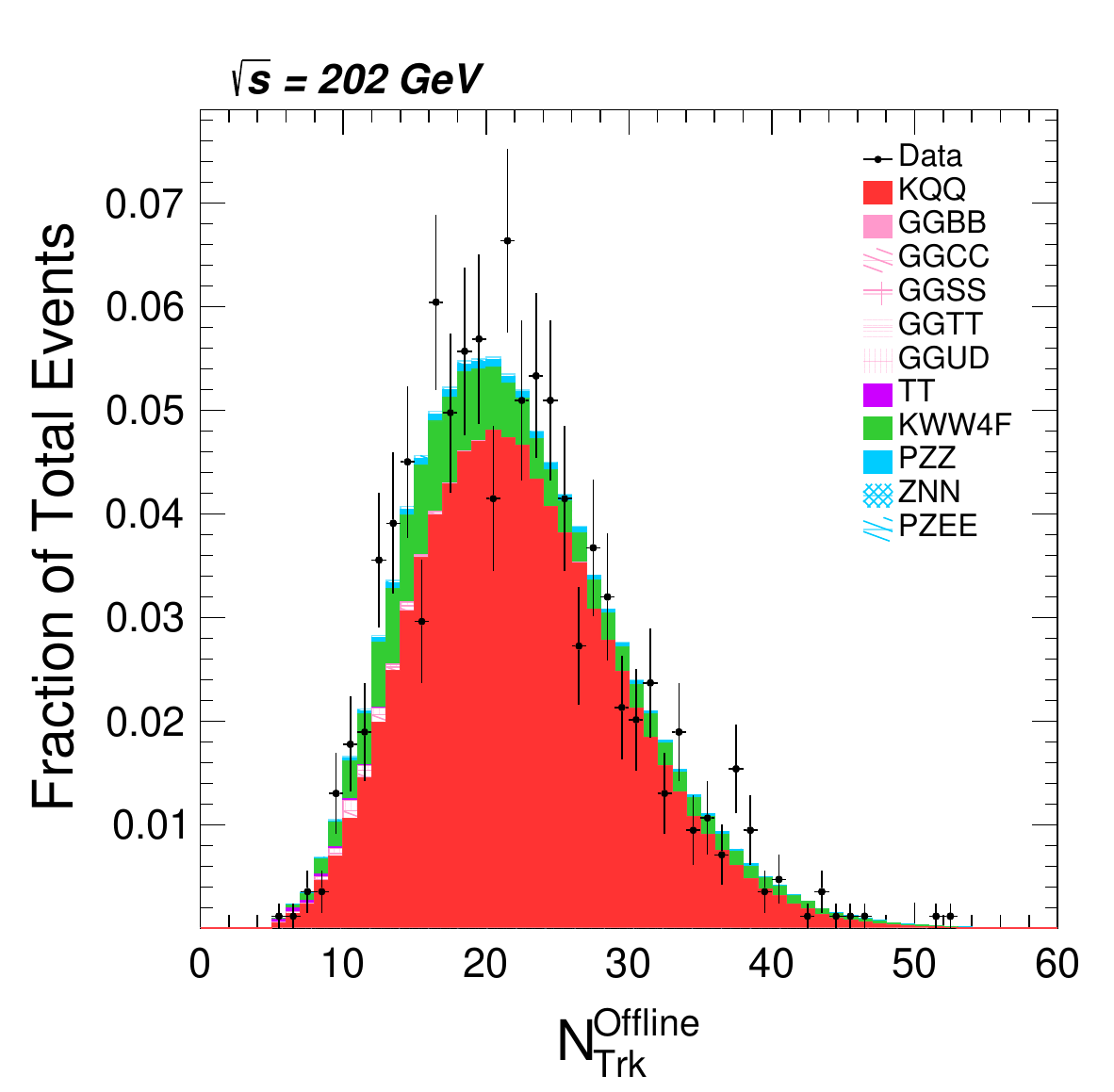}
        \caption{$\sqrt{s}=202$~GeV (Year 1999)}
    \end{subfigure}
    \begin{subfigure}[b]{0.32\textwidth}
        \includegraphics[width=\textwidth,angle=0]{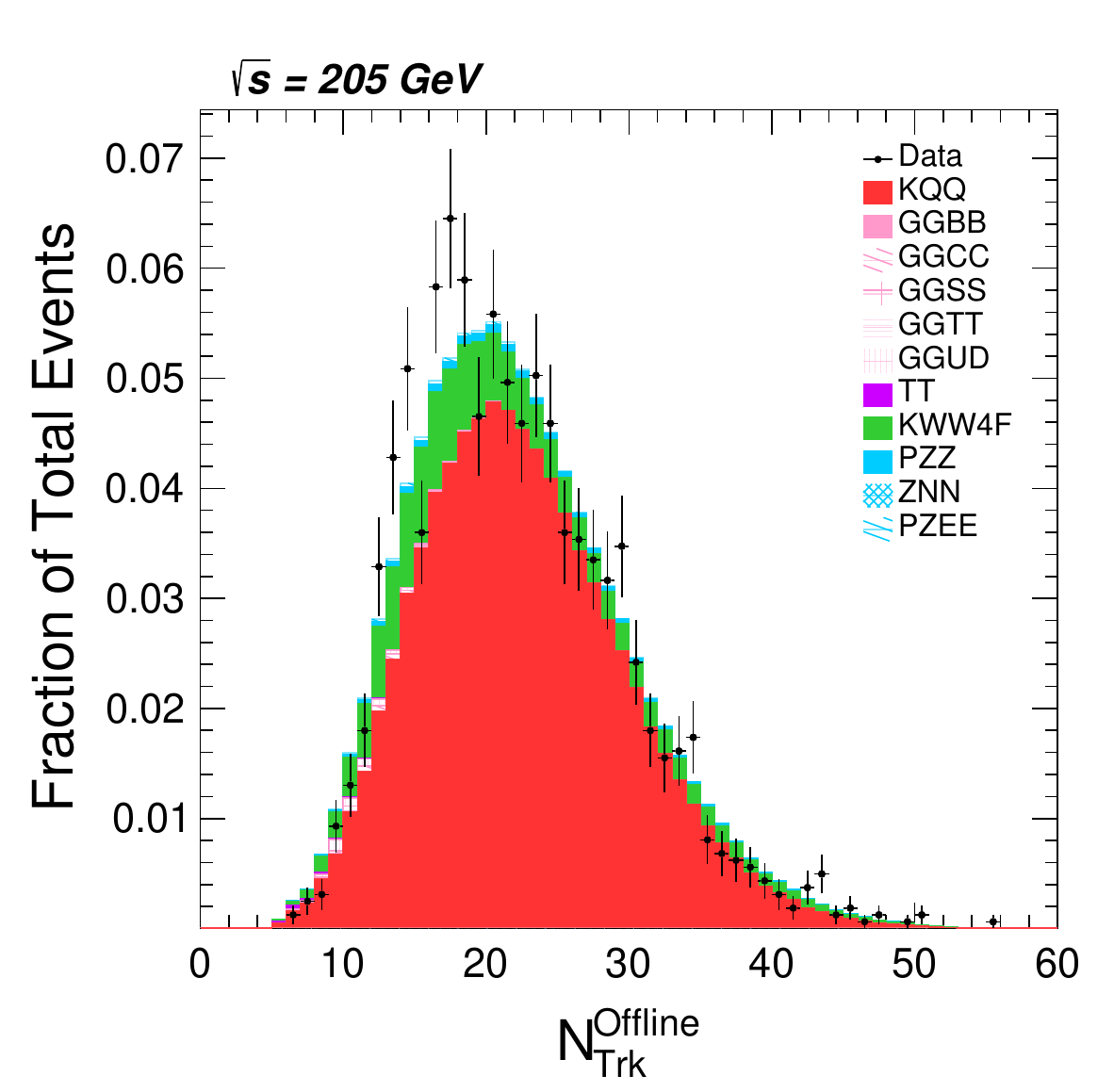}
        \caption{$\sqrt{s}=205$~GeV (Year 2000)}
    \end{subfigure}
    \begin{subfigure}[b]{0.32\textwidth}
        \includegraphics[width=\textwidth,angle=0]{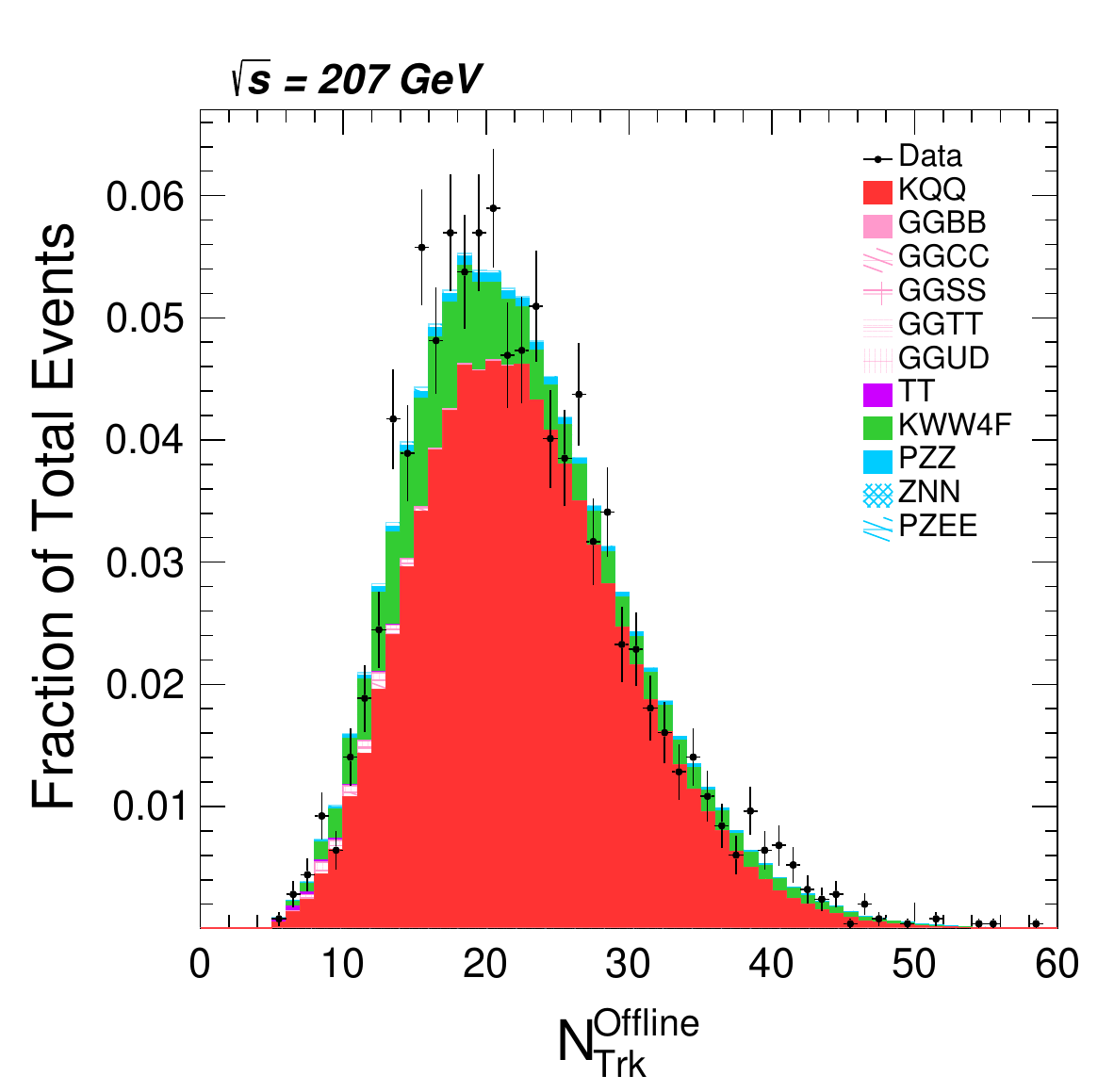}
        \caption{$\sqrt{s}=207$~GeV (Year 2000)}
    \end{subfigure}
\caption{The offline multiplicity (\ntrkoff) distributions for high-energy events from different years. 
Data are presented as black error bars; stacked histograms show the decomposition of different simulated processes scaled according to the calculated cross-sections.
}
\label{fig:NtrkOffline}
\end{figure}

Figure~\ref{fig:NtrkOffline} displays the offline multiplicity distributions in data collected at various collision energies. In the low-multiplicity region, there is a residual contribution from QED ($e^+e^- \to \tau^+\tau^-$) and two-photon events. However, these are significantly suppressed due to the hadronic event selection. Above the $W^{+}W^{-}$ production threshold ($\sqrt{s} > 160$~GeV), four-fermion processes mediated by either single or double $W$ or $Z$ bosons serve as sub-dominant channels in hadronic decays. At the highest multiplicity, four-fermion processes, especially $W^{+}W^{-}$ production, emerge as the dominant channels.

The multiplicity classes used in this study, along with their corresponding fraction of data, and the mapping of average offline multiplicities $\langle {\rm N}_{\rm Trk}^{\rm Offline}\rangle$ to average multiplicities after efficiency correction $\langle {\rm N}_{\rm trk}^{\rm corr}\rangle$ are listed in Table~\ref{tab:NtrkCorr}.

\begin{table}[ht]
\caption{Summary table for particle selections and event selections.  Neutral particles are selected mainly for the event thrust calculation.}
\begin{center}
\begin{tabularx}{\textwidth}{l|l}
\hline\hline
\multicolumn{2}{l}{Charged particles}  \\
\hline
Acceptance              & $|\cos\theta|<0.94$ \\
High quality tracks     & $p_{\rm T} \ge 0.2$ GeV\\
                        & at least 4 TPC hits \\
Impact parameter        & $d_0<2$~cm, $z_0<10$~cm \\
\hline
\multicolumn{2}{l}{Neutral particles (for thrust calculation)}  \\
\hline
Acceptance              & $|\cos\theta|<0.98$ \\
Energy cut              & $E>0.4$~GeV \\
\hline
\multicolumn{2}{l}{Event selection}  \\
\hline
ISR                     & $\sqrt{s'} \ge 0.9 \sqrt{s}$ \\
                        & $M_{\rm vis} \ge 0.7 \sqrt{s}$ \\
Hadronic events         & at least five good tracks \\
                        & total reconstructed charged-particle energy $\ge 15$~GeV \\
Acceptance              & $7\pi/36 \le \theta_{\rm sphericity} \le 29\pi/36$\\
\hline\hline
\end{tabularx} 
\label{tab:SelectionSummary}
\end{center}
\end{table}

\begin{figure}[ht]
\centering
     \begin{subfigure}[b]{0.32\textwidth}
        \includegraphics[width=\textwidth,angle=0]{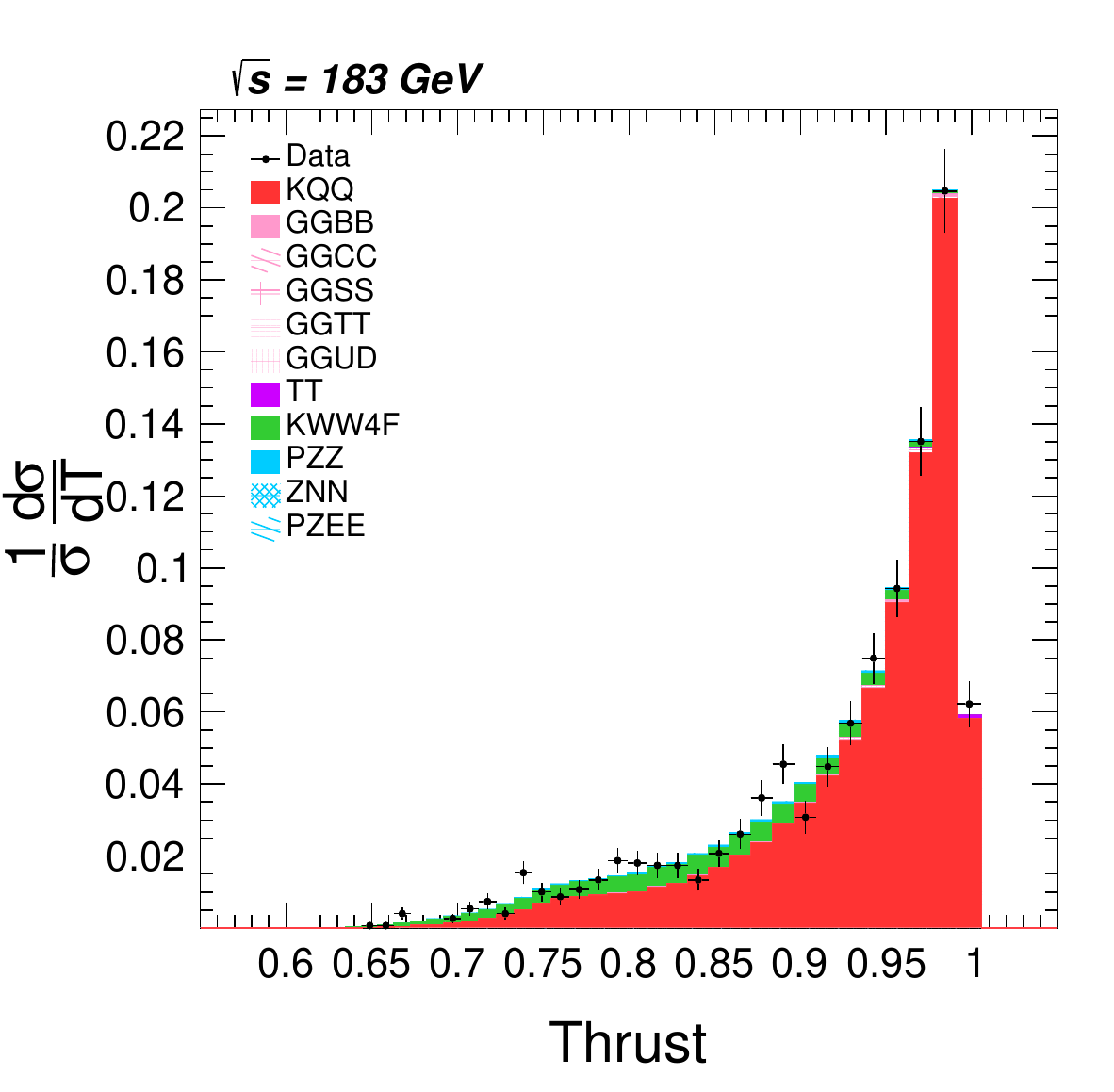}
        \caption{$\sqrt{s}=183$~GeV (Year 1997)}
    \end{subfigure}
    \begin{subfigure}[b]{0.32\textwidth}
        \includegraphics[width=\textwidth,angle=0]{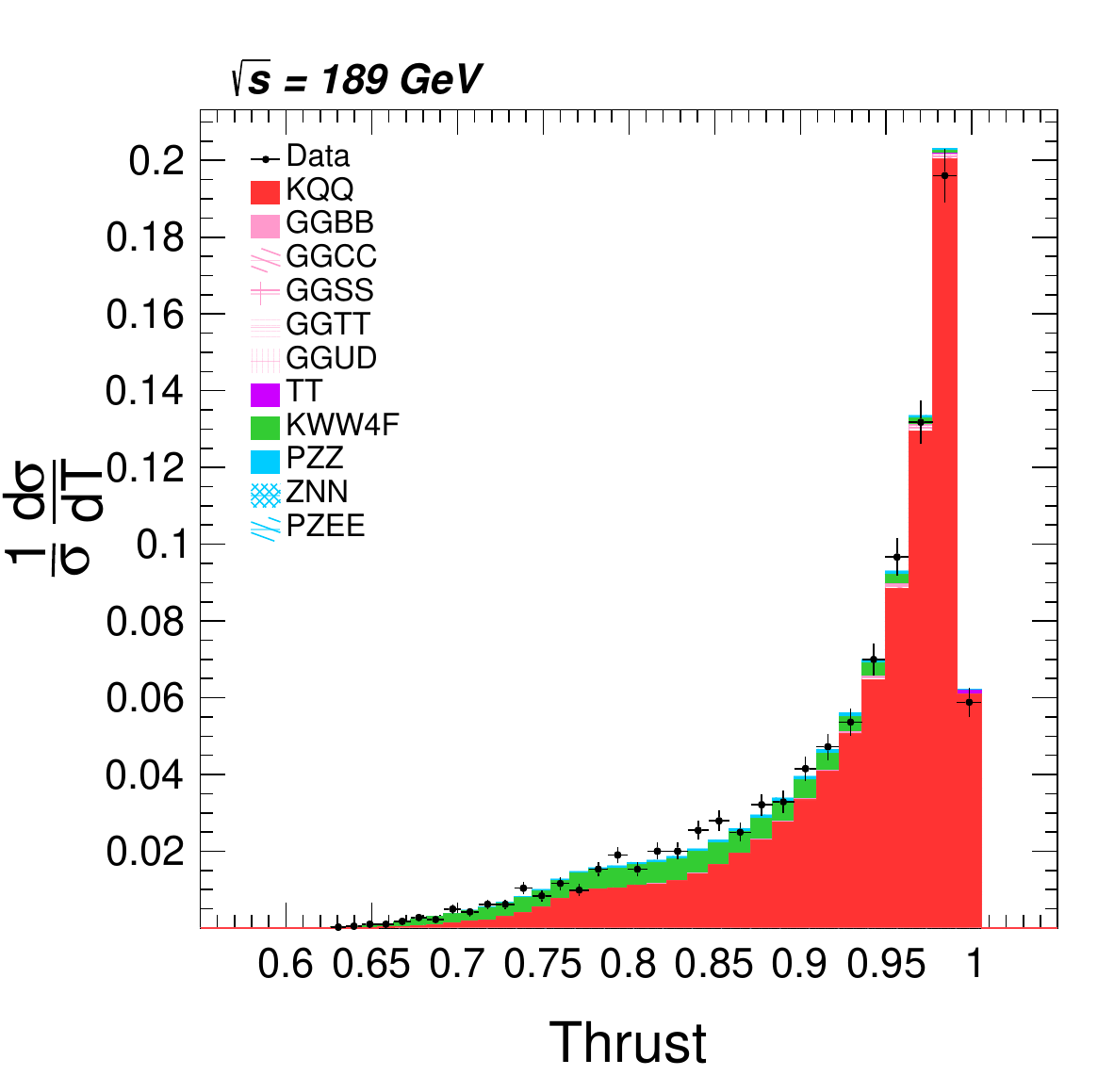}
        \caption{$\sqrt{s}=189$~GeV (Year 1998)}
    \end{subfigure}
    \begin{subfigure}[b]{0.32\textwidth}
        \includegraphics[width=\textwidth,angle=0]{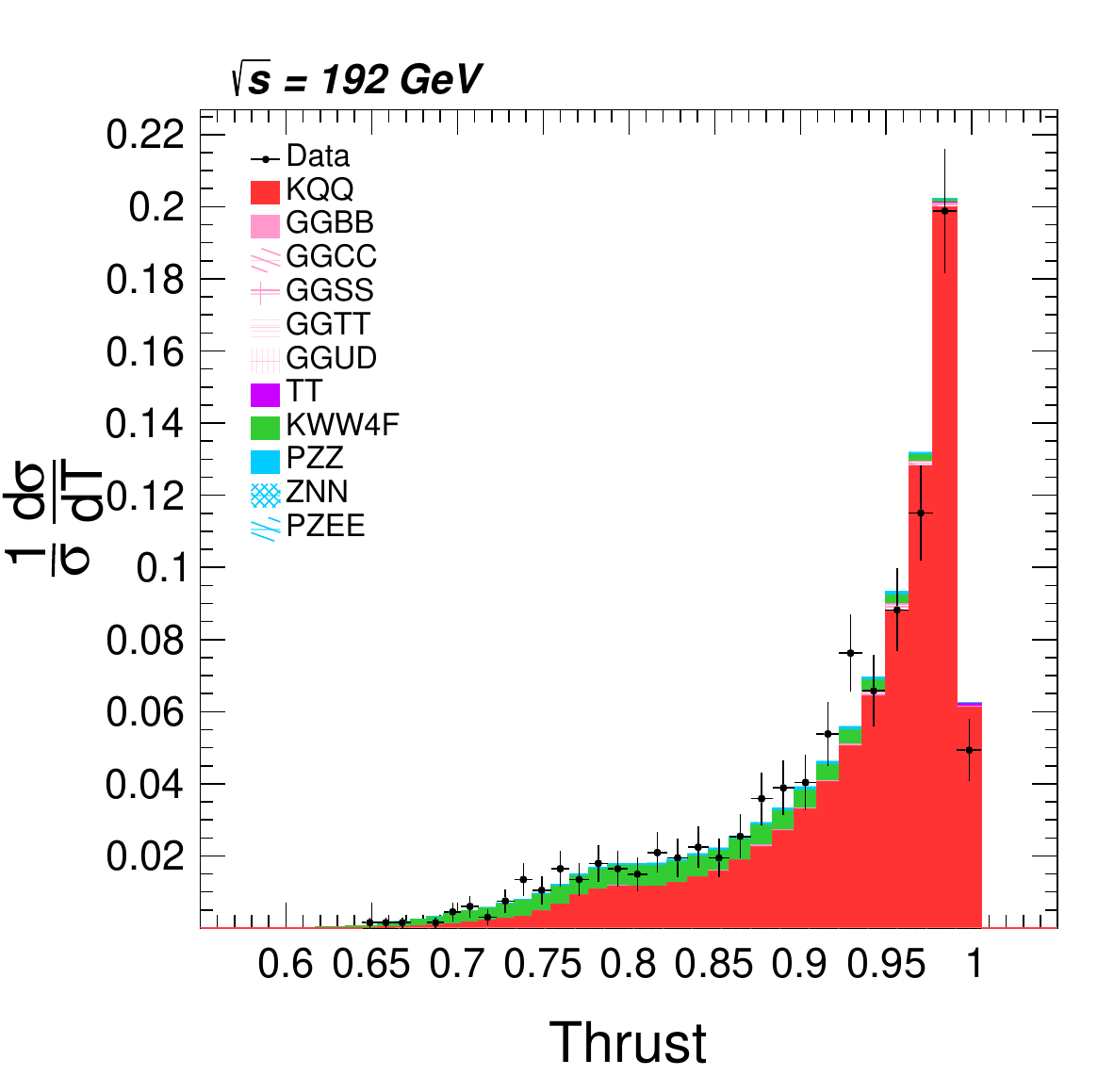}
        \caption{$\sqrt{s}=192$~GeV (Year 1999)}
    \end{subfigure}
    \begin{subfigure}[b]{0.32\textwidth}
        \includegraphics[width=\textwidth,angle=0]{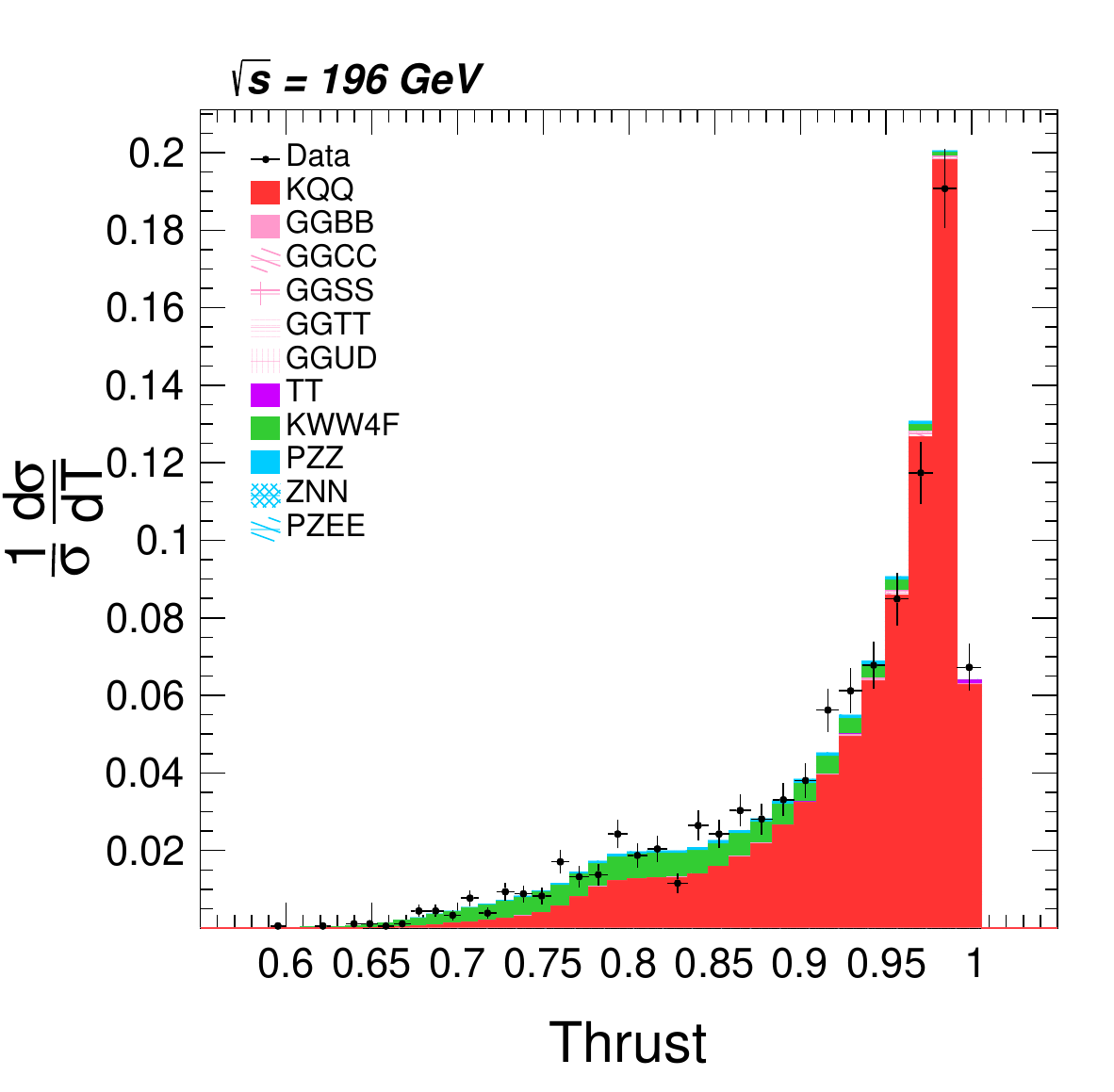}
        \caption{$\sqrt{s}=196$~GeV (Year 1999)}
    \end{subfigure}
    \begin{subfigure}[b]{0.32\textwidth}
        \includegraphics[width=\textwidth,angle=0]{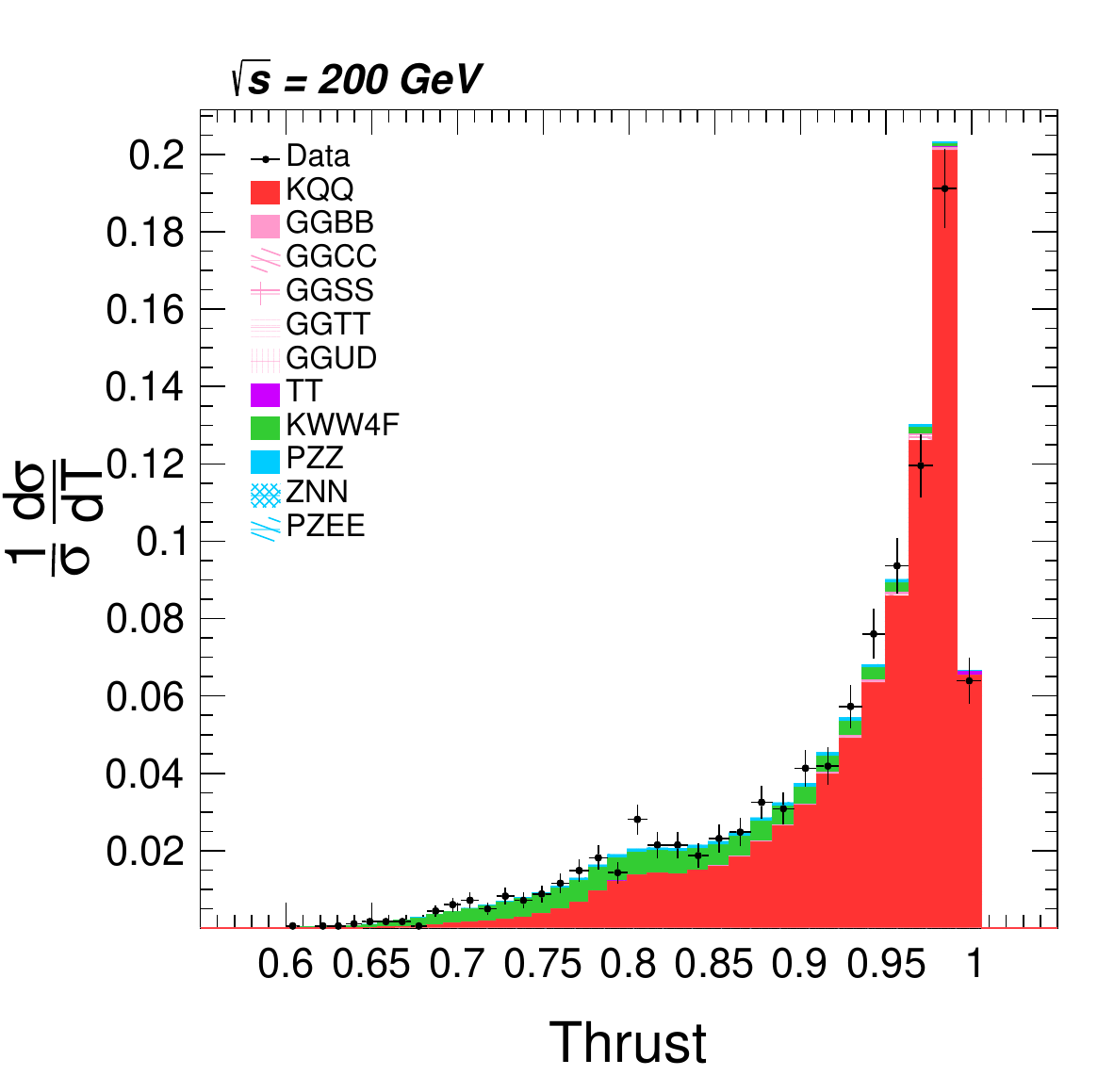}
        \caption{$\sqrt{s}=200$~GeV (Year 1999)}
    \end{subfigure}
    \begin{subfigure}[b]{0.32\textwidth}
        \includegraphics[width=\textwidth,angle=0]{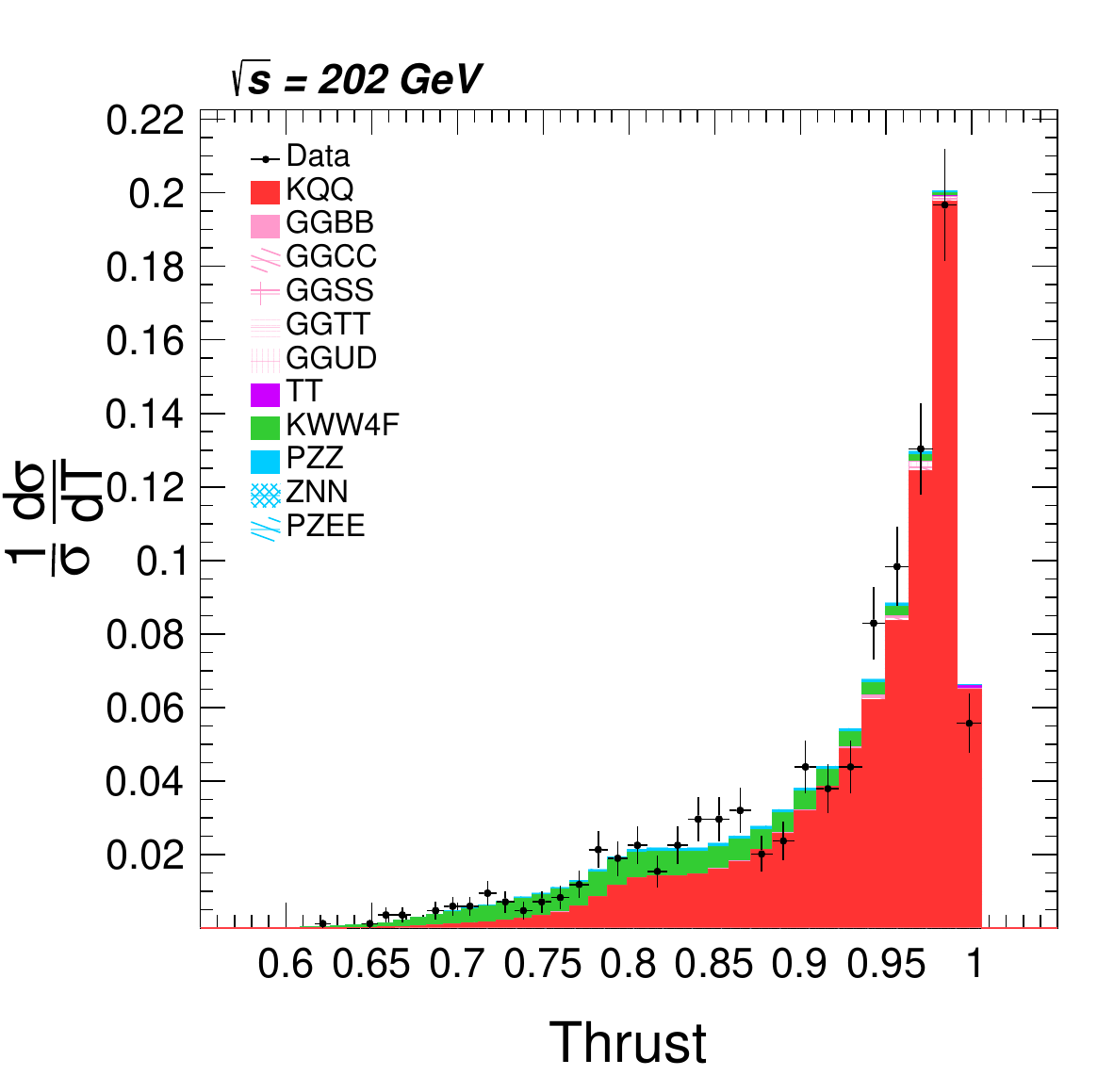}
        \caption{$\sqrt{s}=202$~GeV (Year 1999)}
    \end{subfigure}
    \begin{subfigure}[b]{0.32\textwidth}
        \includegraphics[width=\textwidth,angle=0]{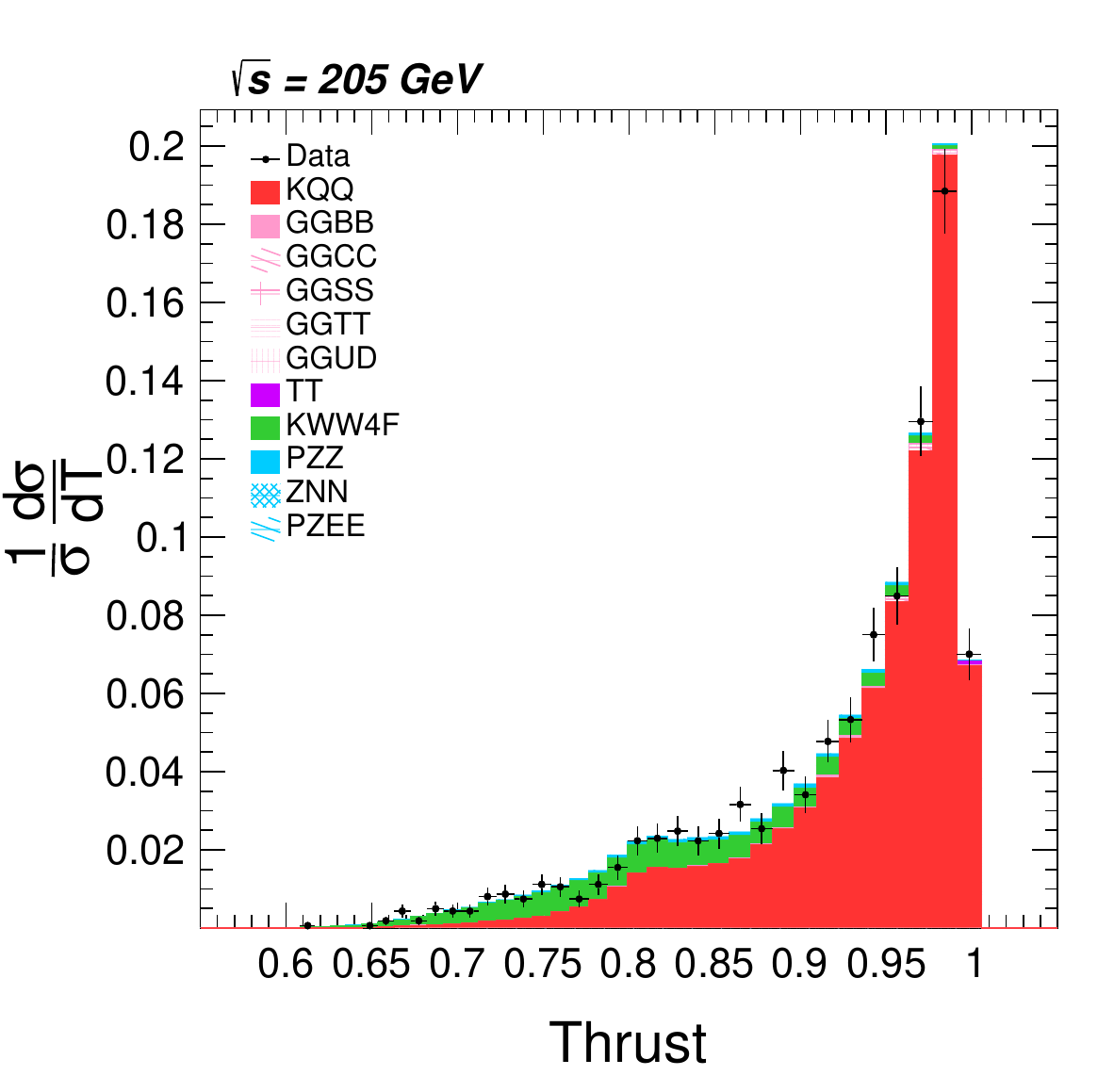}
        \caption{$\sqrt{s}=205$~GeV (Year 2000)}
    \end{subfigure}
    \begin{subfigure}[b]{0.32\textwidth}
        \includegraphics[width=\textwidth,angle=0]{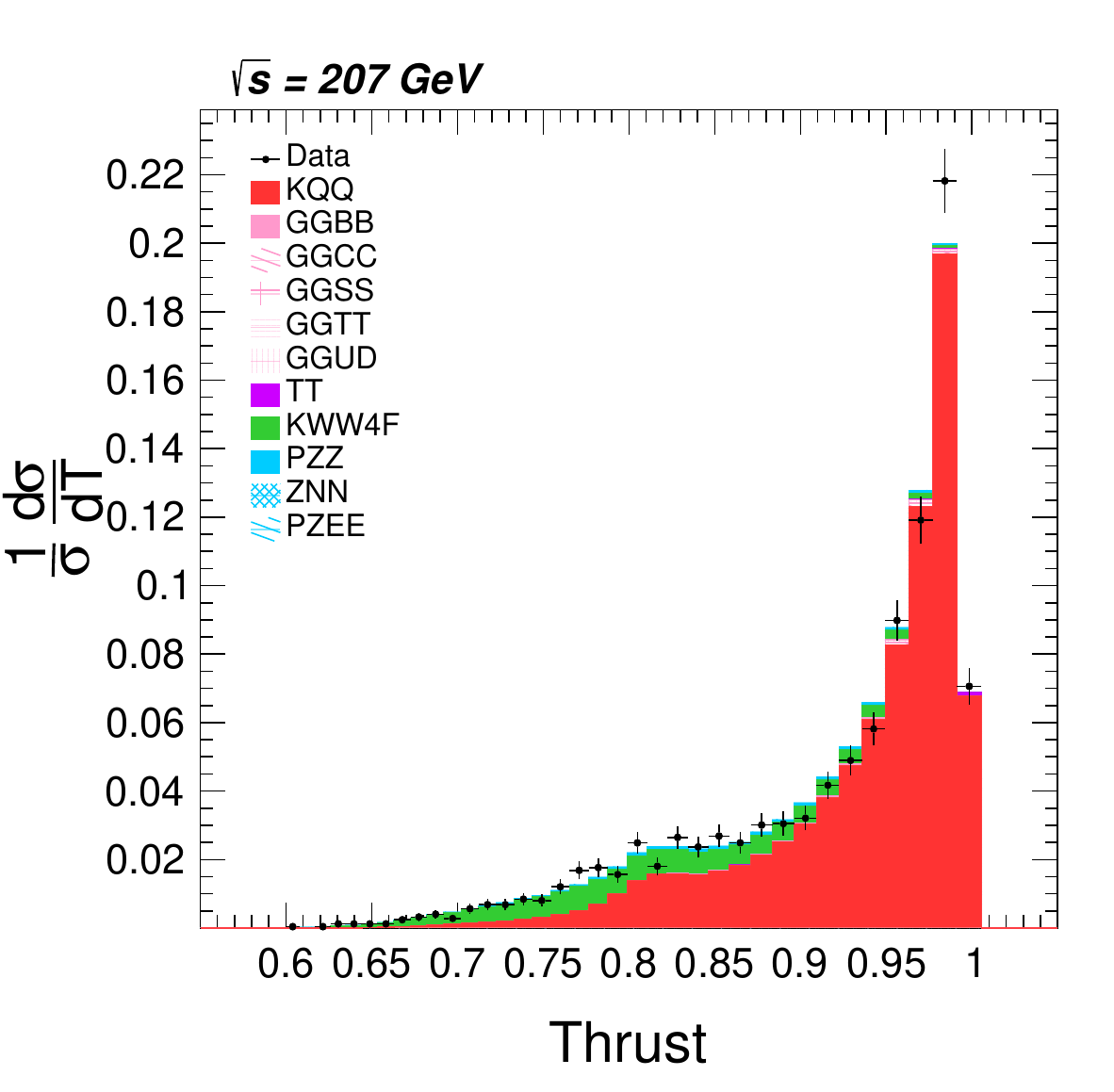}
        \caption{$\sqrt{s}=207$~GeV (Year 2000)}
    \end{subfigure}
\caption{The thrust distributions for high-energy events from different years. 
Data are presented as black error bars; stacked histograms show the decomposition of different simulated processes scaled according to the calculated cross-sections.
}
\label{fig:Thrust}
\end{figure}
\begin{figure}[ht]
\centering
     \begin{subfigure}[b]{\textwidth}
     \centering
        \includegraphics[width=0.40\textwidth]{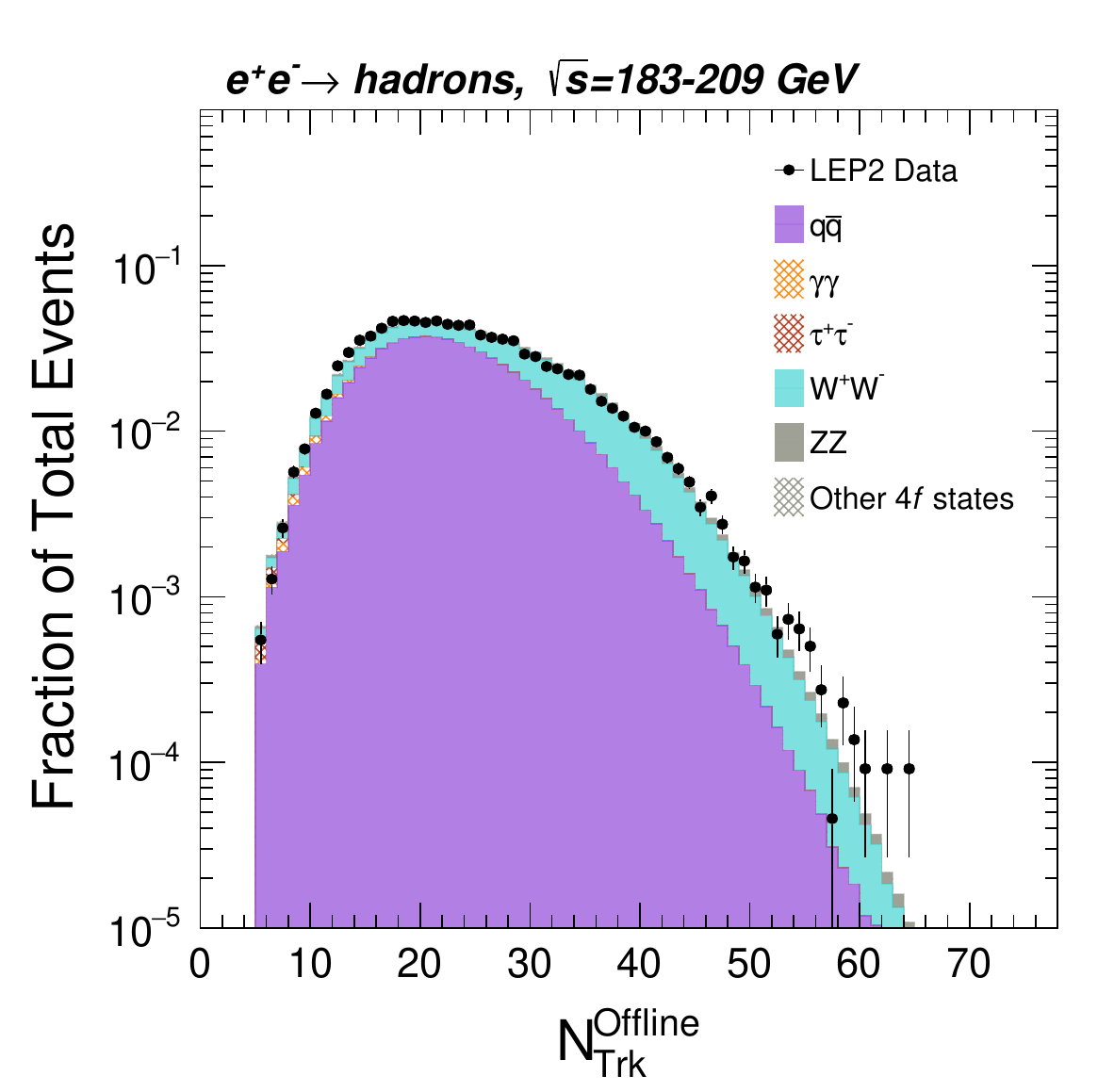}
        \includegraphics[width=0.40\textwidth]{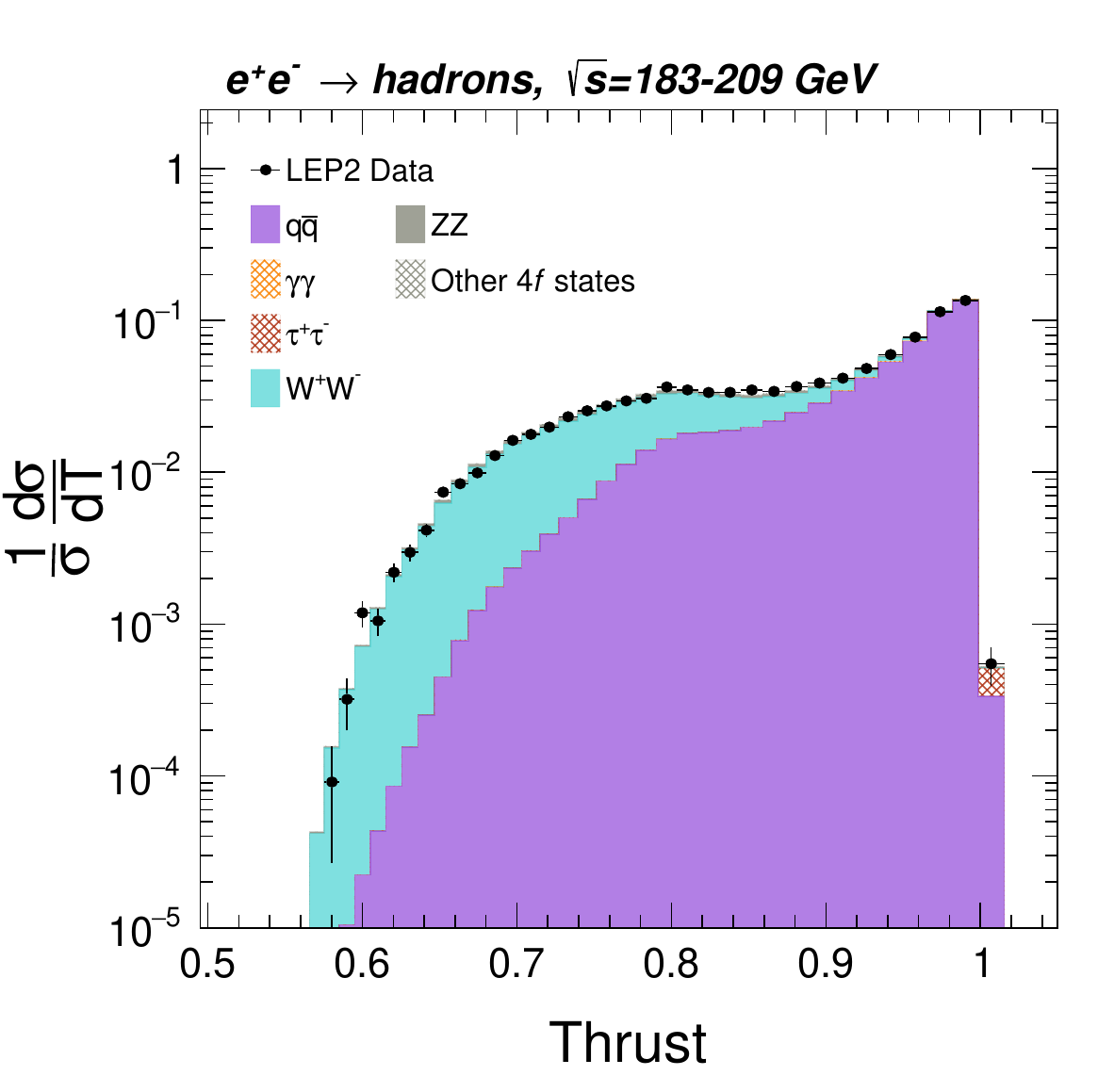}
        \caption{Merged LEP-II sample}
    \end{subfigure}
\caption{The multiplicity and thrust distributions of high-energy events of years 1997-2000. 
Data are presented as black error bars; stacked histograms show the decomposition of different simulated processes scaled according to the calculated cross-sections.
}
\label{fig:HighEnergySampleBySources}
\end{figure}

\begin{figure}[ht]
\centering
     \centering
        \includegraphics[width=0.40\textwidth]{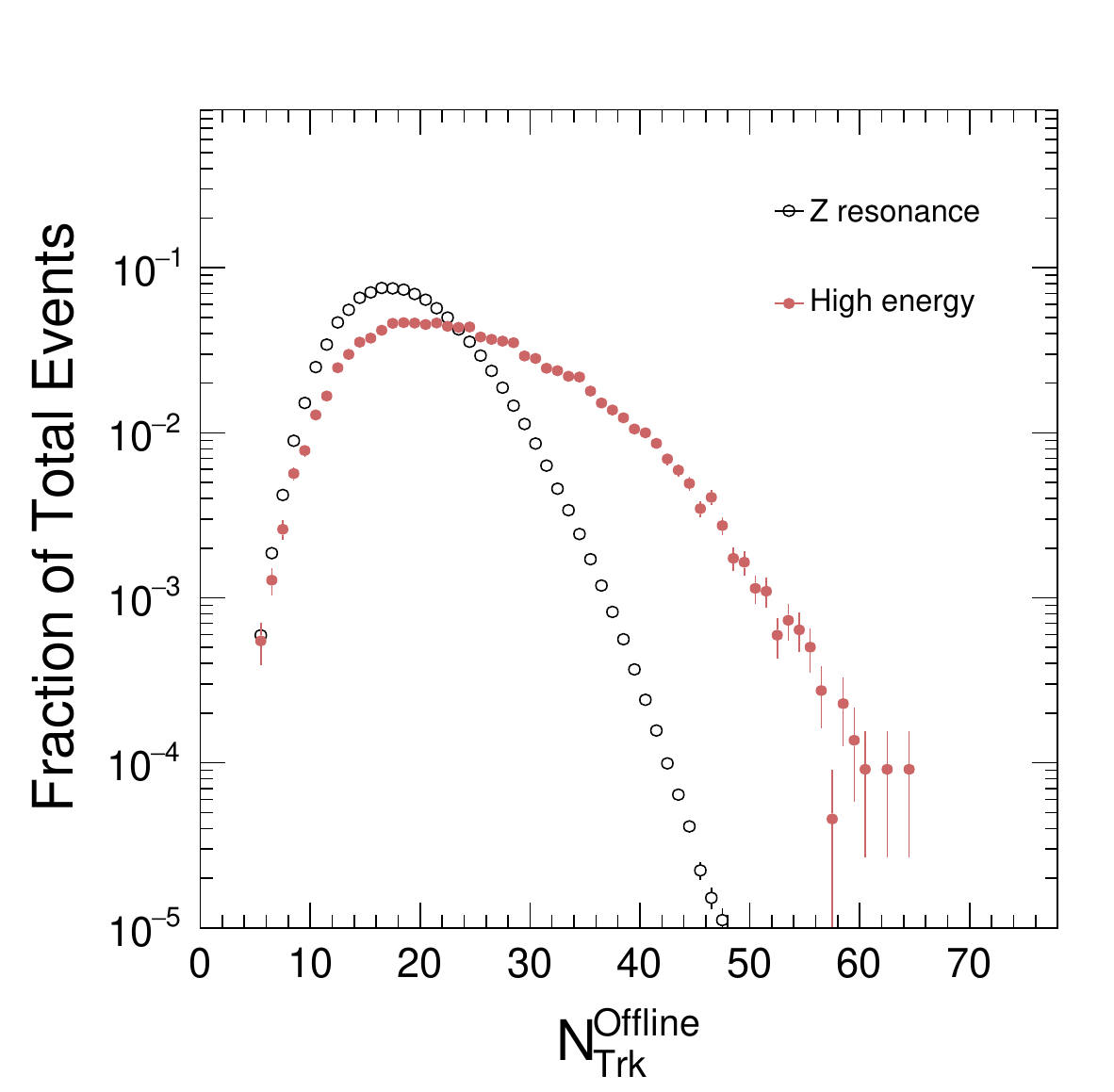}
        \includegraphics[width=0.40\textwidth]{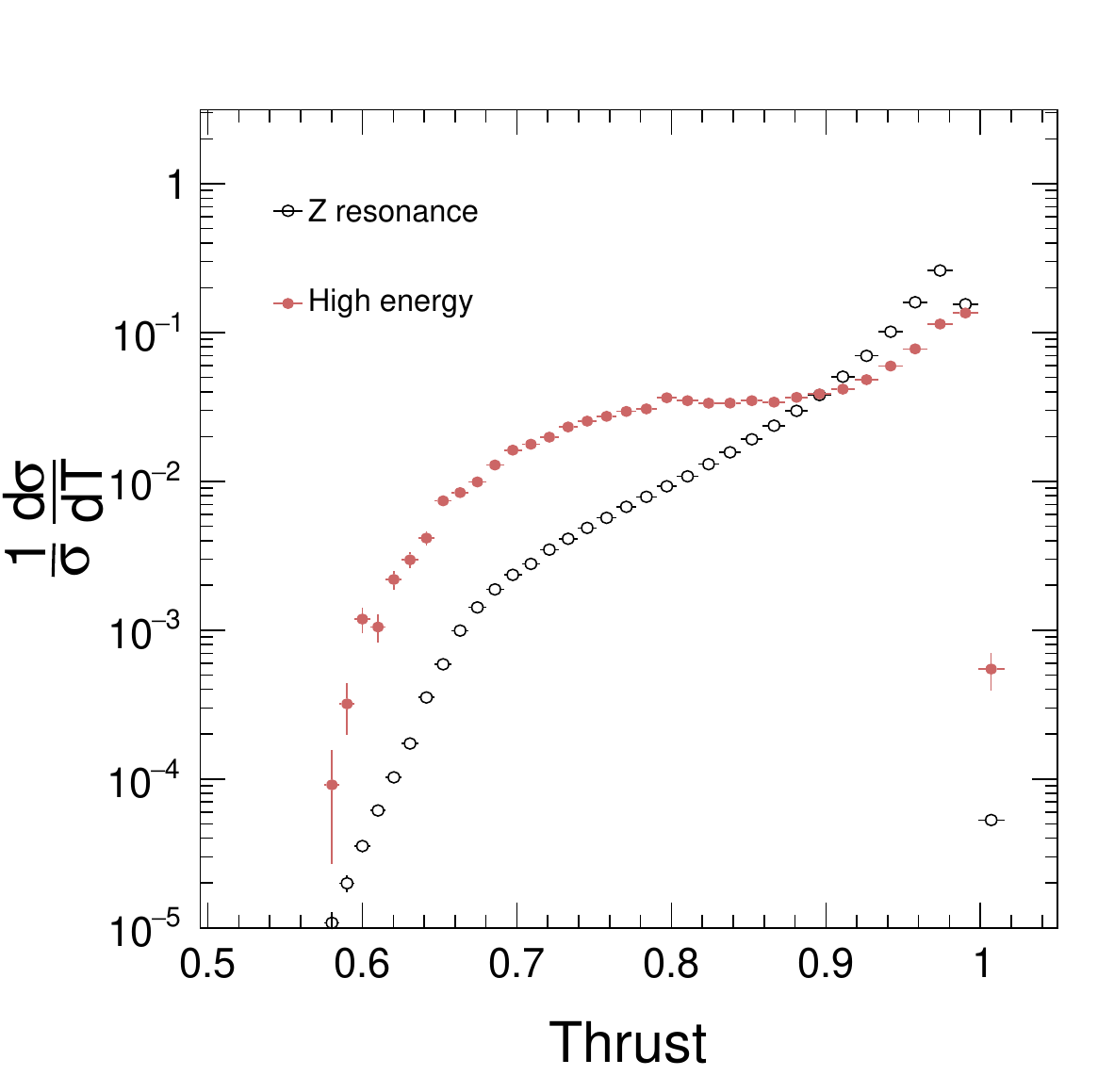}
\caption{The comparison of $Z$-resonance's and high-energy's multiplicity and thrust distributions. 
}
\label{fig:LEP1cfLEP2}
\end{figure}

\begin{table}[ht]
\caption{Average multiplicities and corrected multiplicities of different \ntrkoff intervals for the LEP-I and LEP-II datasets.}
\begin{center}
\footnotesize
\begin{tabularx}{\textwidth}{
    >{\centering}p{0.08\textwidth} | 
    >{\centering}p{0.12\textwidth}
    >{\centering}p{0.09\textwidth}
    >{\centering}p{0.09\textwidth} |
    >{\centering}p{0.08\textwidth} | 
    >{\centering}p{0.12\textwidth}
    >{\centering}p{0.09\textwidth}
    X
}
\hline\hline
\multirow{2}{*}{\parbox{0.09\textwidth}{\ntrkoff interval}} &
\multicolumn{3}{c|}{LEP-I} & 
\multicolumn{1}{c|}{\multirow{2}{*}{\parbox{0.09\textwidth}{\ntrkoff interval}}} &
\multicolumn{3}{c}{LEP-II} \\
\cline{2-4} \cline{6-8}
&  Fraction (\%) & $\left<{\rm N}_{\rm Trk}^{\rm Offline}\right>$ & $\left<{\rm N}_{\rm trk}^{\rm corr}\right>$ &
&   Fraction (\%) & $\left<{\rm N}_{\rm Trk}^{\rm Offline}\right>$ & $\left<{\rm N}_{\rm trk}^{\rm corr}\right>$ \\
\hline 
$[5,10)$        & 3.1 & 8.2 & 8.9 & $[10,20)$       & 58.6 & 15.2 & 17.3\\
$[10,20)$       & 59.2 & 15.2 & 15.8 & $[20,30)$       & 33.1 & 23.1 & 25.7 \\
$[20,30)$       & 34.6 & 23.1 & 23.4 & $[30,40)$       & 3.7 & 32.6 & 35.9 \\
$[30,\infty)$   & 3.1 & 32.4 & 32.6 & $[40,50)$       & 0.4 & 42.8 & 47.1 \\
$[35,\infty)$   & 0.5 & 36.9 & 37.2 & $[50,\infty)$   & $<$0.1 & 53.0 & 58.4\\
\hline\hline
\end{tabularx}
\label{tab:NtrkCorr}
\end{center}
\end{table} 

\subsection{Event thrust}
\label{sec:Thrust}
The event thrust axis~\cite{PhysRevLett.39.1587} is obtained through maximizing the sum of the projected particle momenta on itself in the center-of-mass frame, formulating as
\begin{equation}
\hat{n} \equiv max_{\hat{n}} \frac{\sum\nolimits_{i} \left| \Vec{p_i} \cdot \hat{n} \right|}{\sum\nolimits_{i} \left| \Vec{p_i} \right|},
\label{eqn:Thrust}
\end{equation}
where $\hat{n}$ is the resulting event thrust axis and $\Vec{p_i}$ is the momentum of the $i$-th particle, extending over charged, neutral particles and the missing transverse momentum. 

Neutral particles are energy flow candidates from the ECAL and HCAL. Requirements on the detector acceptance of $|\cos\theta| < 0.98$ and a cluster energy greater than 0.4~GeV are applied.
The invisible objects, such as neutrinos, and non-reconstructed particles at the reconstruction level, are taken into accounts as the missing momentum in the calculation of the event thrust, given by
\begin{equation}
\Vec{p}_{\rm MET} = -\sum\nolimits_{\rm neu, chg} \Vec{p}.
\label{eqn:MissP}
\end{equation}

In Figure~\ref{fig:HighEnergySampleBySources} we compare the offline multiplicity distributions and the thrust distributions for the merged LEP-II sample. 

The MC simulation provides an excellent description of the data. While the $q\bar{q}$ final states dominant the lower multiplicity range, $W^{+}W^{-}$ contribution increases with \ntrkoff and becomes the most significant contribution at high multiplicity.

Figure~\ref{fig:LEP1cfLEP2} shows the offline multiplicity and the thrust distributions in data collected at different collision energies. The high-energy sample provides a significantly higher multiplicity reach compared to the $Z$-resonance sample. The thrust distribution of the high-energy sample is also comparably more spherical as opposed to the $Z$-resonance sample due to the larger contribution of multi-jets in the final state in LEP-II than LEP-I.

\clearpage

\section{Two-particle correlation function Analysis}
\label{sec:TwoParticleCorrelationFunction}
Two-particle correlations are then calculated by following  heavy ion collisions and hadron collisions measurements, which have been introduced in ref.~\cite{Belle:2022fvl,The:2022lun} and given as
\begin{equation}
\begin{aligned}
\frac{1}{{\rm N}_{\rm trk}^{\rm corr}}\frac{d^2{\rm N}^{\rm pair}}{d\Delta\eta d\Delta\phi} &= C(\Delta \eta, \Delta \phi ) = B(0,0) \times \frac{S(\Delta\eta, \Delta\phi)}{B(\Delta\eta, \Delta\phi)},\\
\end{aligned}
\label{eqn:2PC}
\end{equation}
with ${\rm N}_{\rm trk}^{\rm corr}$ denotes the number of tracks after efficiency correction in the event, and ${\rm N}^{\rm pair}$ is the number of pairing yields associated with trigger particles.
The signal and background correlation functions are expressed as $S(\Delta\eta, \Delta\phi)$ and $B(\Delta\eta, \Delta\phi)$, counting the per-trigger-particle pairing yields within a single event and with a ``mixed event''~\cite{Khachatryan:2010gv}, formulated as 
\begin{equation}
\begin{aligned}
S(\Delta\eta, \Delta\phi) &= \frac{1}{{\rm N}_{\rm trk}^{\rm corr}}\frac{d^2{\rm N}^{\rm same}} {d\Delta\eta d\Delta\phi},\\
B(\Delta\eta, \Delta\phi) &= \frac{1}{{\rm N}_{\rm trk}^{\rm corr}}\frac{d^2{\rm N}^{\rm mix}}{d\Delta\eta d\Delta\phi}.
\end{aligned}
\label{eqn:2PCSigAndBkg}
\end{equation}
In this analysis, 48 events that are within the same \ntrkoff multiplicity class are used to form a mixed events. For events with $\ntrkoff$ greater than a high-multiplicity threshold ($\ntrkoff \ge 35$), we consider them in the same event-mixing pool, that mutual-mixing is allowed. The treatment is to avoid the resampling bias within a stringent high-multiplicity class. In this construction, the background correlation $B(\Delta\eta, \Delta\phi)$ counts the random combination correlation for the trigger track and the associated track that originate from different events. The $B(0,0)$ factor in eq.~\ref{eqn:2PC} is a normalization choice related with the artificially-constructed $B(\Delta \eta, \Delta \phi)$. For the two-particle correlation reported in the highest multiplicity interval ($\ntrkoff \ge 50$), 96 events are used for a mixed event generation. The more statistics is helpful for mitigating the statistical fluctuation in the background correlation function, and hence the ratio correlation function $C(\Delta \eta, \Delta \phi)$.

Following the previous two-particle correlation studies in \ee systems~\cite{Badea:2019vey,Belle:2022fvl,The:2022lun}, we analyze correlation functions with two reference coordinates -- the ``beam-axis'' and the ``thrust-axis'' coordinates. The former fixes the reference axis to the $z$-axis for all events, while the latter chooses for each event its thrust axis, which approximates the outgoing $q\bar{q}$ direction at the reconstruction level, as the reference axis. In the thrust-axis analysis, the new reference $z$-axis is the thrust axis $\hat{n}$, and the choice of $\phi=0$ (or new reference $x$-axis) is assigned with $\hat{n}\times(\hat{n}\times\hat{z})$.
Particles' pseudorapidity and azimuth positions are then re-defined with respect to the new reference frame, yielding ($\eta_T, \phi_T$) that are used for calculating two-particle angular separations in the thrust-axis analysis.
The thrust axis coordinate is a more intuitive reference frame to understand the fragmenting activity surrounding the outgoing leading dijet direction in the \ee system. 

However, since the outgoing direction of each event is randomly orientated, a thrust-mixing reweighting correction on the event topology ($\eta_T, \phi_T$) is adopted in order to match the mixed events' fragmenting topology to the physical ones.
The reweighting factor is obtained from a histogram division of the physical spectrum by the mixed-event spectrum and applied as track weights, written out as
\begin{equation}
m(|\eta_T|, \phi_T, {\rm N}_{\rm Trk}^{\rm Offline}) = \left[\frac{d^2{\rm N}^{\rm phys}}{d|\eta_T| d\phi_T}/\frac{d^2{\rm N}^{\rm mix}}{d|\eta_T| d\phi_T}\right]_{{\rm N}_{\rm Trk}^{\rm Offline}}.
\label{eqn:ThrustMixingCorrection}
\end{equation}
This ensures that the event topology of mixed events conform to that of physical events, and leaves the pairing of the trigger particle with tracks from the mixed event uncorrelated at the same time. 

To study the correlation function structures in greater details, we explore the azimuthal differential associated yields (one-dimensional correlations in \dphi) 
\begin{equation}
Y(\Delta\phi) = \frac{1}{{\rm N}_{\rm trk}^{\rm corr}}\frac{d{\rm N}^{\rm pair}}{d\Delta\phi}= \frac{1}{\Delta\eta_{\rm max}-\Delta\eta_{\rm min}}\int\limits_{\Delta\eta_{\rm min}}^{\Delta\eta_{\rm max}} \frac{1}{{\rm N}_{\rm trk}^{\rm corr}}\frac{d^2{\rm N}^{\rm pair}}{d\Delta\eta d\Delta\phi}d\Delta\eta,
\label{eqn:DeltaPhiAssociatedYield}
\end{equation}
under the specific range of $\Delta\eta$ projection.
Three $\Delta\eta$ regions are studied: short range ($0 \le |\Delta\eta| < 1$), middle range ($1 \le |\Delta\eta| < 1.6$) and long range ($1.6 \le |\Delta\eta| < 3.2$). The long-range azimuthal differential associated yield is denoted as $Y_l(\Delta \phi)$ in the discussions that follow. 

In this work, we study two-particle correlations in different multiplicity classes, as outlined in Table~\ref{tab:NtrkCorr}. We also explore two-particle correlations with requirements on the transverse momenta $p_T$ of both particles. The trigger particle and the associated particle considered in eq.~\ref{eqn:2PC} are therefore required to fall within the same $p_T$ class: $[0, 0.5)$, $[0.5, 1)$, $[1,2)$, $[2,3)$ and $[3, \infty)$ (in GeV).



\subsection{Detector effect corrections}
\label{sec:Corrections}

\subsubsection*{Tracking efficiency correction}
\label{sec:EfficiencyCorrection}

Using the MC sample introduced in Section~\ref{sec:Sample}, we study the reconstruction effects caused by nonuniform detection efficiency and mis-reconstruction bias. A reweighting factor is applied to the reconstructed tracks; this factor is the inverse of the tracking efficiency. This adjustment brings the spectra of the reconstructed tracks closer to reflecting the geometrical acceptance of the generator. The efficiency is given by
\begin{equation}
\varepsilon(p_{\rm T}, \theta, \phi, {\rm N}_{\rm Trk}^{\rm Offline}) = \left[\frac{d^3 {\rm N}^{\rm reco}}{dp_{\rm T} d\theta d\phi}/\frac{d^3 {\rm N}^{\rm gen}}{dp_{\rm T} d\theta d\phi}\right]_{{\rm N}_{\rm Trk}^{\rm Offline}},
\end{equation}
where \( {\rm N}^{\rm reco} \) denotes the number of charged particles at the reconstruction level, and \( {\rm N}^{\rm gen} \) denotes the same at the generator level. The efficiency correction factor establishes a correspondence between the reconstruction level and the generator level across the \( p_{\rm T}, \theta, \) and \( \phi \) spectra.

\subsubsection*{Residual MC correction}
\label{sec:residMCCorr}

After corrections are made as described above, remaining possible reconstruction effects are handled with the bin-by-bin correction method~\cite{Choudalakis:2011rr}.
The correction factor is derived from the ratio of MC correlation functions at the reconstruction and generator level and is defined as ${\rm C}(\Delta \phi) = \frac{Y(\Delta \phi)_{{\rm gen}, i_g}}{Y(\Delta \phi)_{{\rm reco}, i_r}}$, where indices $i_g$ and $i_r$ are \ntrkoff bins counted at the generator and reconstruction level, respectively.
Final data correlation results are obtained by multiplying the original correlation function by the bin-by-bin correction factor.

\subsection{Bayesian analysis}

The analysis employs Bayesian inference to detail the statistical uncertainties of both correlation yields and flow coefficients, which will be elaborated upon in Section~\ref{sec:flow}. We have verified that the Bayesian estimation utilized in this analysis is unbiased, with reported central values aligning with observed values. The primary rationale behind adopting the Bayesian analysis is to offer a more detailed estimation of uncertainties, in particular when assuming a Gaussian distribution is not ideal for a data set that features a non-Gaussian distribution. In this way, we will take the non-Gaussian-like tail into account.

We construct the posterior probability $\mathcal{P}(\mu_{x}|x)$ for each observed bin value in the signal and background correlation functions (Eqn.~\ref{eqn:2PCSigAndBkg}) with Bayes' theorem
\begin{equation}
\mathcal{P}(\mu_{x}|x) = \frac{\mathcal{P}(x|\mu_{x})}{\mathcal{P}(x)} \mathcal{P}(\mu_{x}),
\label{eqn:BayesTheorem}
\end{equation}
where the random variable $x$ represents the observable of interest; in this case it is the bin value in a correlation function.
We consider the likelihood function for each bin distribution $\mathcal{P}(x|\mu_{x})$ as a ``weighted Poisson distribution~\cite{Bohm:2013gla},'' 
since the bin value ($x$) is obtained by weighting the count of pairing yields ($n$) with an efficiency correction factor ($w$), where $n$ follows the Poisson distribution with a Poisson mean $\mu$
\begin{equation}
n \sim \mathcal{P}_{N}(n|\mu) = \frac{e^{-\mu} \mu^n}{n!}.
\label{eqn:pois}
\end{equation}
The relation between weighted and unweighted correlation yields is linear rescaling, $x = w n$. Therefore, the likelihood function of $x$ given $\mu$ can be easily obtained by a transformation of variables
\begin{equation}
\mathcal{P}_{X}(x|\mu) = \frac{1}{w} \mathcal{P}_{N}(x/w|\mu) = \frac{1}{w} \frac{e^{-\mu} \mu^{(x/w)}}{(x/w)!},
\label{eqn:pois_x_mu}
\end{equation}
or, if expressed using the ``weighted truth parameter'' $\mu_{x} = w \mu$ (in this case $\mu_{x}$ is the truth parameter of the weighted Poisson distribution), one gets
\begin{equation}
\mathcal{P}_{X}(x|\mu_x) = \frac{1}{w} \frac{e^{-(\mu_x/w)} (\mu_x/w)^{(x/w)}}{(x/w)!}.
\label{eqn:pois_x_mu2}
\end{equation}
For large $n~(n \ge 10)$, to boost the calculation speed, we approximate the Poisson distribution as a Gaussian.
In Eqn.~\ref{eqn:BayesTheorem}, we adopt a uniform prior $\mathcal{P}(\mu_{x})$ in this parameter estimation. The denominator $\mathcal{P}(x)$ is the normalization factor.

The obtained posterior probability $\mathcal{P}(\mu_{x}|x)$ of a signal or background correlation function bin value is used for toy-sample generation, whereby the more complicated probability distributions of other parameters, such as the ratio correlation yields $C(\Delta \eta, \Delta \phi)$ and the long-range yields $Y(\Delta \phi)$, can be derived with their operational definitions. For every signal and background correlation function bin, we generate a pseudo dataset of 50k $\mu_x$'s, thus 50k toy signal and background correlation functions. The toy-sample generations of individual bins are independent of one another. We further impose a constraint, requiring the toy correlation function to conserve the total pairing yields: $\int S(\Delta \eta, \Delta \phi )d \Delta \eta d\Delta \phi = {\rm const.}_{\rm sig.}$ and $\int B(\Delta \eta, \Delta \phi )d \Delta \eta d\Delta \phi = {\rm const.}_{\rm bkg.}$, where the constants are set using the observed data.

For each toy $S(\Delta \eta, \Delta \phi )$ and $B(\Delta \eta, \Delta \phi )$, we calculate using the Eqn.~\ref{eqn:2PC} to obtain the toy ratio correlation function $C(\Delta \eta, \Delta \phi )$, and using the Eqn.~\ref{eqn:DeltaPhiAssociatedYield} to obtain the one-dimensional azimuthal differential associated yields $Y(\Delta \phi)$. With these 50k 2-D and 1-D correlation toy samples, we estimate the truth parameter (the correlation yield) via the ``maximum a posteriori (MAP)'' method
\begin{equation}
\hat{\mu_x}_{\rm MAP}(\mu_x) = \underset{\mu_x}{\mathrm{arg\,max}}~\mathcal{P}(\mu_x | x).
\label{eqn:map}
\end{equation}
We scan for the maximum of the posterior probability $\mathcal{P}(\mu_x | x)$ within the 10-sigma-deviation interval around the observed value $x$: $[x-10\sigma_x, x+10\sigma_x]$. Dividing this interval by 1000 steps is the scanning granularity.
The associated 1-$\sigma$ uncertainty range of the estimated parameter, or the credible interval $[{\mu_x}_L, {\mu_x}^U]$, is determined such that it covers 68\% of the cumulative distribution function $F_{\pmb{\mu_x}}(\mu_x) = \mathcal{P}({\pmb{\mu_x}} \le \mu_x | x)$, expressed as
\begin{equation}
\int_{{\mu_x}_L}^{{\mu_x}^U} \mathcal{P}(\mu_x | x) d \mu_x = 68\%, \quad {\rm or} \quad F_{\pmb{\mu_x}}({\mu_x}^U) - F_{\pmb{\mu_x}}({\mu_x}_L) = 68\%.
\label{eqn:credibleInterval}
\end{equation}
For the case that the estimated $\hat{\mu_x}_{\rm MAP}$ is close to the boundary of the posterior function, we quote the one-sided interval as the uncertainty range
\begin{equation}
\def\arraystretch{1.}
\left\{ \begin{array}{l}
F_{\pmb{\mu_x}}({\mu_x}^U) = 68\%, \\
F_{\pmb{\mu_x}}({\mu_x}_L) = 0\%
\end{array} \right., \quad {\rm or} \quad 
\left\{ \begin{array}{l}
F_{\pmb{\mu_x}}({\mu_x}^U) = 100\%, \\
F_{\pmb{\mu_x}}({\mu_x}_L) = 32\%.
\end{array} \right.
\label{eqn:credibleInterval_oneSided}
\end{equation}
This happens, for example, when one observes zero yield ($x=0$), the estimated $\hat{\mu_x}_{\rm MAP}$ will be close to 0; thus, the one-sided interval will be reported.
Otherwise, the uncertainty range $[{\mu_x}_L, {\mu_x}^U]$ is determined with
\begin{equation}
\def\arraystretch{1.}
\left\{ \begin{array}{l}
F_{\pmb{\mu_x}}({\mu_x}^U) = F_{\pmb{\mu_x}}(\hat{\mu_x}_{\rm MAP}) + 34\%, \\
F_{\pmb{\mu_x}}({\mu_x}_L) = F_{\pmb{\mu_x}}(\hat{\mu_x}_{\rm MAP}) - 34\%.
\end{array} \right.
\label{eqn:credibleInterval_twoSided}
\end{equation}


\subsection{Results}
\label{sec:Rst}
\begin{figure}[ht]
\centering
\includegraphics[width=.45\textwidth]{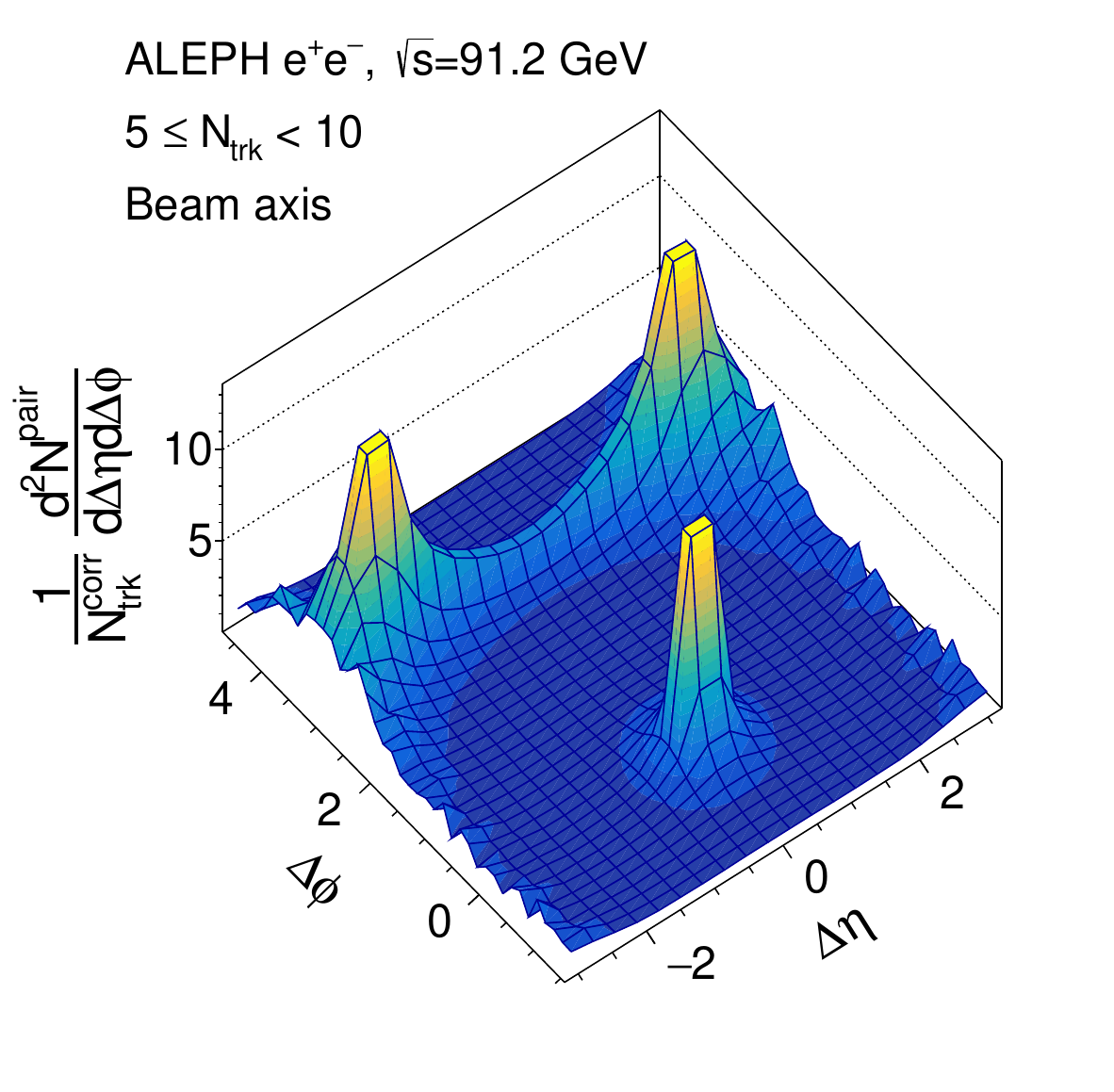}
\includegraphics[width=.45\textwidth]{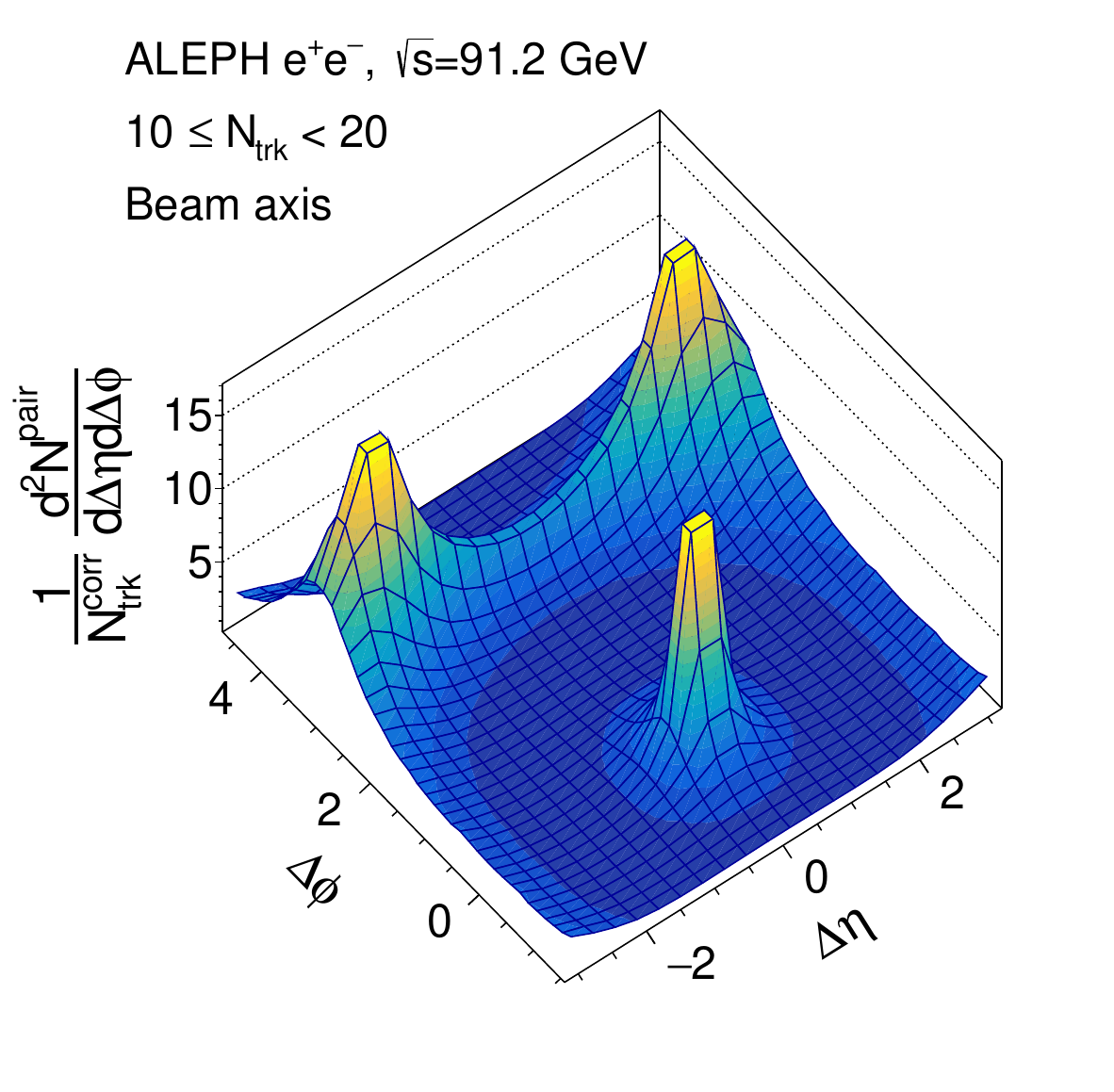}
\includegraphics[width=.45\textwidth]{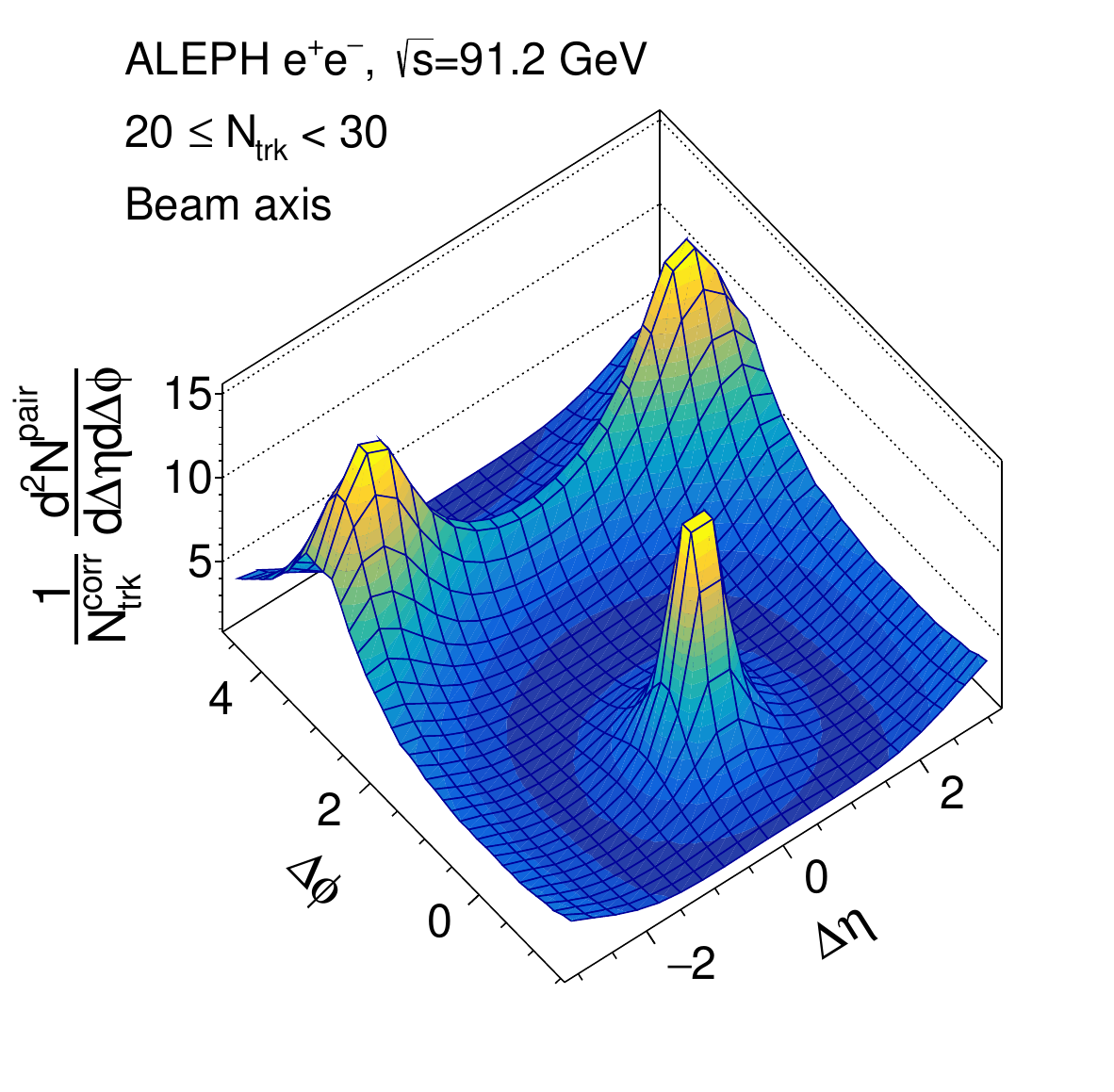}
\includegraphics[width=.45\textwidth]{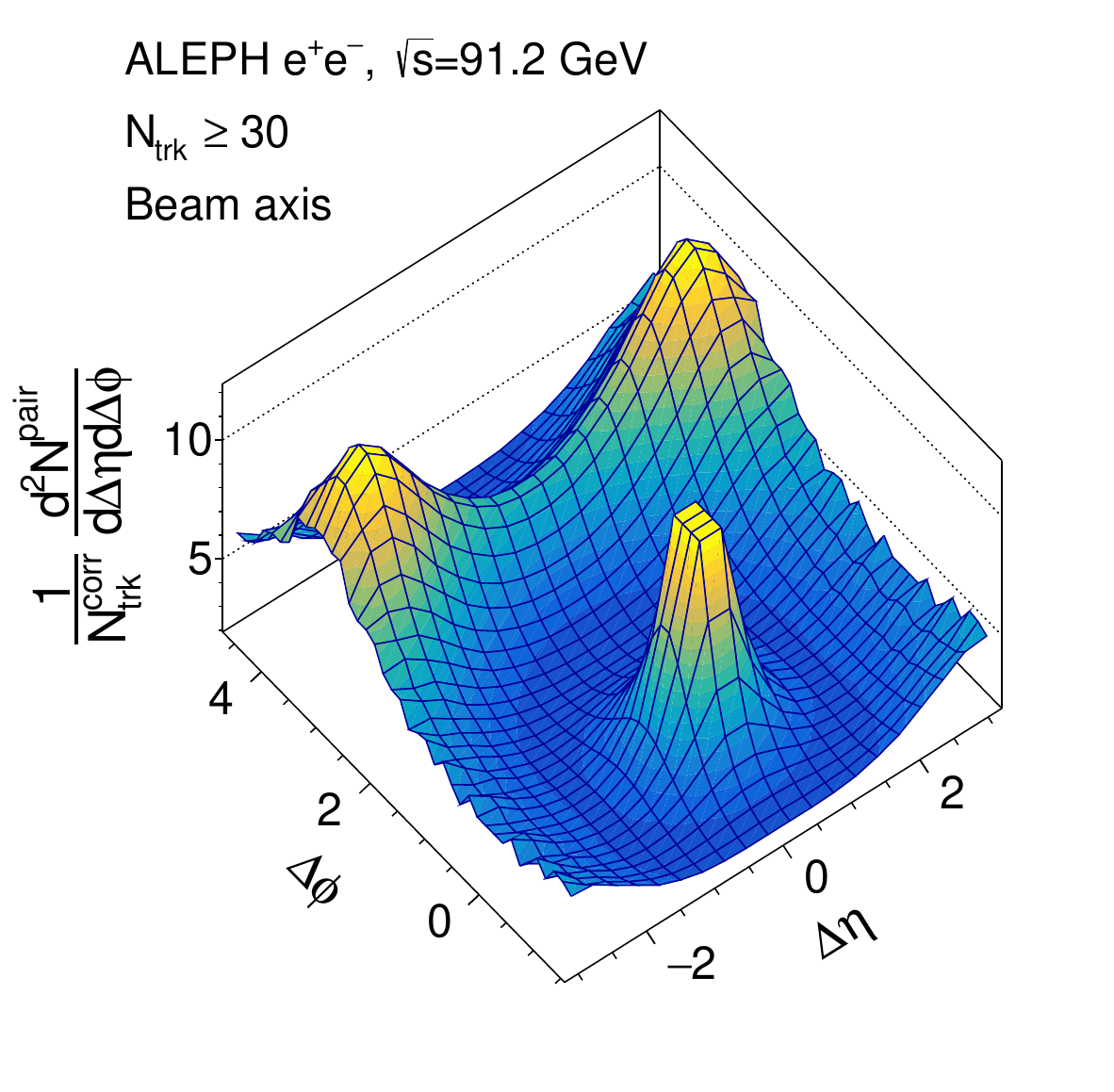}
\includegraphics[width=.45\textwidth]{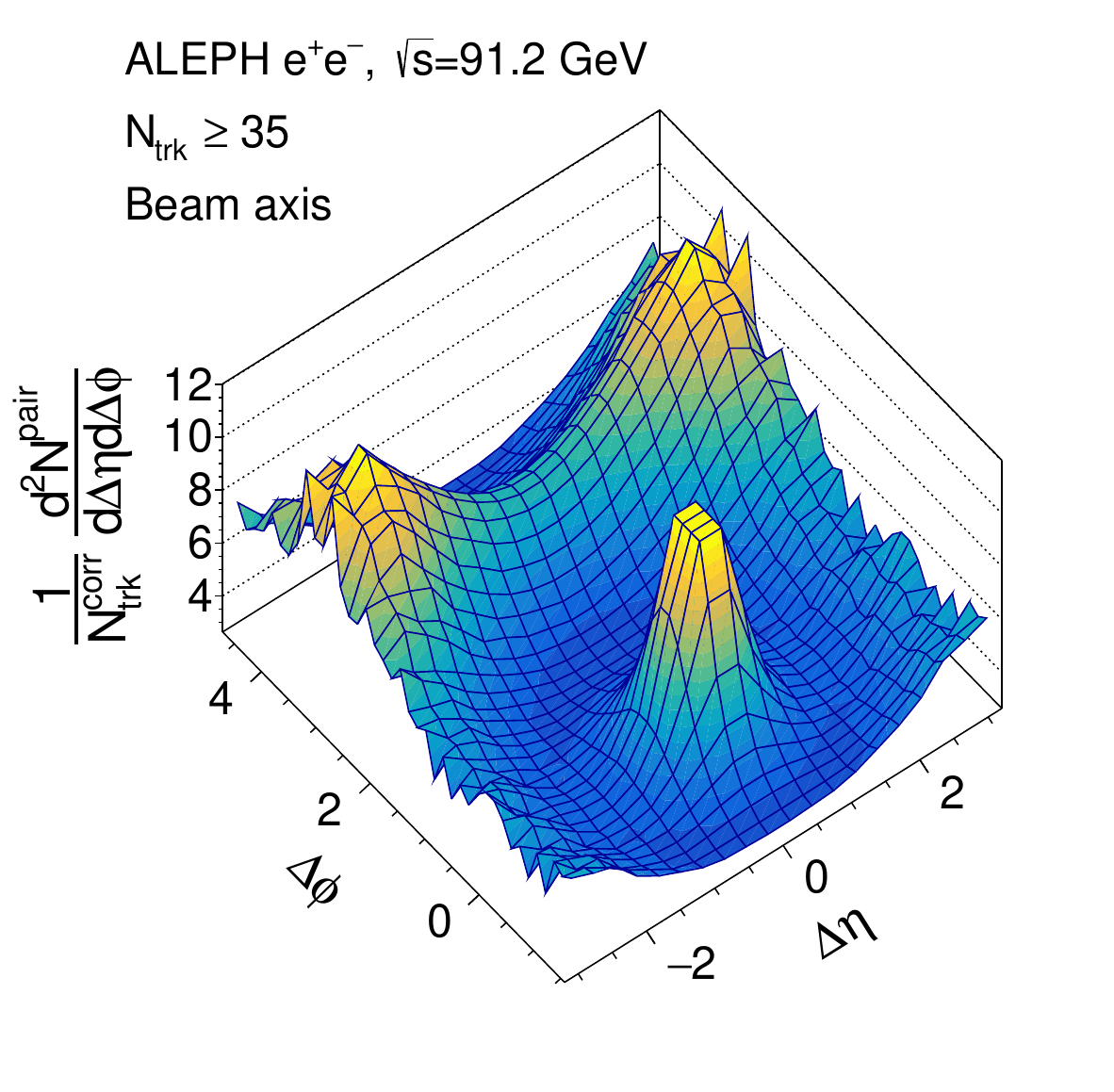}
\includegraphics[width=.45\textwidth]{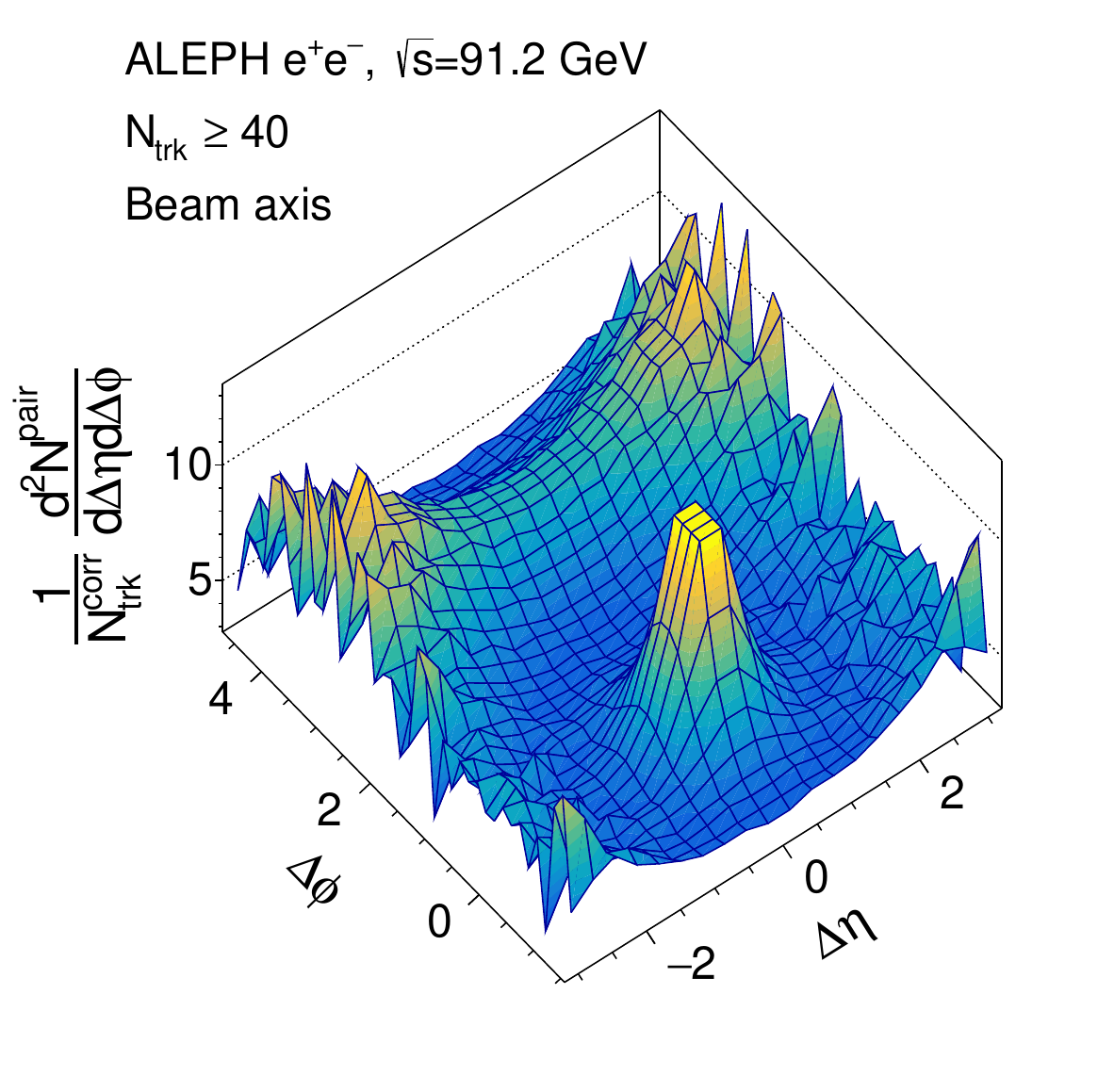}

\caption{Two-particle correlation function for the $Z$-resonance sample, analyzed along the beam axis with aggregated LEP-I and LEP-II statistics. The results are delineated based on the offline multiplicity in intervals such as $[5,10), [10,20), [20,30), [30,\infty), [35,\infty), [40,\infty)$.}
\label{fig:EnergyCut_le_100_Combined_beam}
\end{figure}

\begin{figure}[ht]
\centering
\includegraphics[width=.45\textwidth]{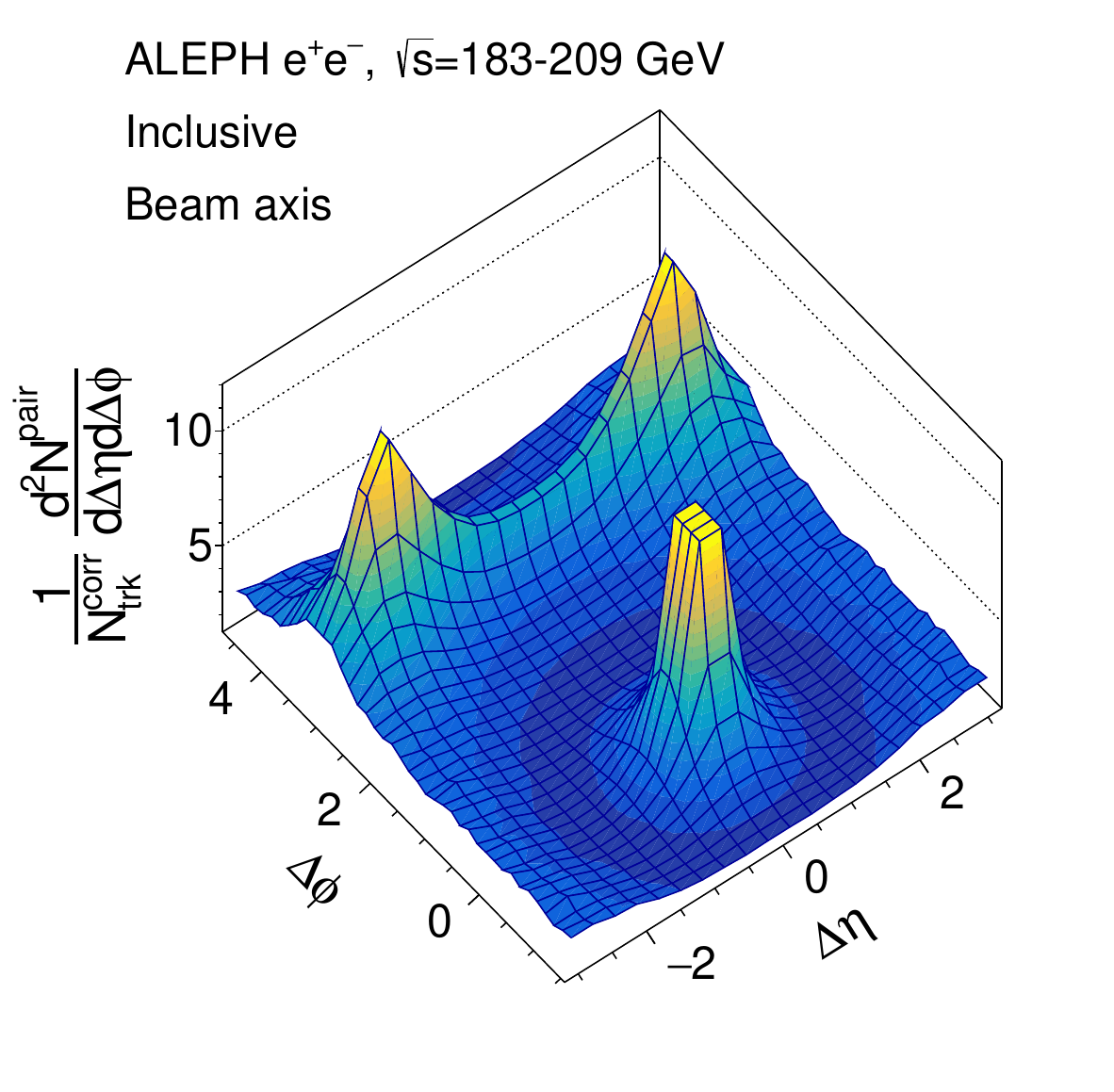}
\includegraphics[width=.45\textwidth]{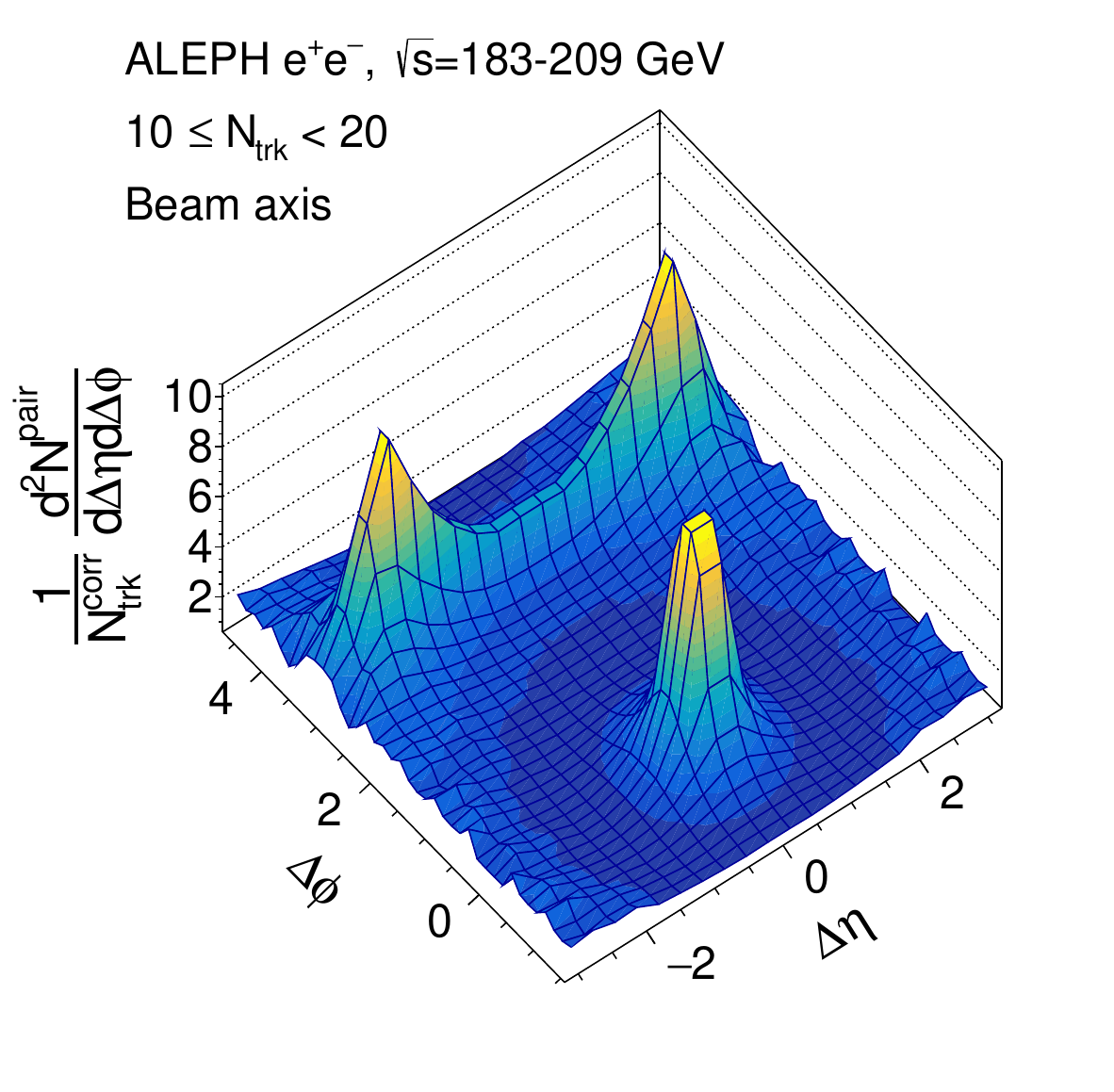}
\includegraphics[width=.45\textwidth]{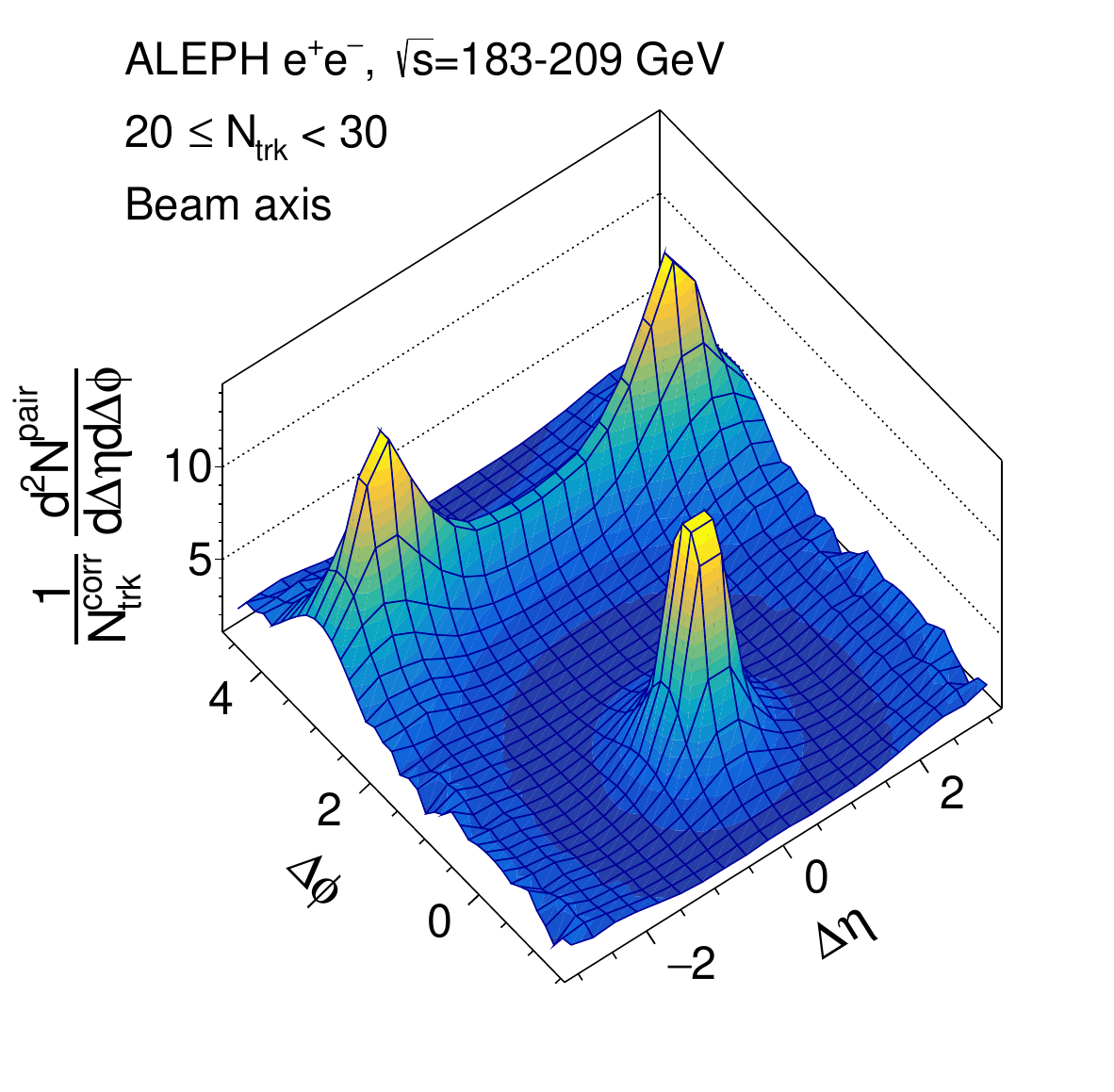}
\includegraphics[width=.45\textwidth]{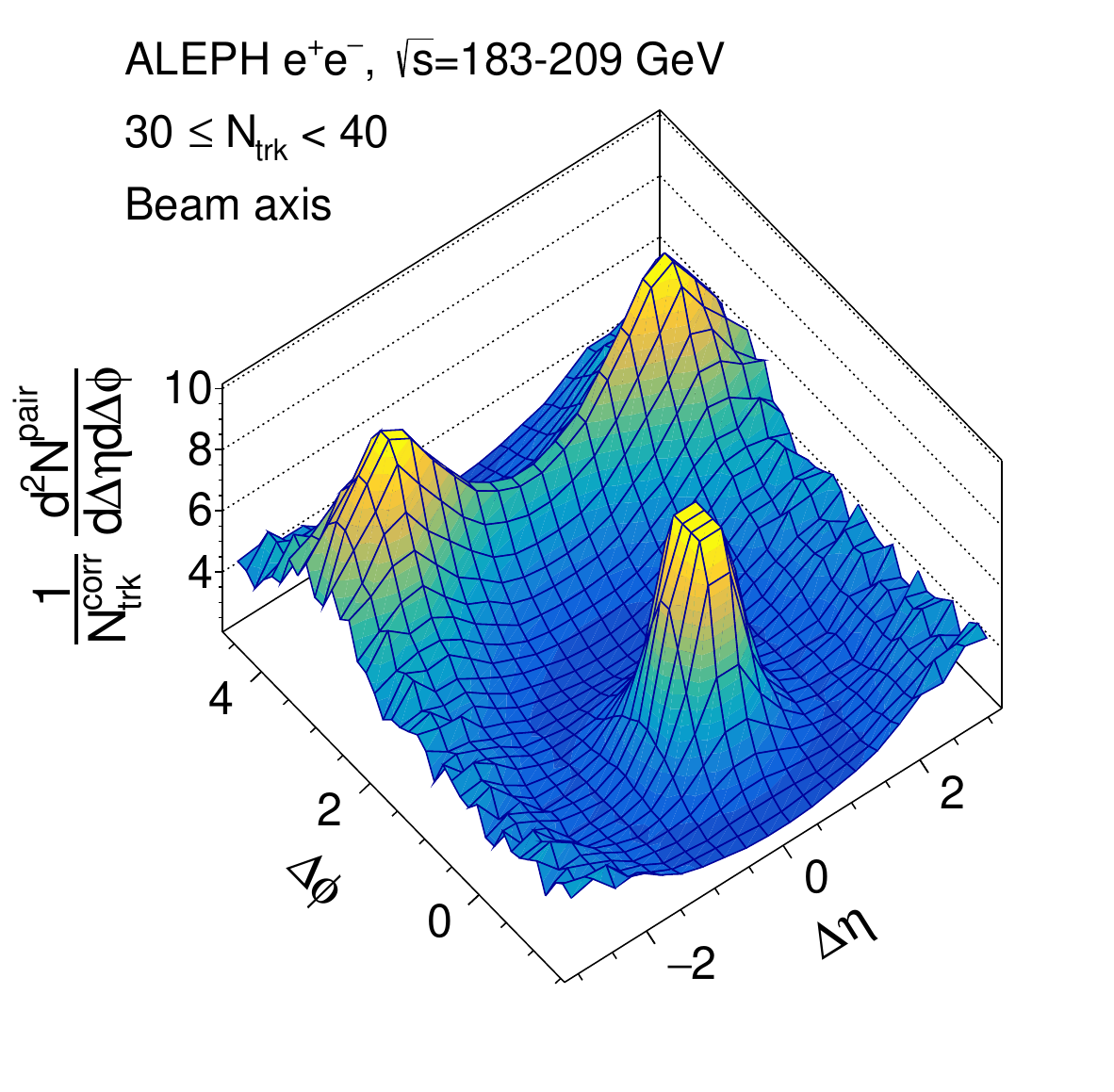}
\includegraphics[width=.45\textwidth]{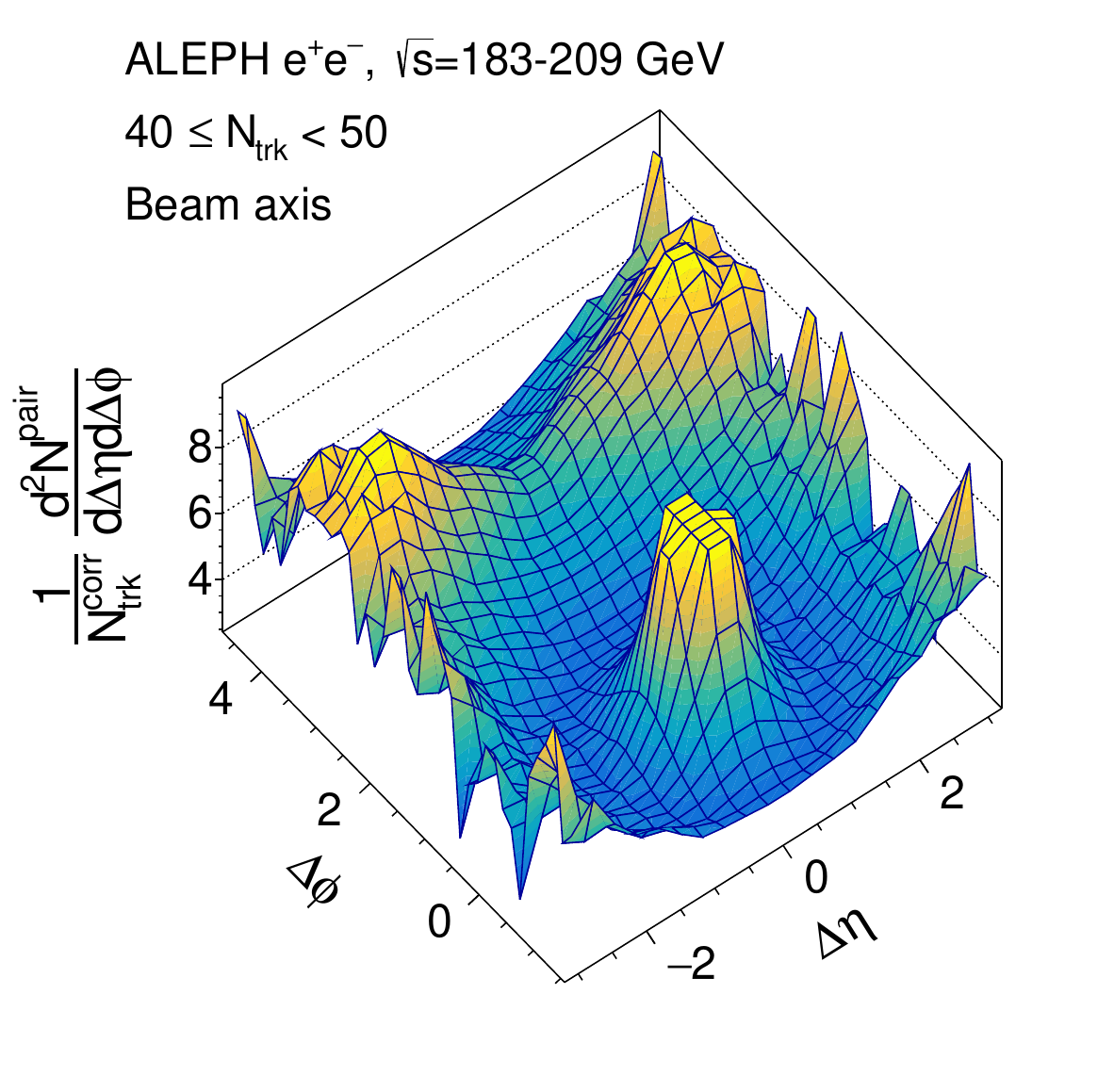}
\includegraphics[width=.45\textwidth]{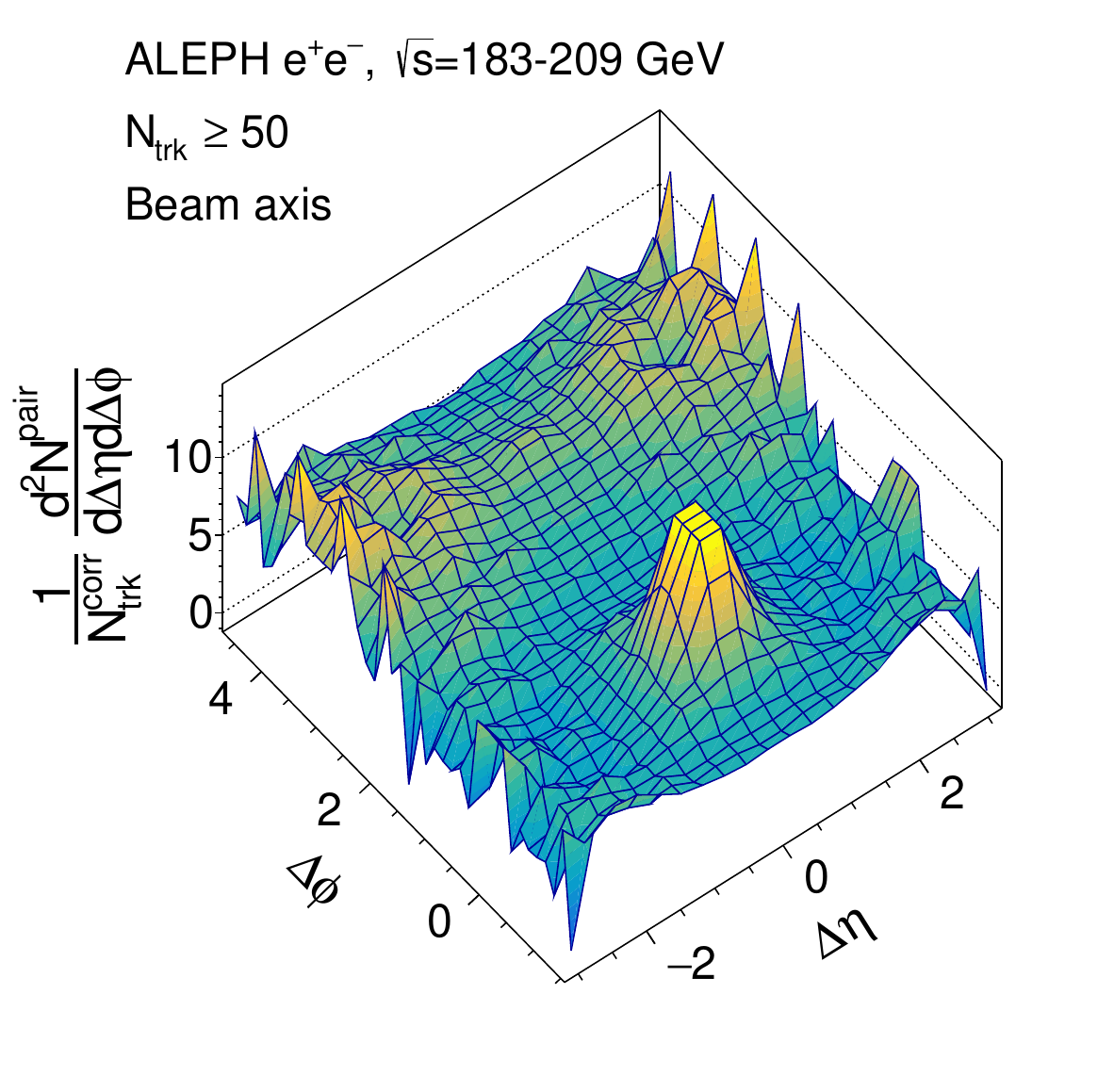}
\caption{Two-particle correlation function for high-energy sample in the beam axis analysis. Results are shown multiplicity-inclusively and as a function of the offline multiplicity in $[10,20), [20,30), [30,40), [40,50), [50, \infty)$ intervals.}
\label{fig:EnergyCut_ge_100_beam}
\end{figure}

In Figures~\ref{fig:EnergyCut_le_100_Combined_beam} and~\ref{fig:EnergyCut_ge_100_beam}, we present the beam-axis two-particle correlations for the $Z$-resonance and high-energy data, respectively, plotted as a function of \ntrk. The observed correlation structure in the beam axis analysis can be elucidated with the following decomposition:

\begin{itemize}
\item The \textit{origin peak} correlation, localized around $(\Delta\eta, \Delta\phi)=(0,0)$, primarily stems from collinear track pairs originating from the same jet.
\item The \textit{away-side} correlation, extended along $\Delta\phi\approx\pi$, is due to back-to-back momentum balancing. Notably, this correlation in $e^+e^-$ annihilation displays an upward tilt towards a larger \deta value, diverging from the away-side pattern in heavy-ion or \pp collisions where the away-side is more localized at $\deta\approx0$. Given that the annihilation products of the \ee collision don't strictly align with the beamline direction, there's an elevated likelihood of encountering back-to-back emitted final states with a pronounced pseudorapidity difference.
\item The \textit{ridge} correlation appears around $\dphi\approx0$ and extends over a broad \deta range~\cite{ALICE:2012eyl}. This \textit{ridge}-like feature is not observed in the low multiplicity $e^+e^-$ data when using the beam axis as a reference.
\end{itemize}


To provide deeper insights, we also inspected the long-range one-dimensional $\Delta\phi$ correlation functions with $1.6\leq|\Delta\eta|<3.2$. This selection facilitates a more straightforward comparison between the data and the archived ALEPH MC.

\begin{figure}[ht]
\centering
    \includegraphics[width=0.45\textwidth]{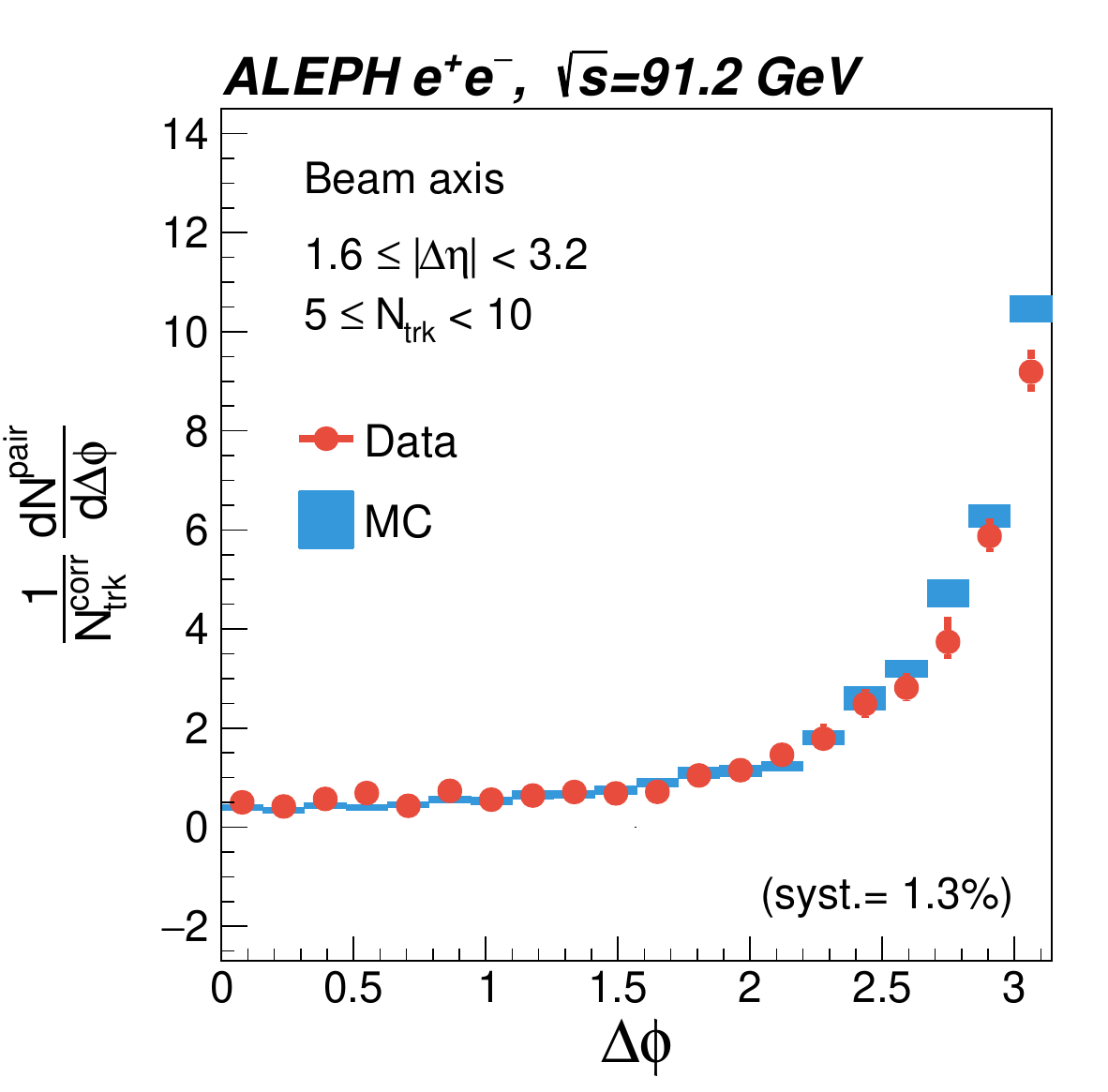}
    \includegraphics[width=0.45\textwidth]{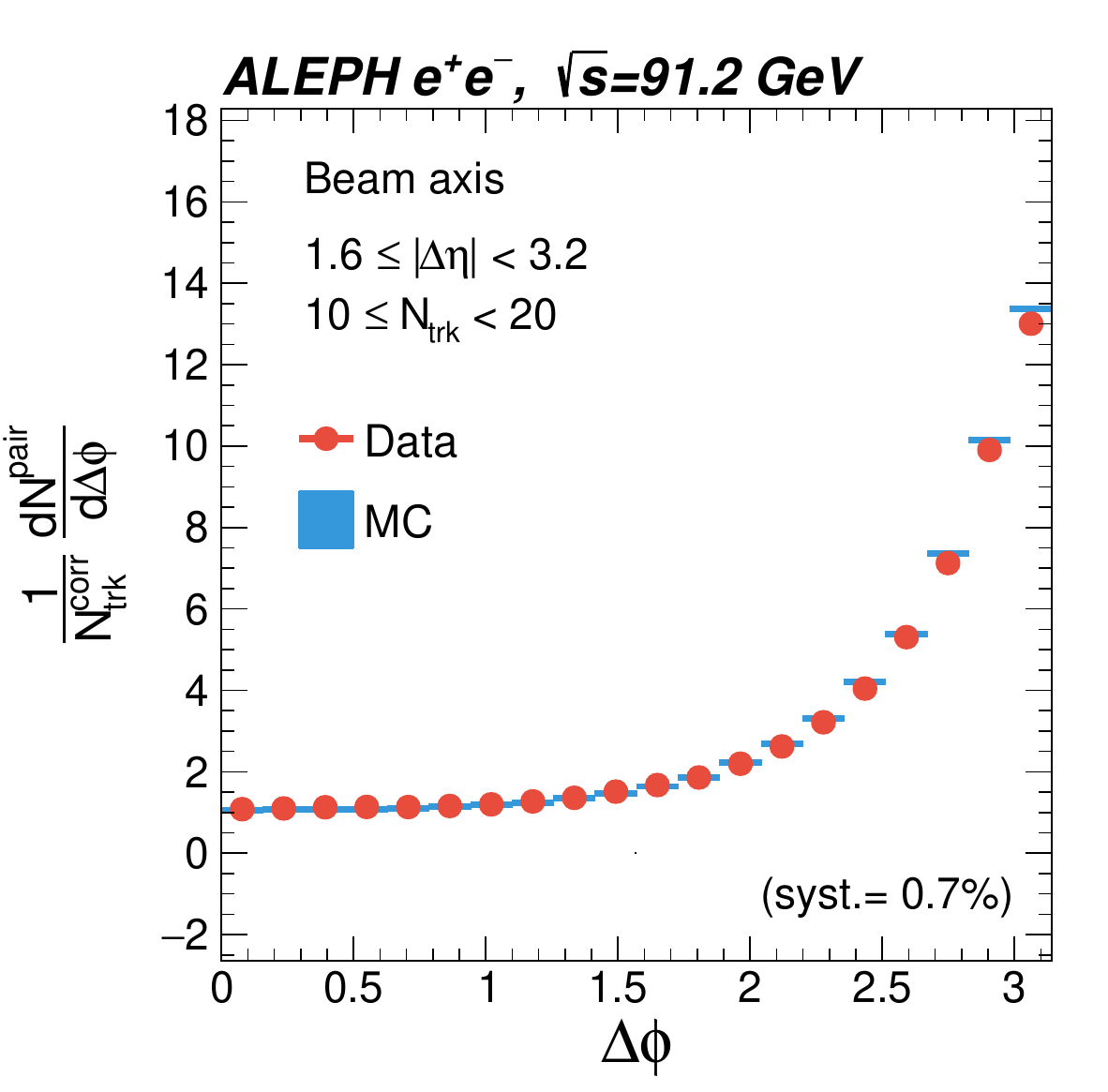}
    \includegraphics[width=0.45\textwidth]{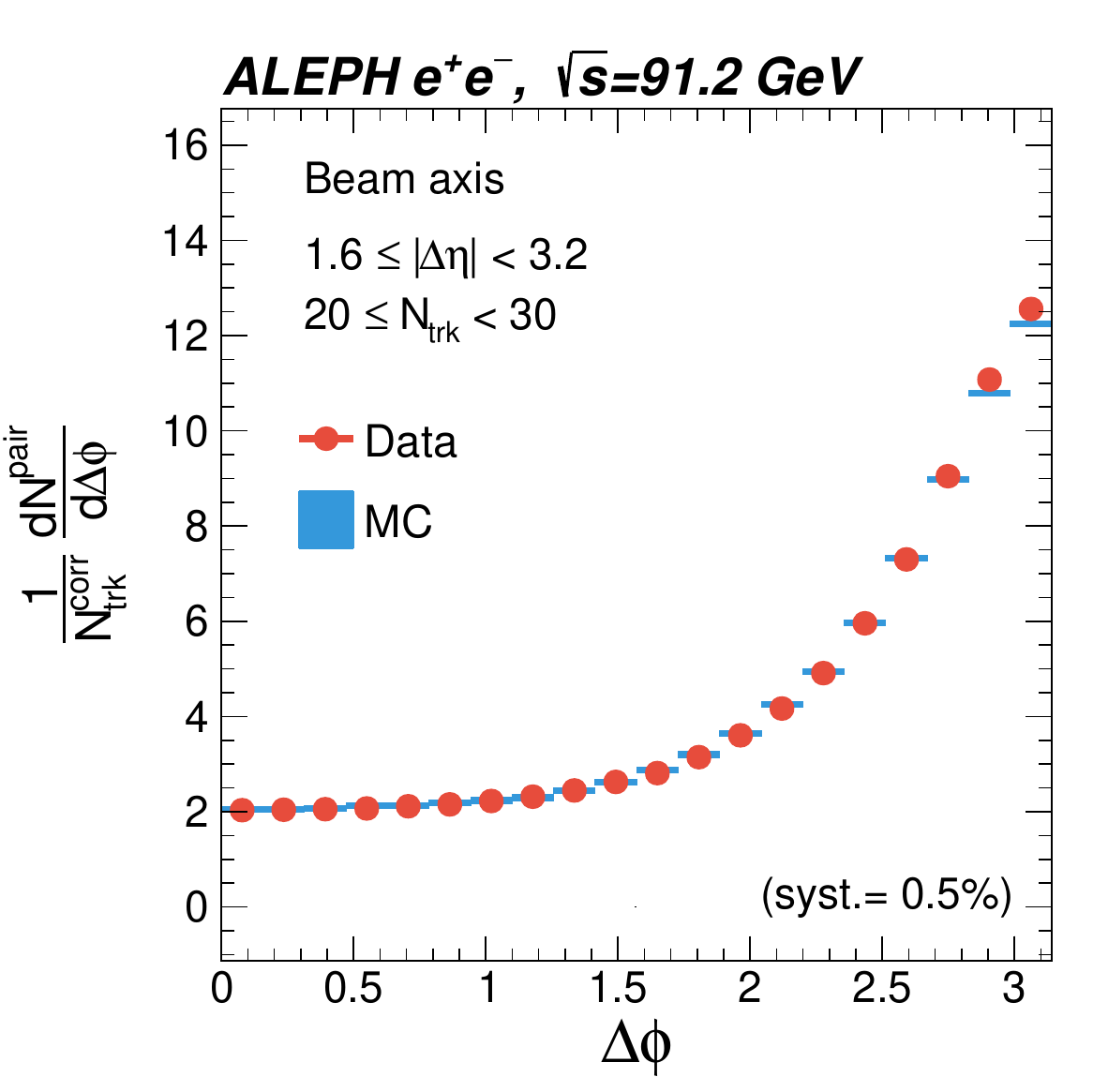}
    \includegraphics[width=0.45\textwidth]{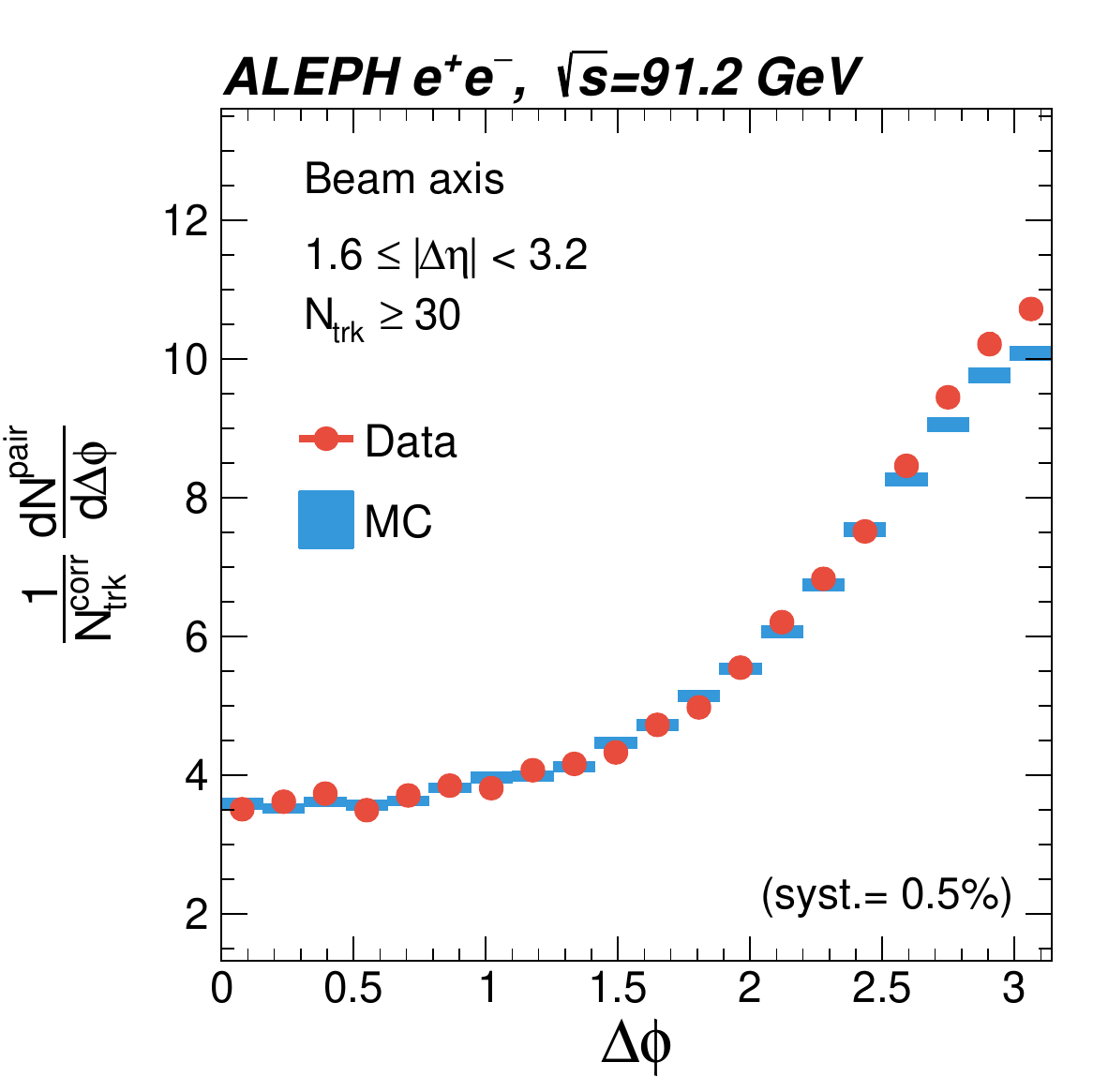}
    \includegraphics[width=0.45\textwidth]{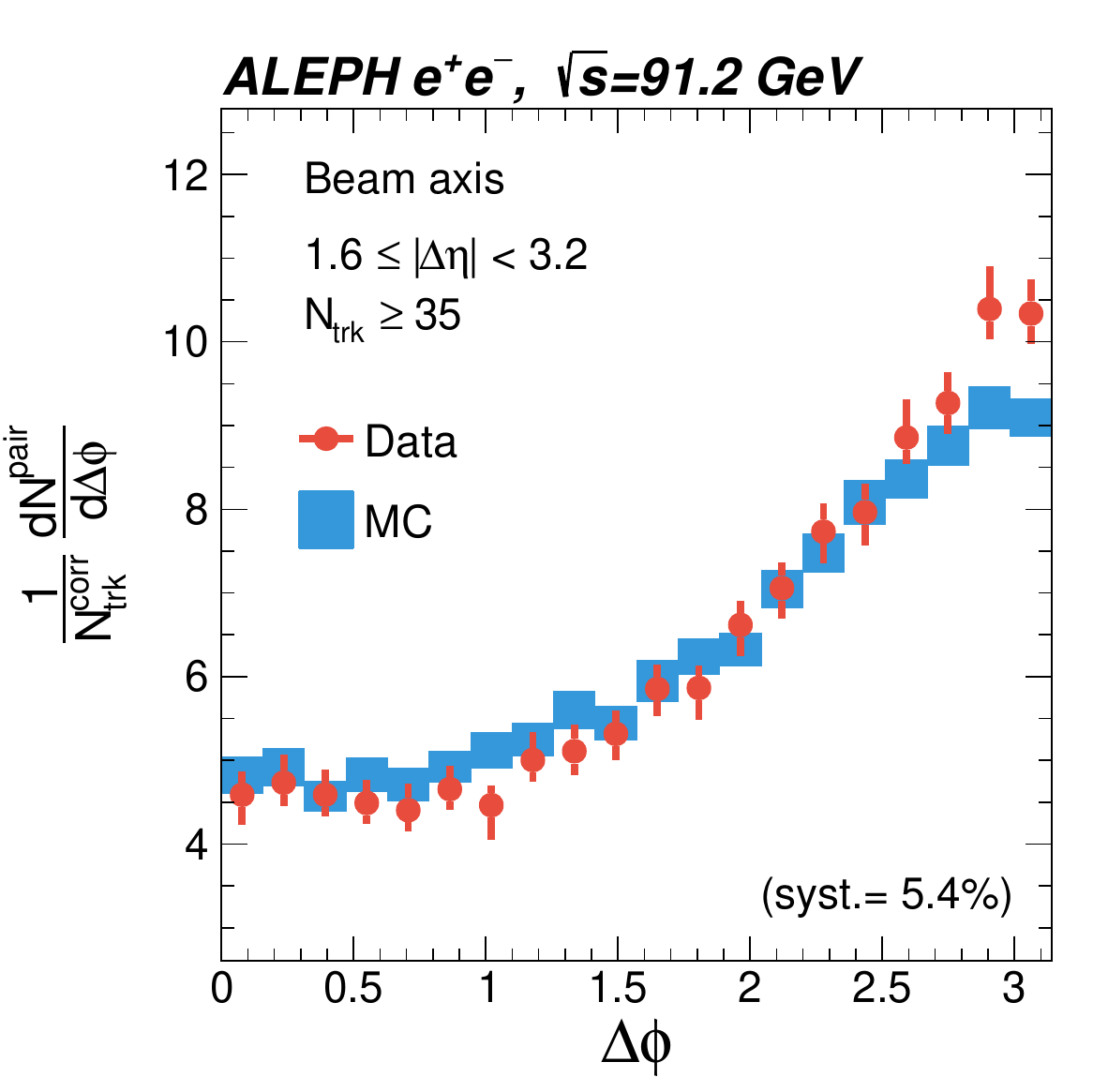}
    \includegraphics[width=0.45\textwidth]{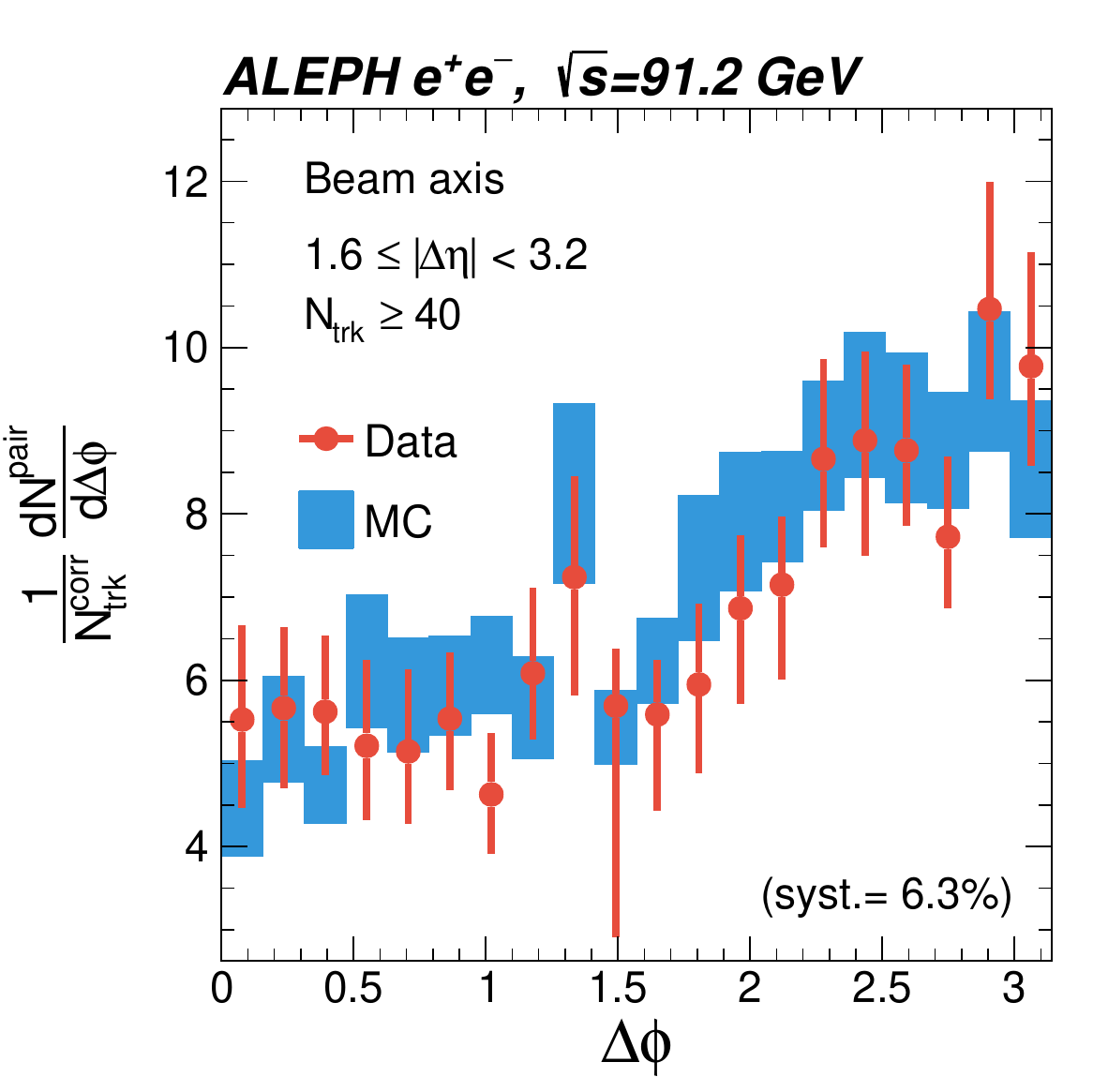}

\caption{Long-range azimuthal differential yields for $Z$-resonance sample in the beam axis analysis with combined LEP-I and LEP-II statistics. Results are shown as a function of the offline multiplicity in $[5,10), [10,20), [20,30), [30,\infty), [35,\infty), [40,\infty)$ intervals.}
\label{fig:EnergyCut_le_100_Combined_beam_1d}
\end{figure}

\begin{figure}[ht]
\centering
\includegraphics[width=.45\textwidth]{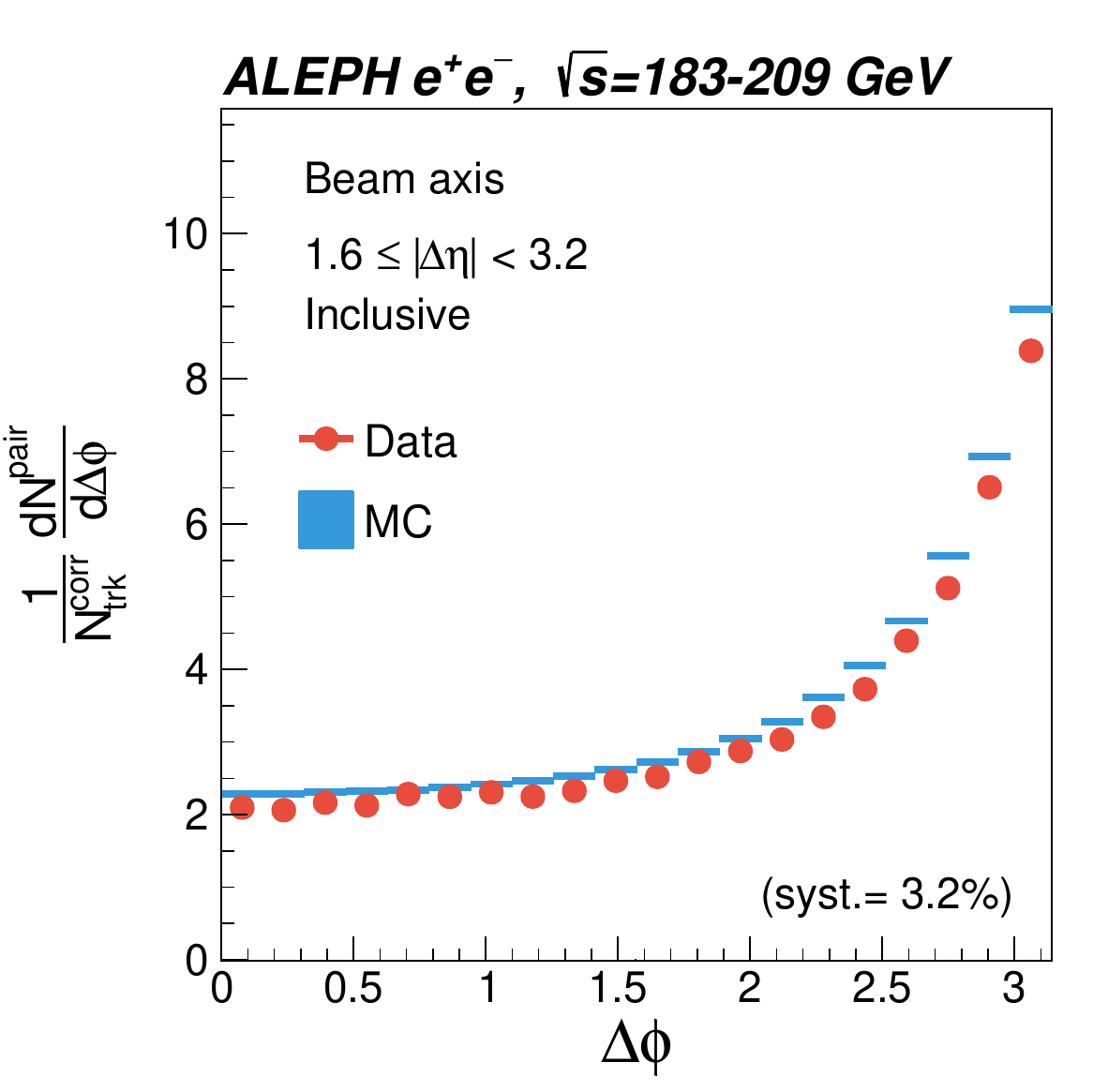}
\includegraphics[width=.45\textwidth]{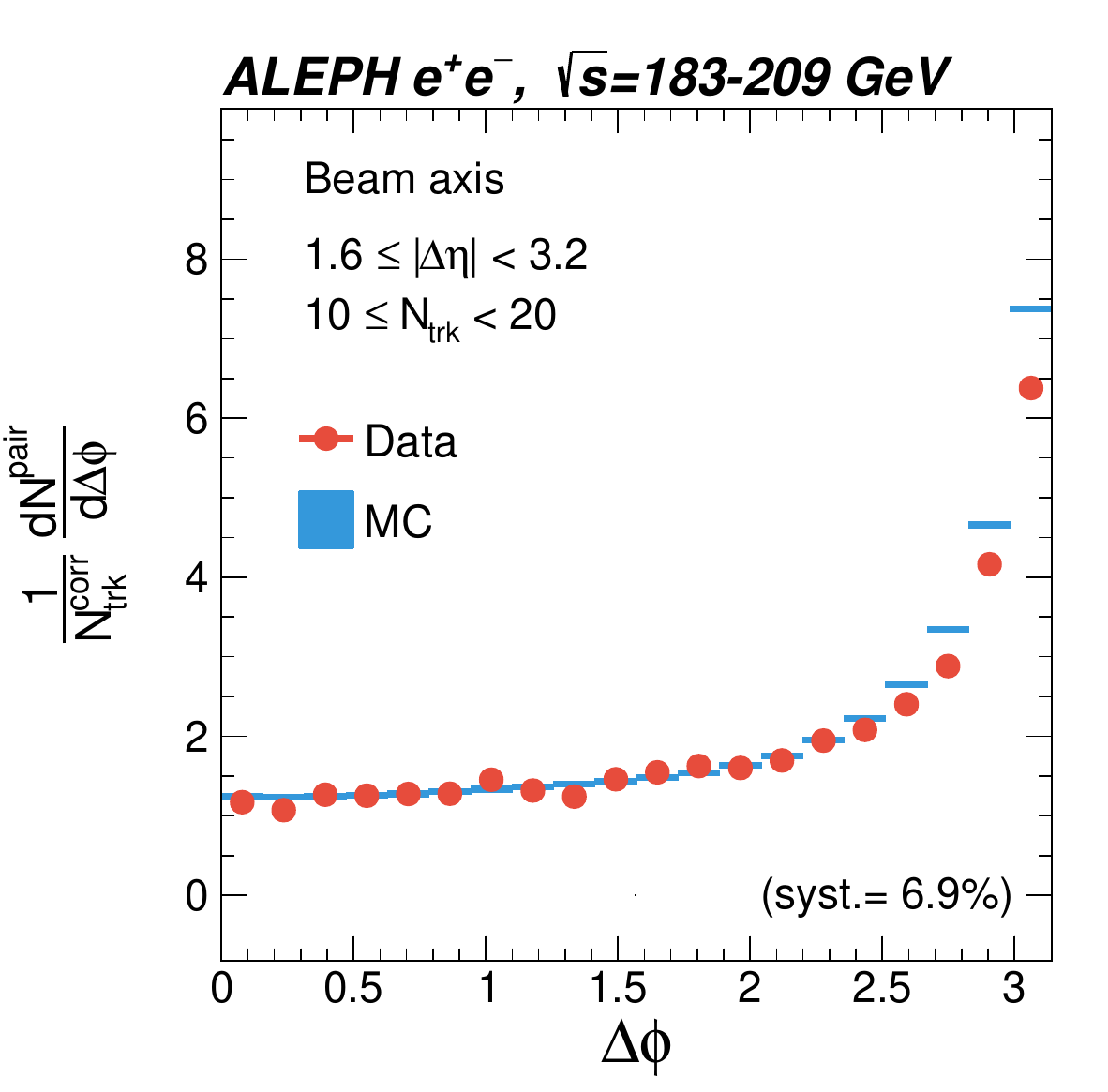}
\includegraphics[width=.45\textwidth]{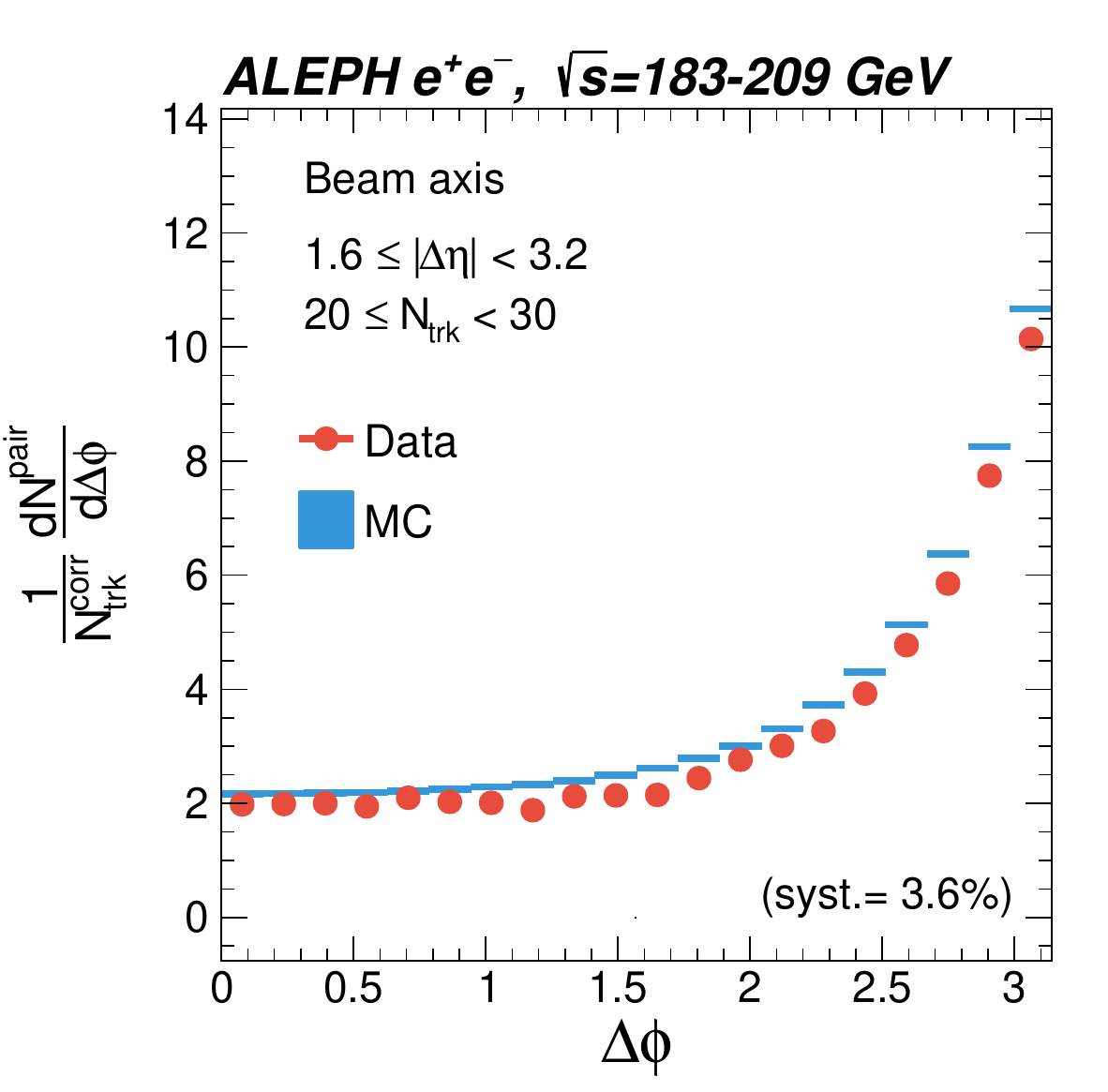}
\includegraphics[width=.45\textwidth]{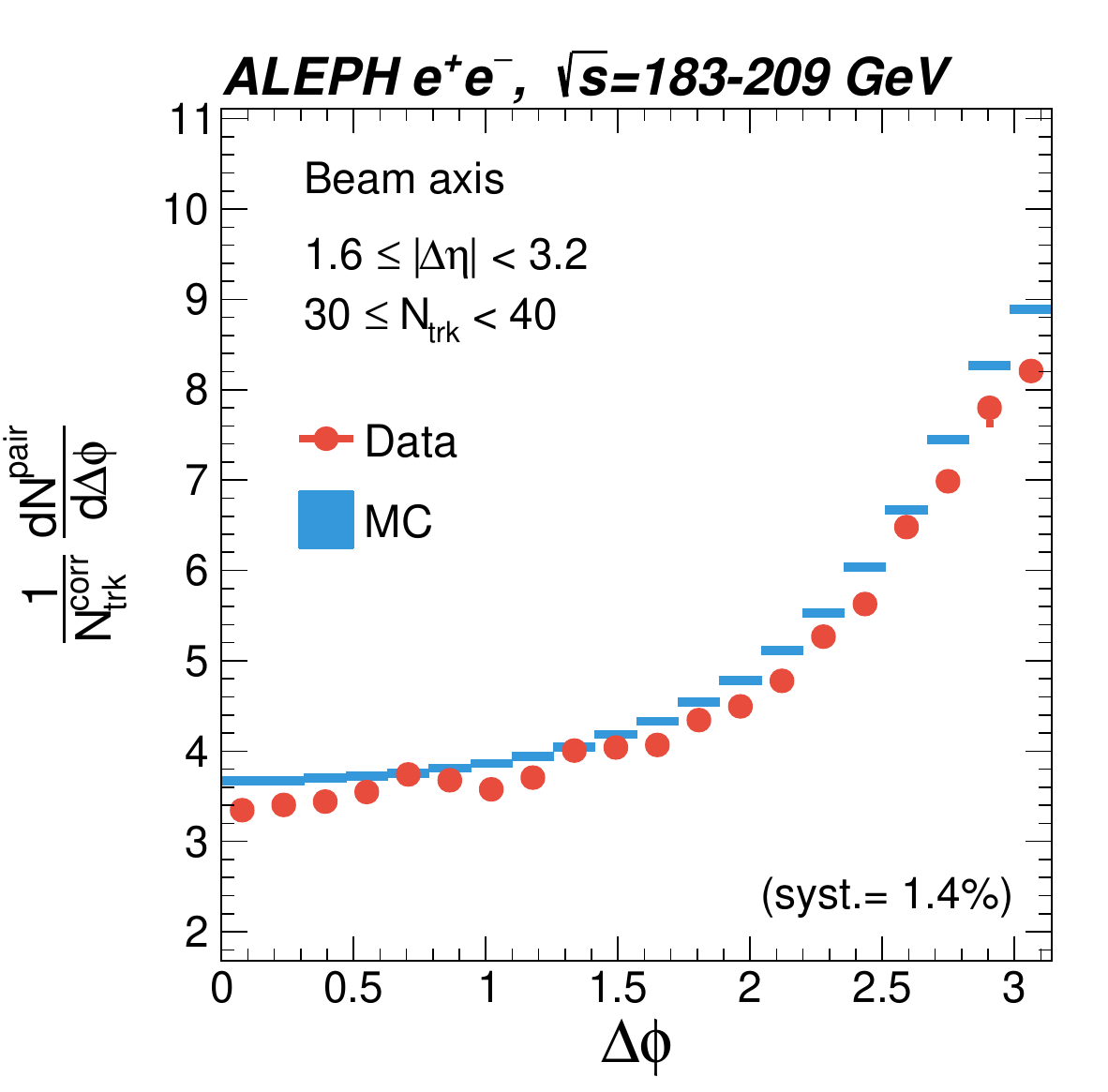}
\includegraphics[width=.45\textwidth]{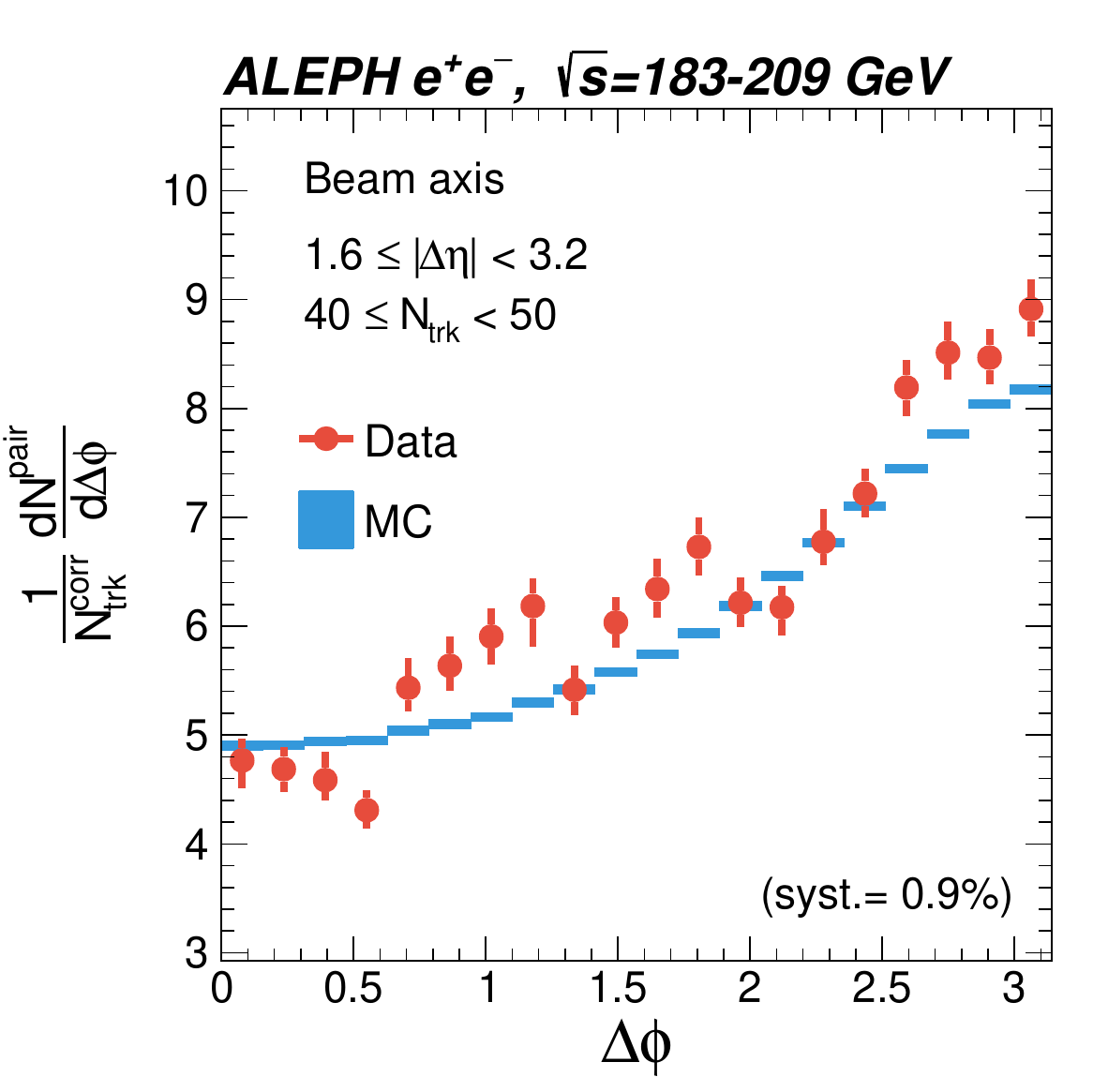}
\includegraphics[width=.45\textwidth]{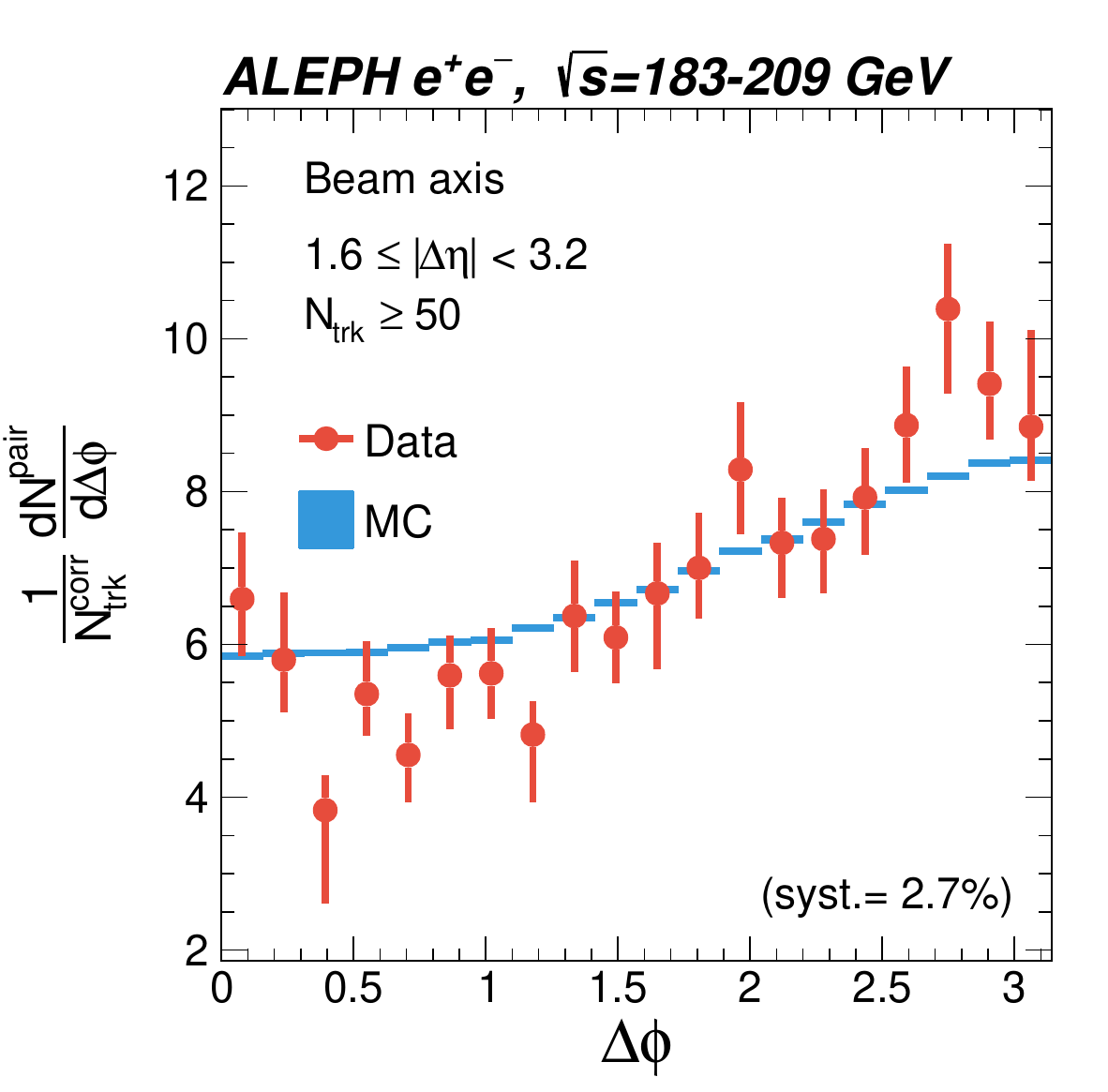}
\caption{Long-range azimuthal differential yields for high-energy sample in the beam axis analysis. Results are shown multiplicity-inclusively and as a function of the offline multiplicity in $[10,20), [20,30), [30,40), [40,50), [50, \infty)$ intervals.}
\label{fig:EnergyCut_ge_100_beam_1d}
\end{figure}

Figures~\ref{fig:EnergyCut_le_100_Combined_beam_1d} and~\ref{fig:EnergyCut_ge_100_beam_1d} show the long-range azimuthal differential associated yield distributions of beam-axis correlation functions, representing the $Z$-resonance and high-energy data, respectively. The $Z$-resonance findings align with a prior publication~\cite{Badea:2019vey} that utilized the LEP-1 dataset. Moreover, results with a higher $\ntrk$ threshold ($>40$) are presented for the first time. Notably, no ridge-like structures were visually detected in the beam axis analysis across all multiplicity brackets.

\clearpage

\begin{figure}[ht]
\centering
    \includegraphics[width=0.45\textwidth]{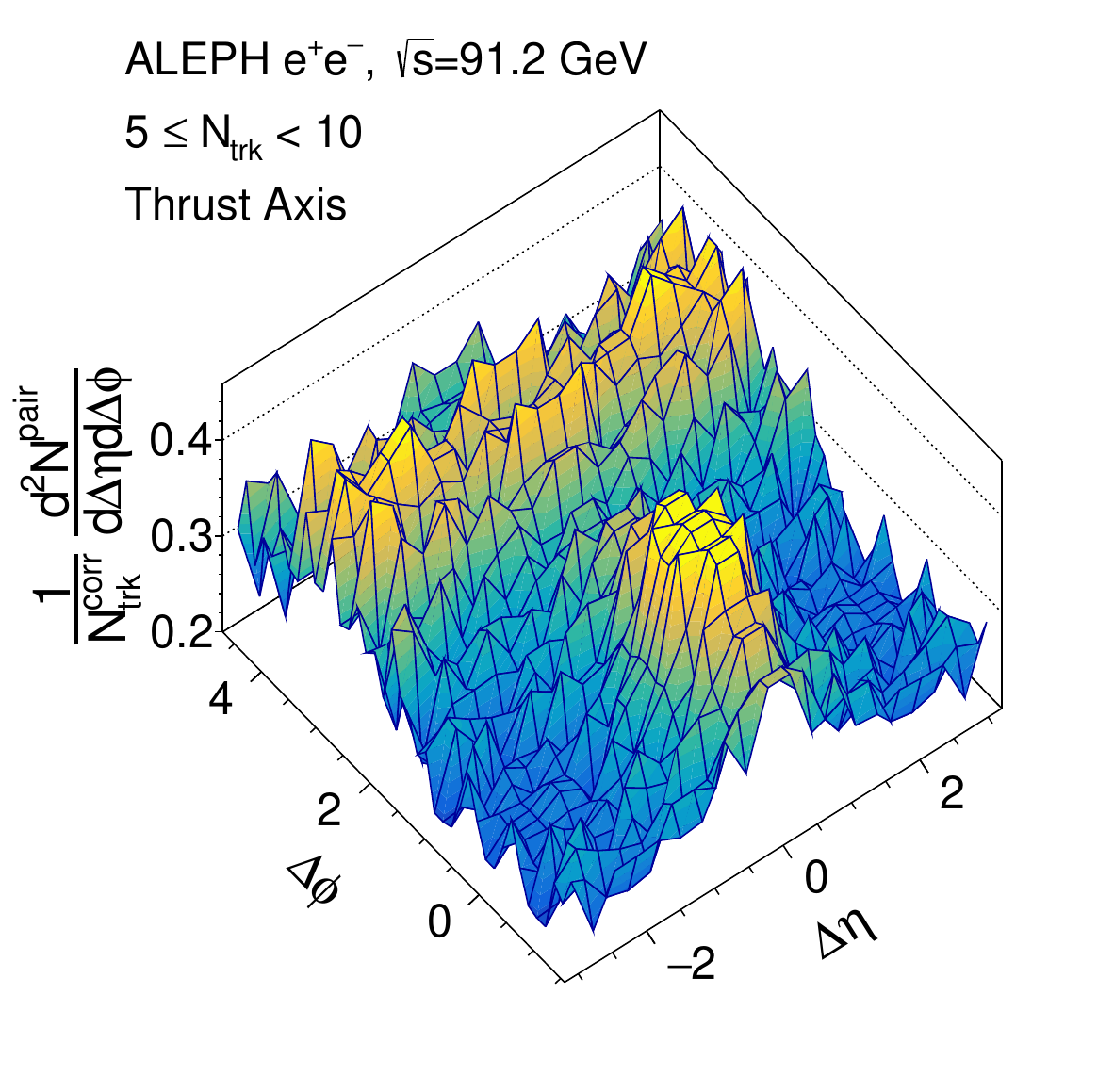}
    \includegraphics[width=0.45\textwidth]{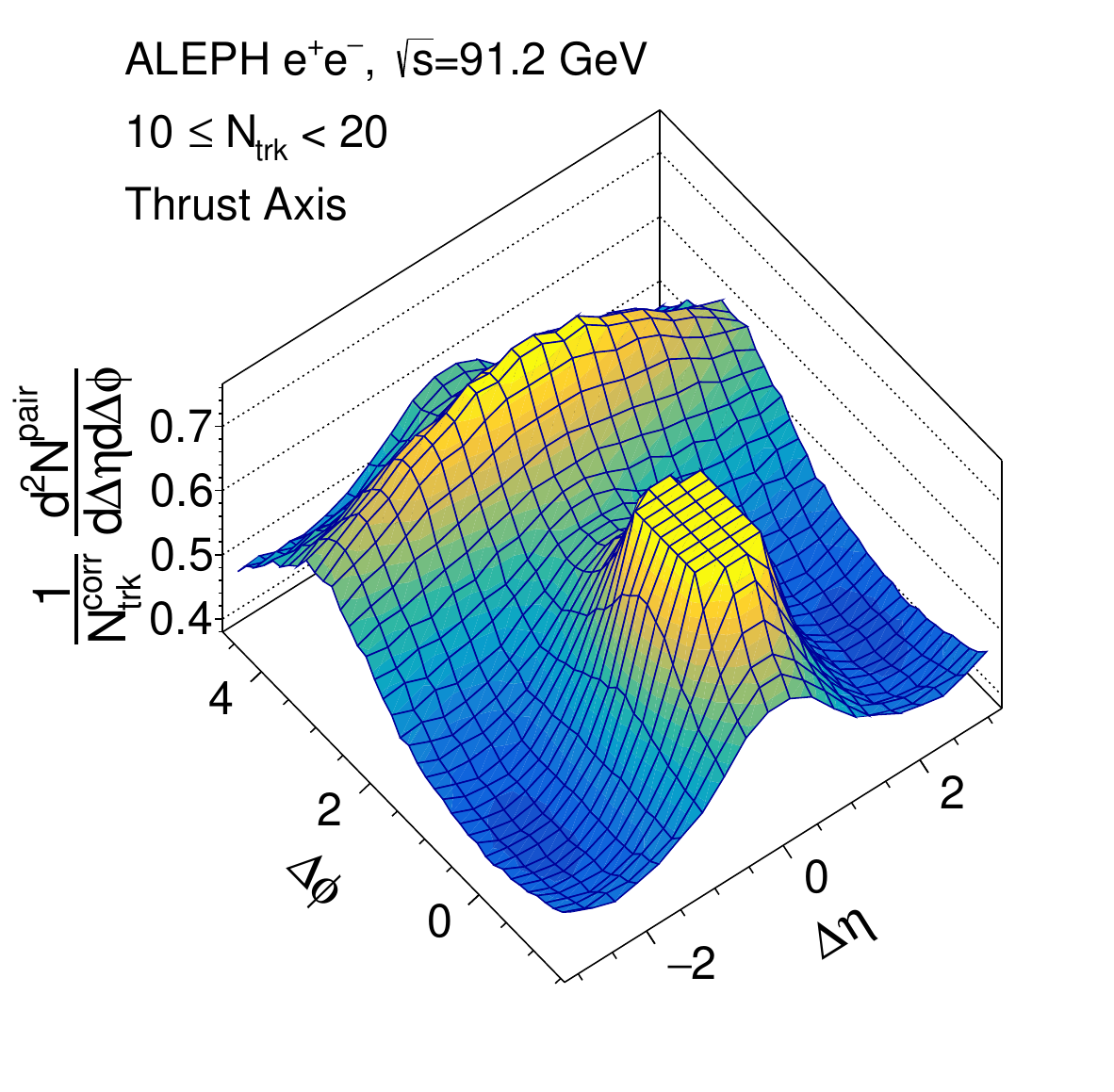}
    \includegraphics[width=0.45\textwidth]{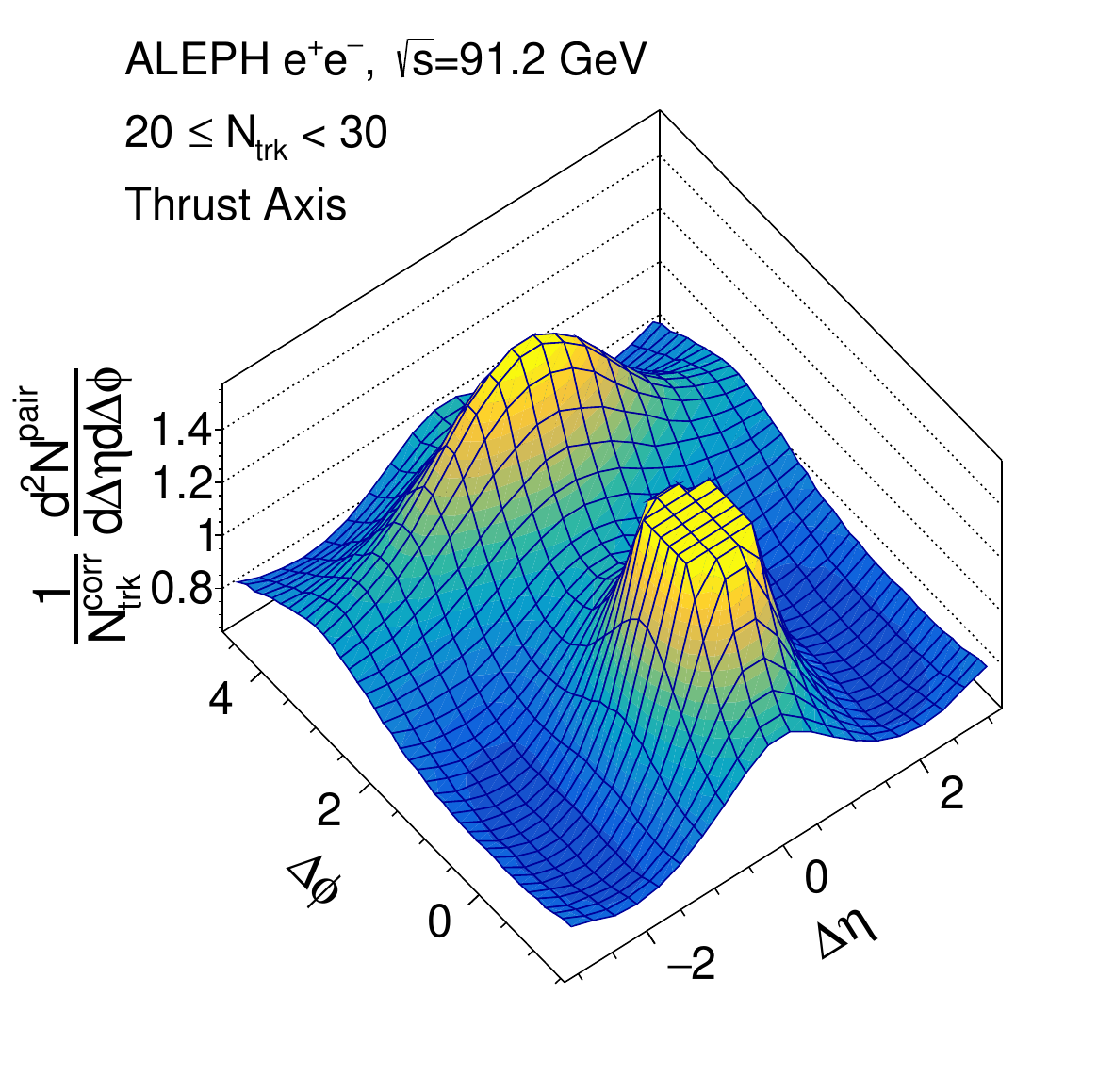}
    \includegraphics[width=0.45\textwidth]{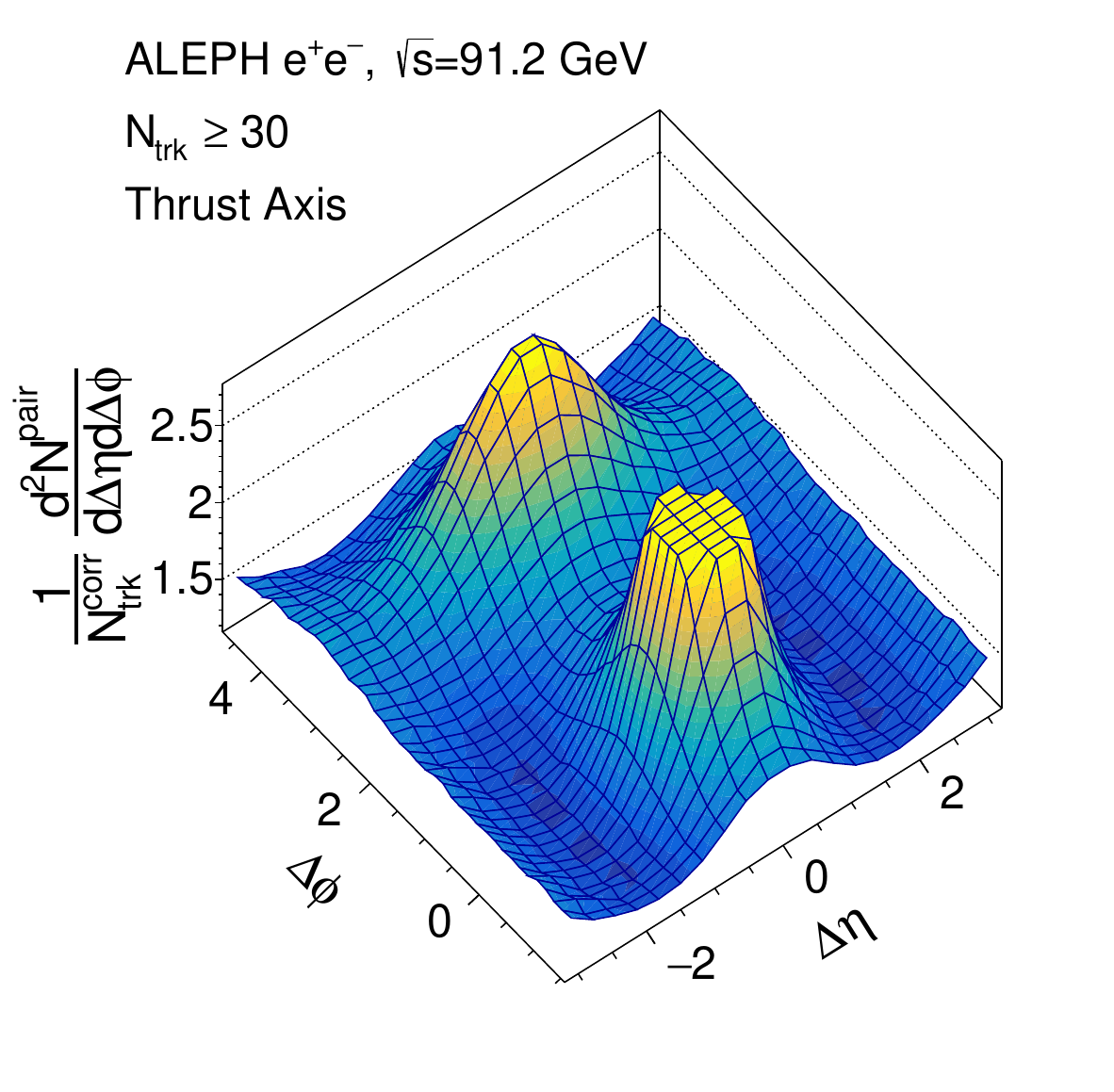}
    \includegraphics[width=0.45\textwidth]{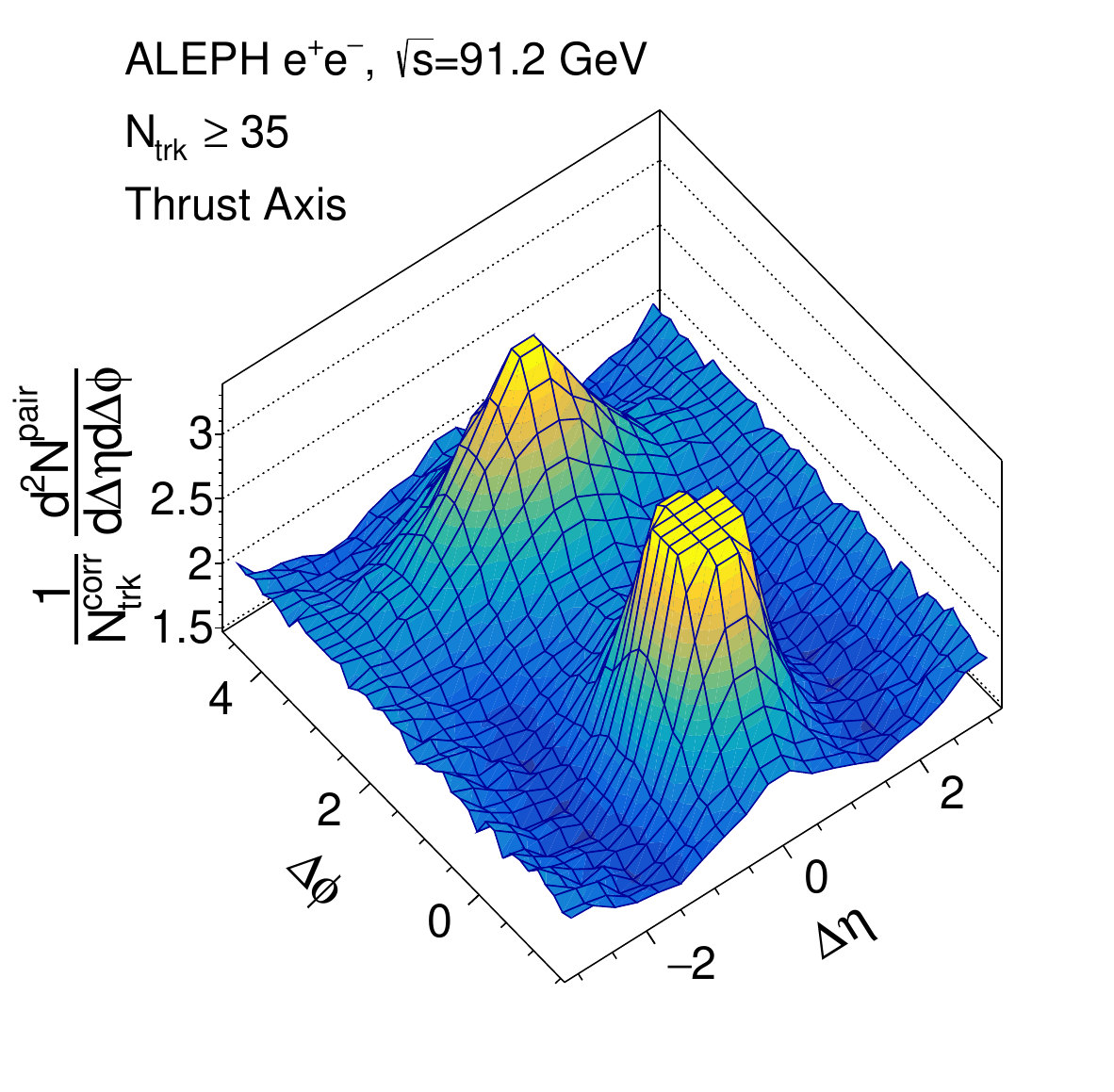}
    \includegraphics[width=0.45\textwidth]{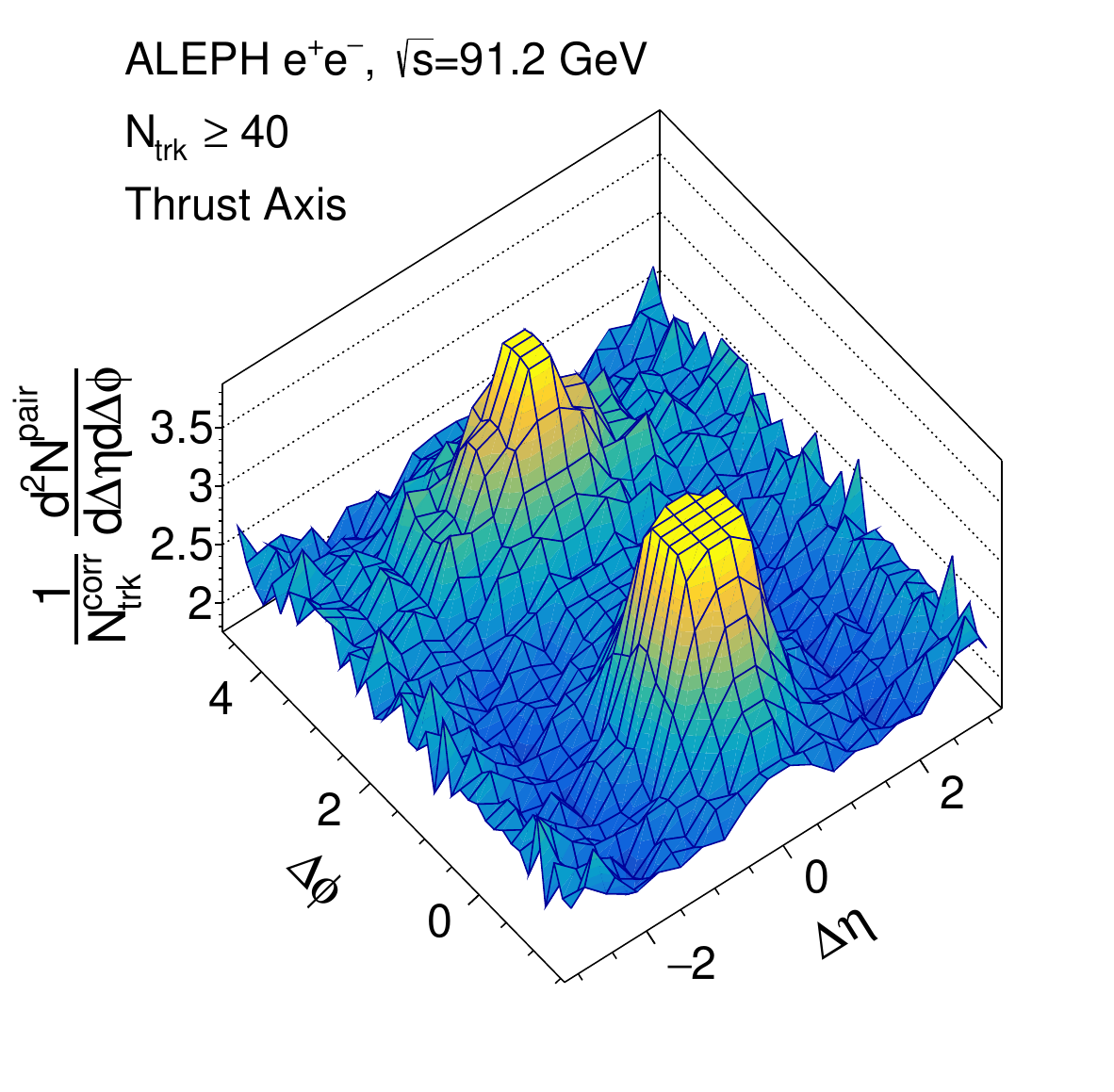}

\caption{Two-particle correlation function for $Z$-resonance sample in the thrust axis analysis with combined LEP-I and LEP-II statistics. Results are shown as a function of the offline multiplicity in $[5,10), [10,20), [20,30), [30,\infty), [35,\infty), [40,\infty)$ intervals.}
\label{fig:EnergyCut_le_100_Combined_thrust}
\end{figure}

\begin{figure}[ht]
\centering
\includegraphics[width=.45\textwidth]{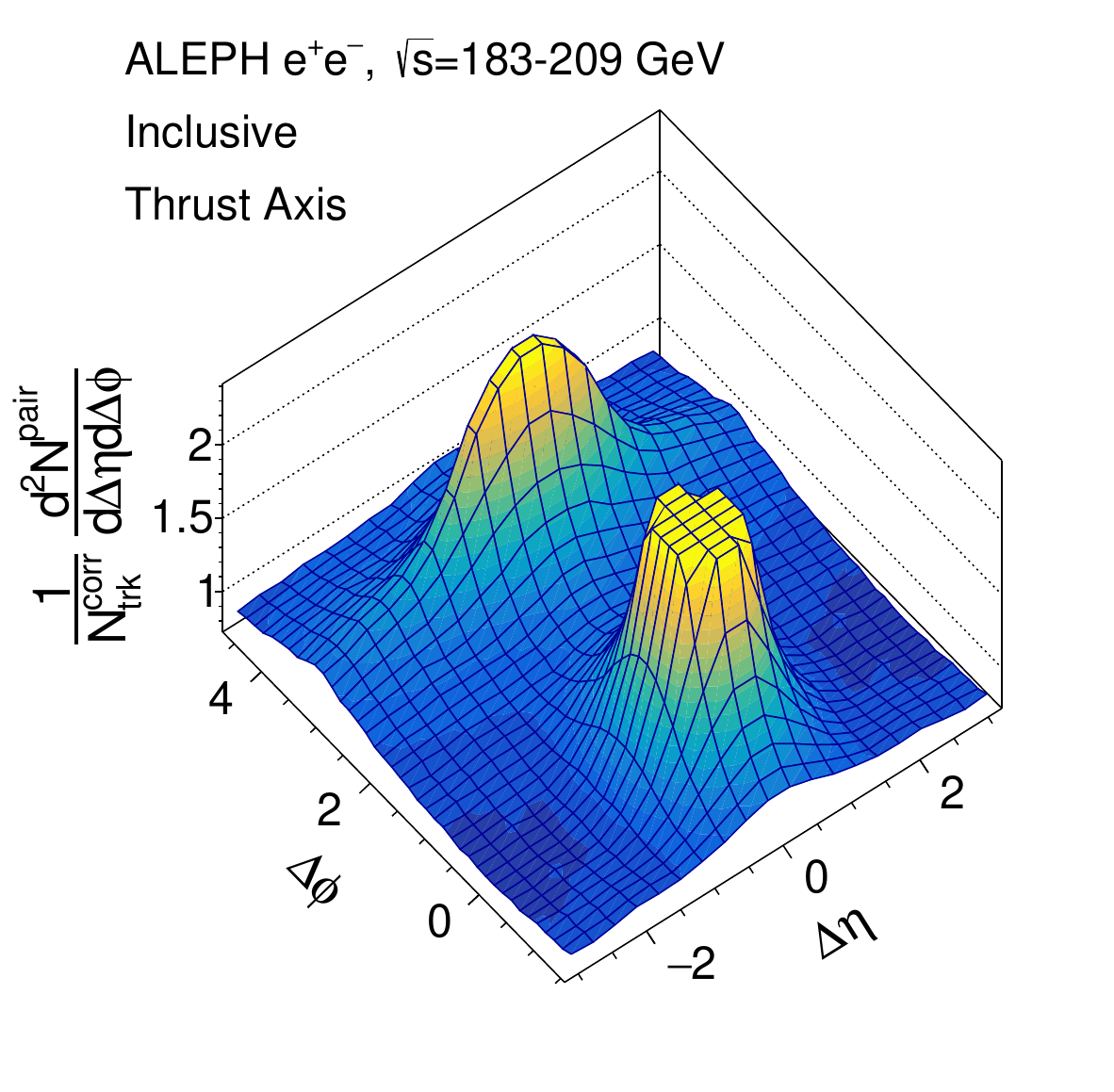}
\includegraphics[width=.45\textwidth]{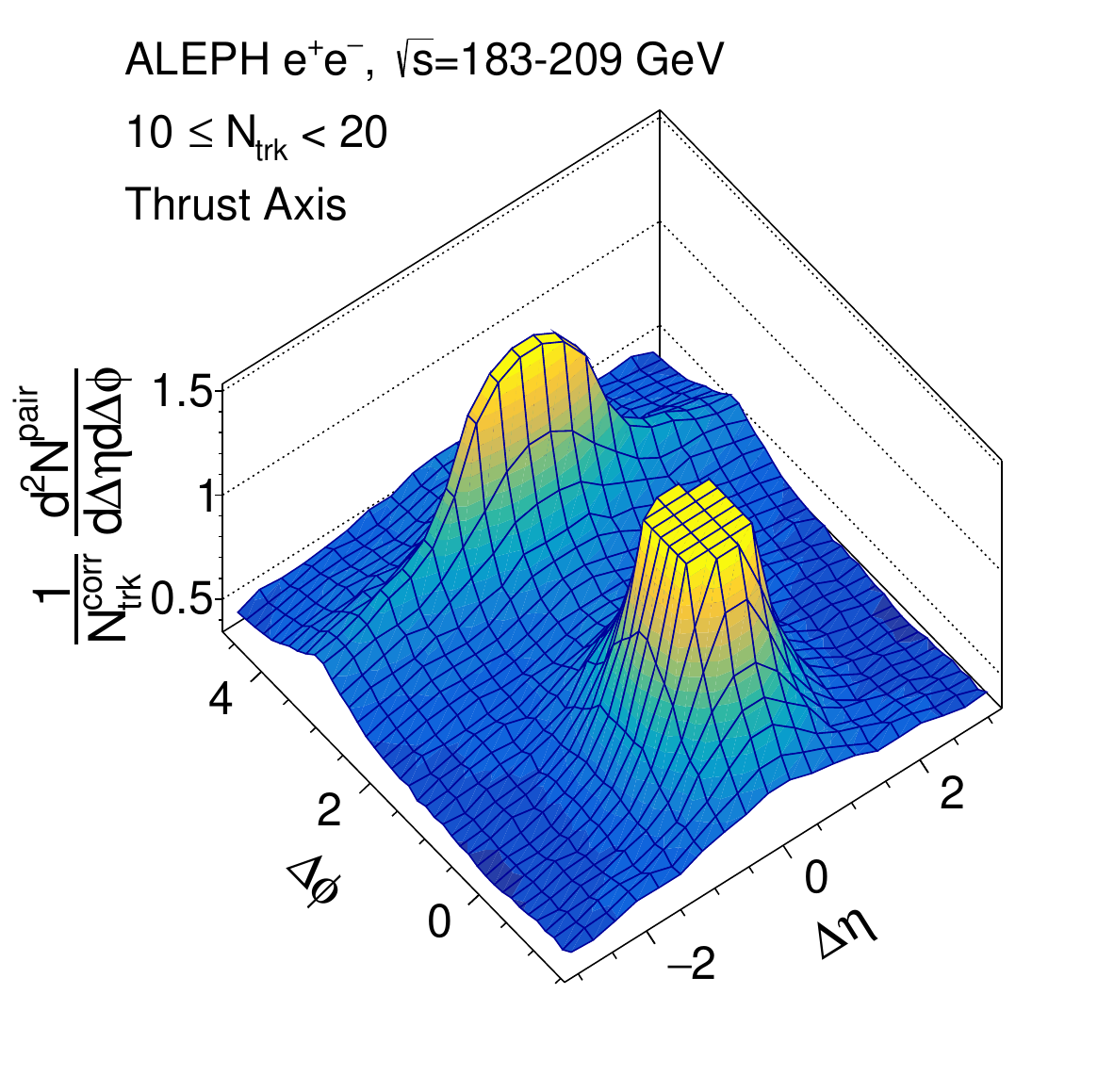}
\includegraphics[width=.45\textwidth]{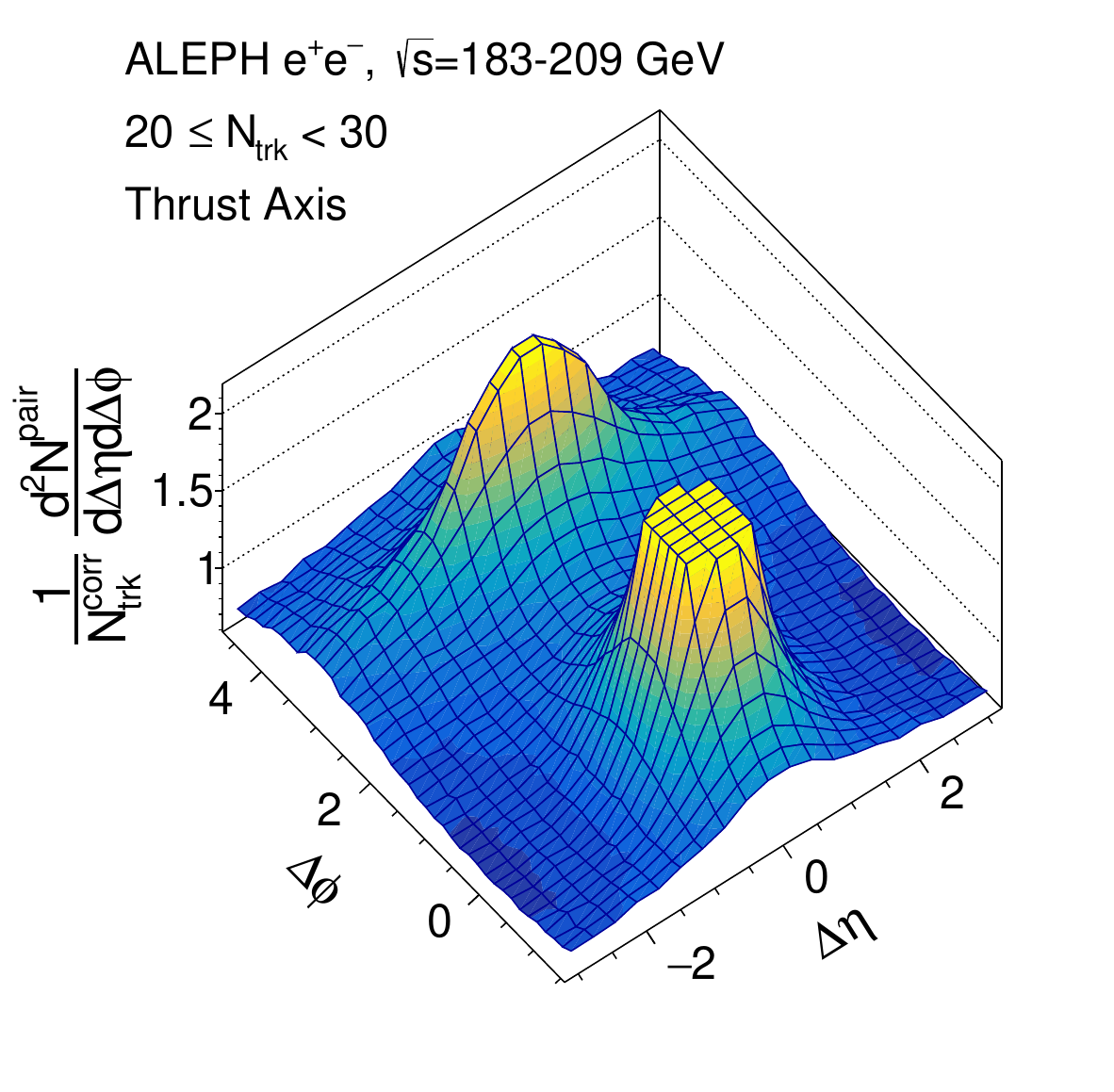}
\includegraphics[width=.45\textwidth]{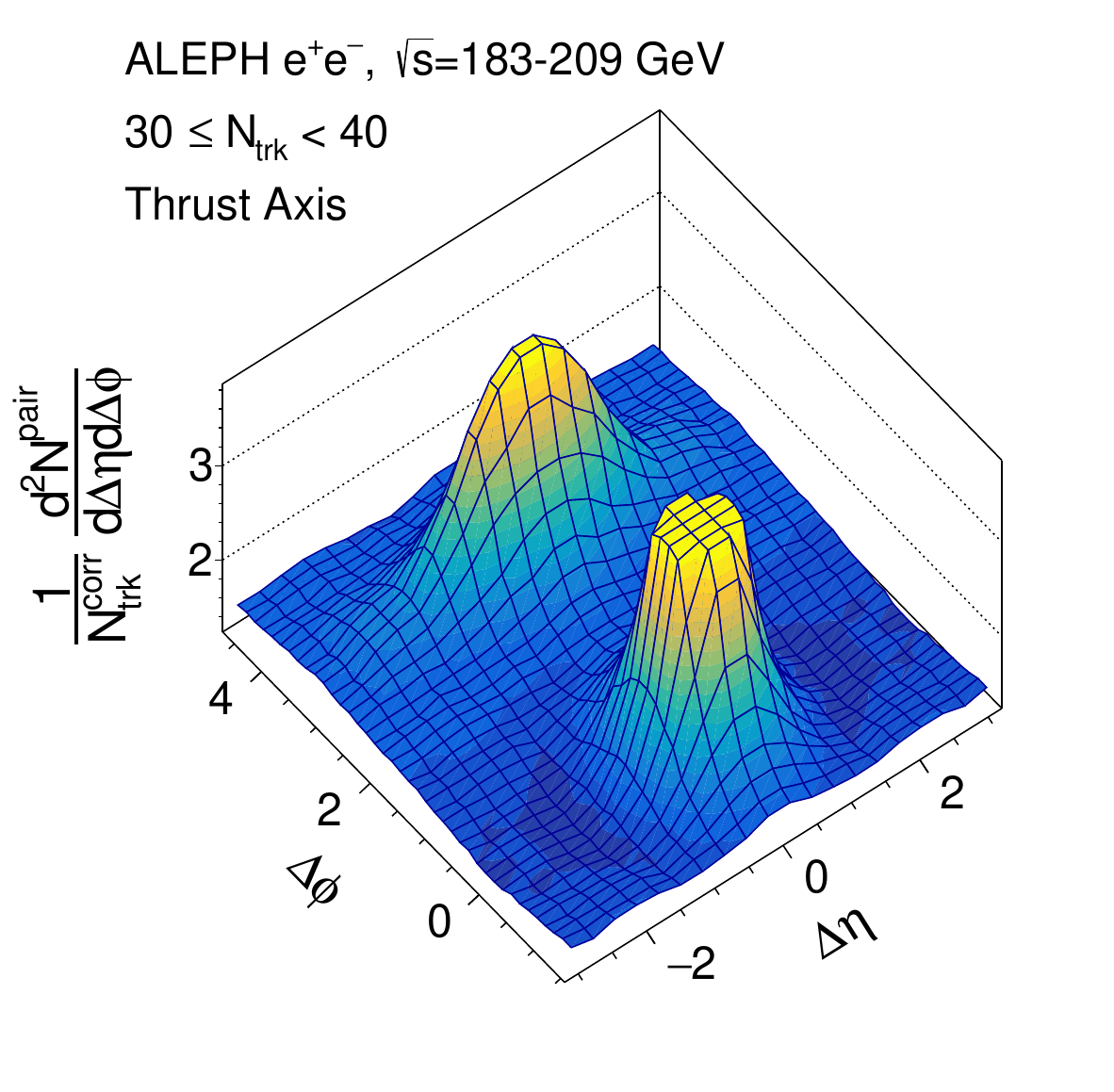}
\includegraphics[width=.45\textwidth]{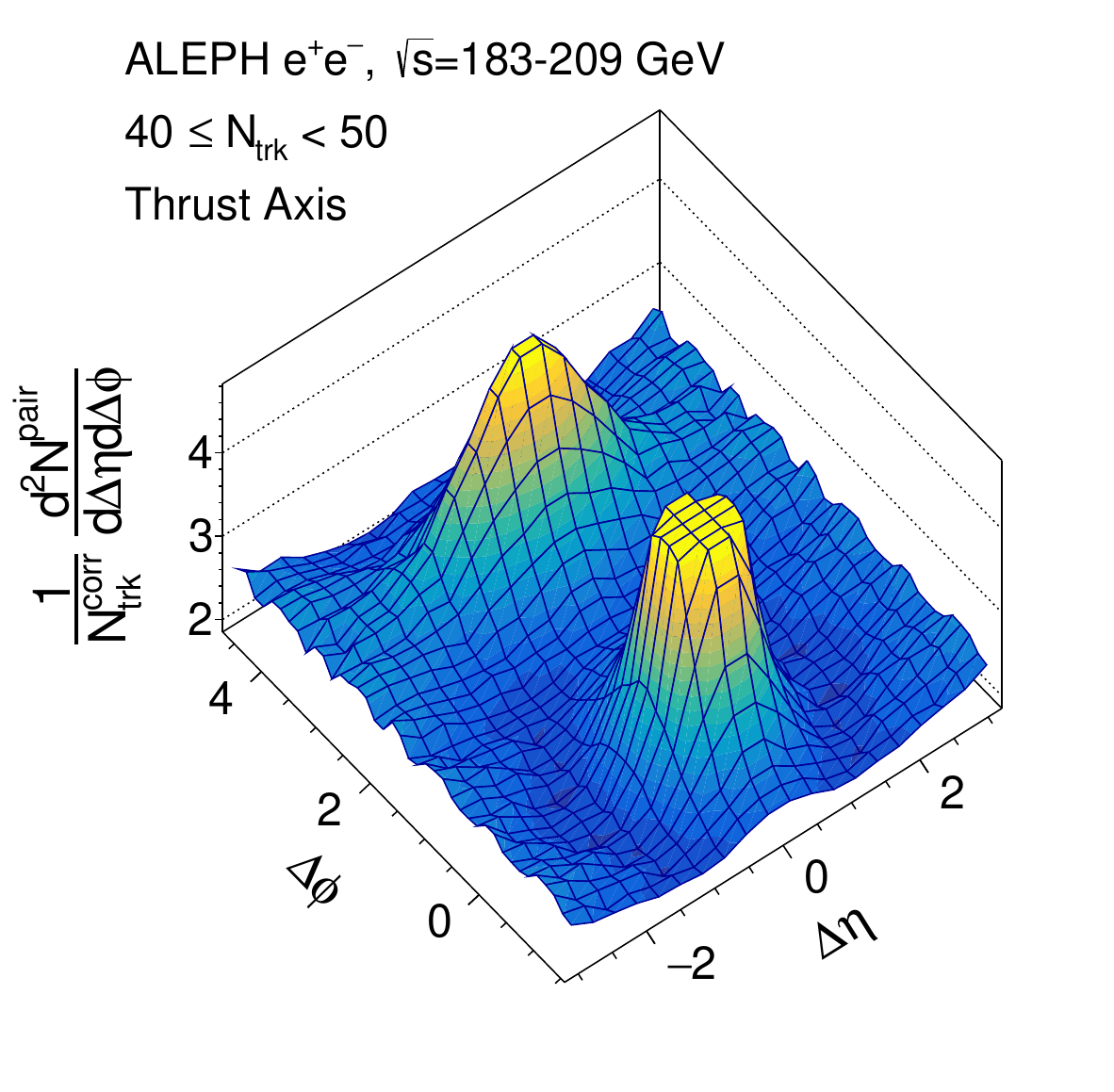}
\includegraphics[width=.45\textwidth]{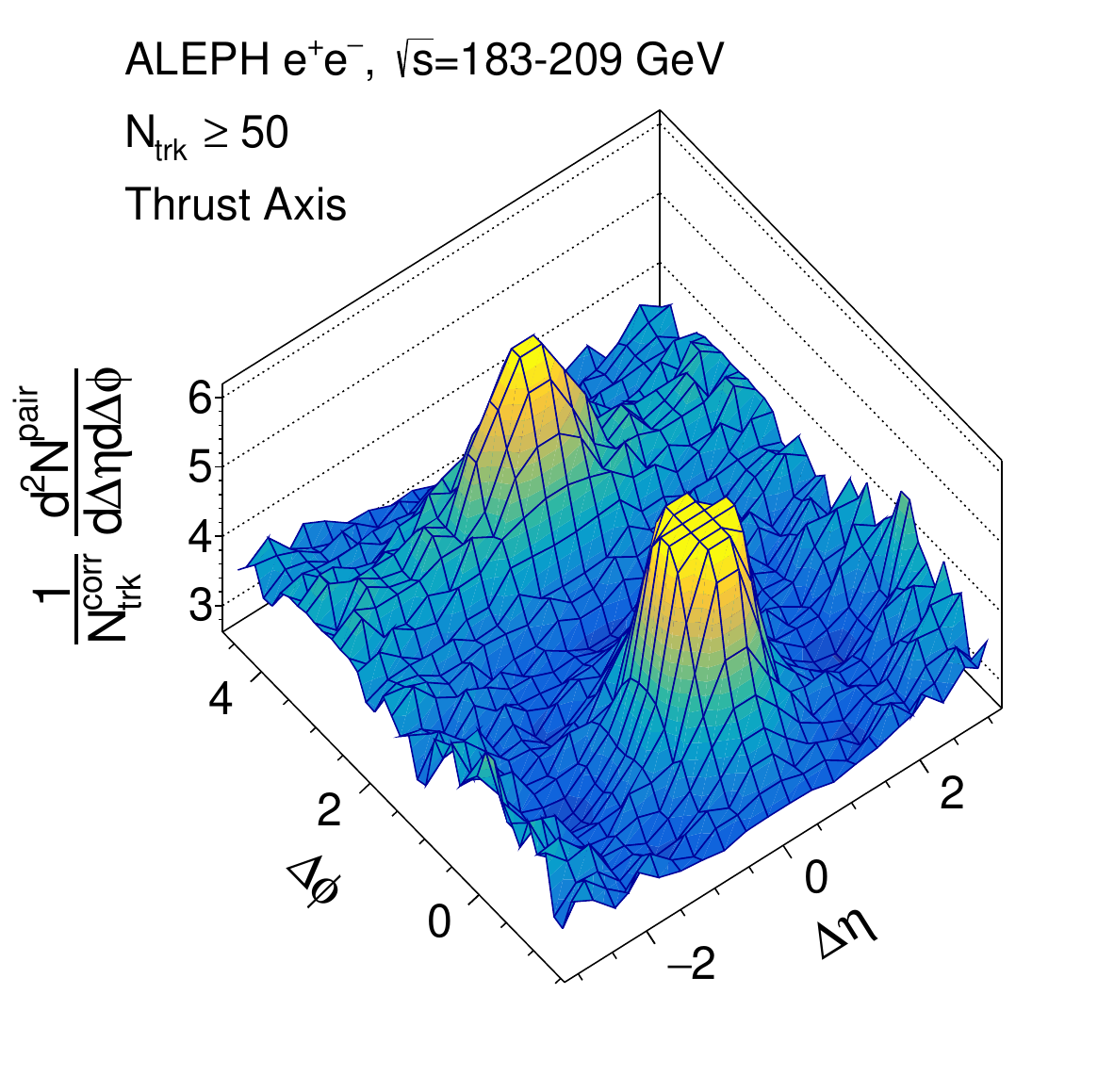}
\caption{Two-particle correlation function for high-energy sample in the thrust axis analysis. Results are shown multiplicity-inclusively and as a function of the offline multiplicity in $[10,20), [20,30), [30,40), [40,50), [50, \infty)$ intervals.}
\label{fig:EnergyCut_ge_100_thrust}
\end{figure}

In Figures~\ref{fig:EnergyCut_le_100_Combined_thrust} and~\ref{fig:EnergyCut_ge_100_thrust}, the thrust-axis two-particle correlations for the $Z$-resonance and high-energy data sets are presented in bins of $\ntrk$. Compared to the beam-axis analysis, there is a notable decrease in the magnitude of correlations. This attenuation is attributed to the exclusion of the predominant back-to-back di-jet correlation generally observed at large pseudorapidity differences. This phenomenon occurs because the thrust axis inherently aligns with the dominant direction of energy flow. At low multiplicity ($\ntrk< 50$) no sigificant ridge-like structure was observed in the correlation function. In the highest multiplicity bin ($\ntrk>50$) an intriguing U shape was revealed at the large $|\Delta\eta|$ and small $\Delta\phi$ phase space, which is studied further in the later sections.

\begin{figure}[ht]
\centering
    \includegraphics[width=0.451\textwidth]{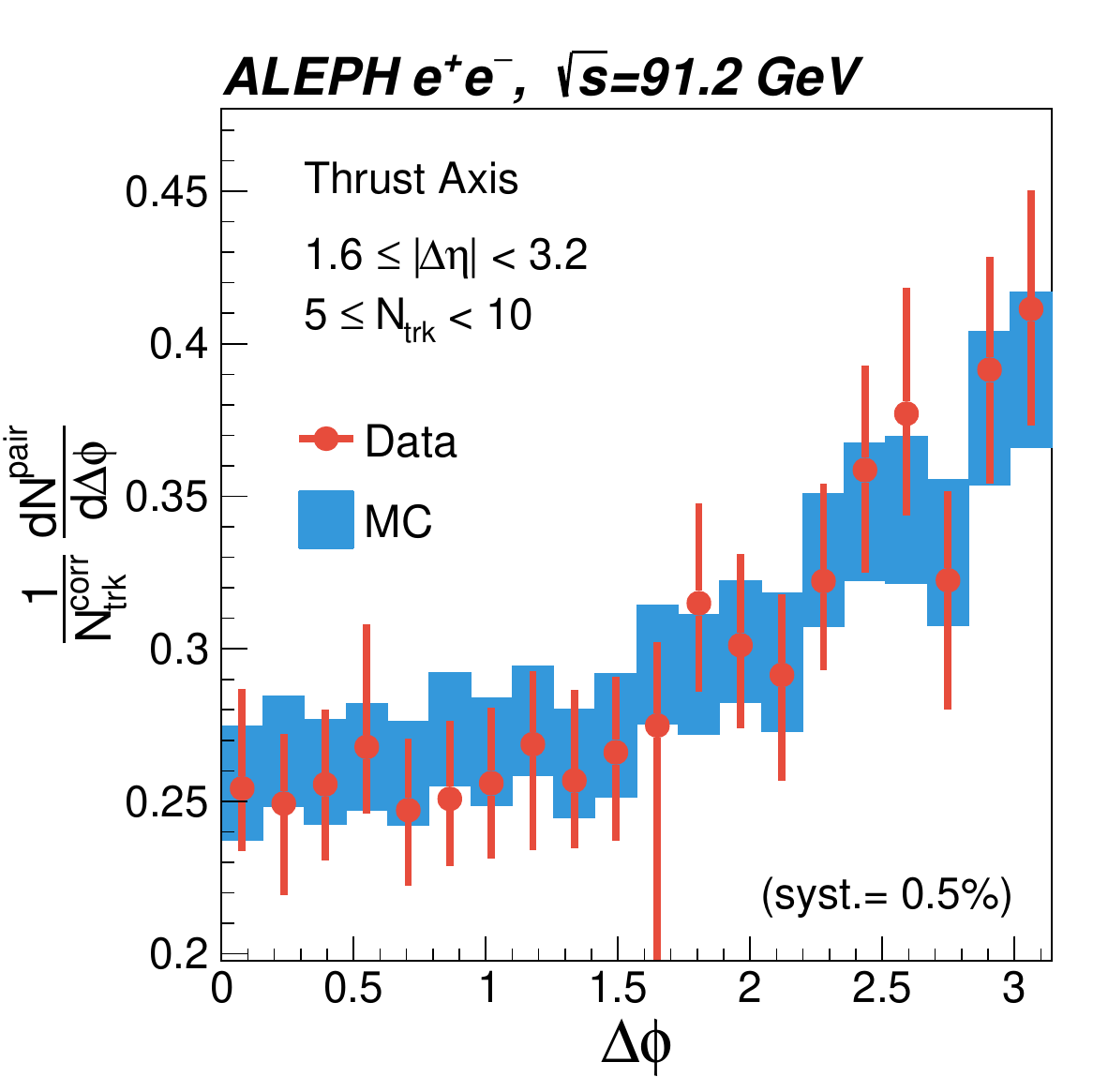}
    \includegraphics[width=0.451\textwidth]{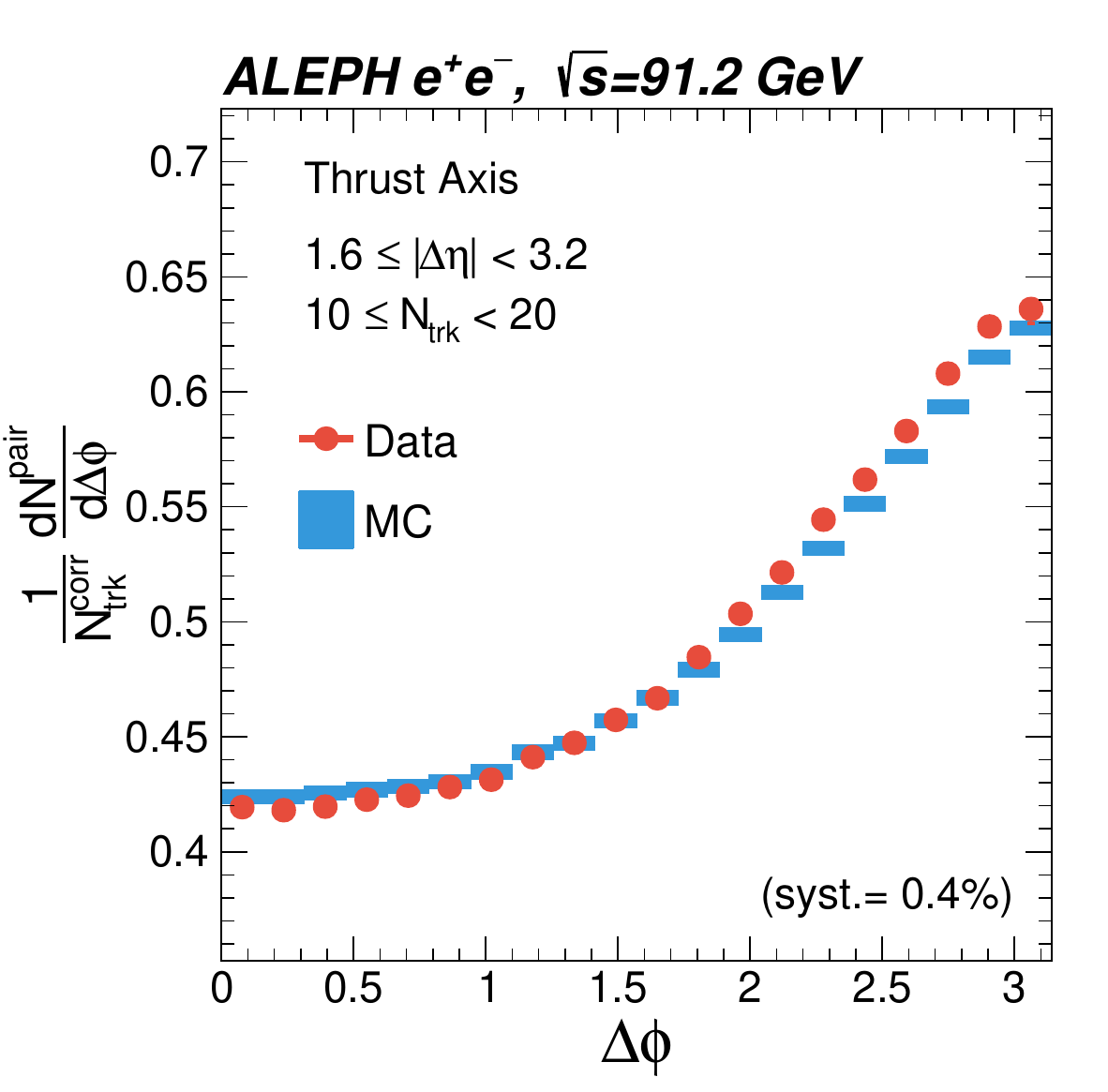}
    \includegraphics[width=0.451\textwidth]{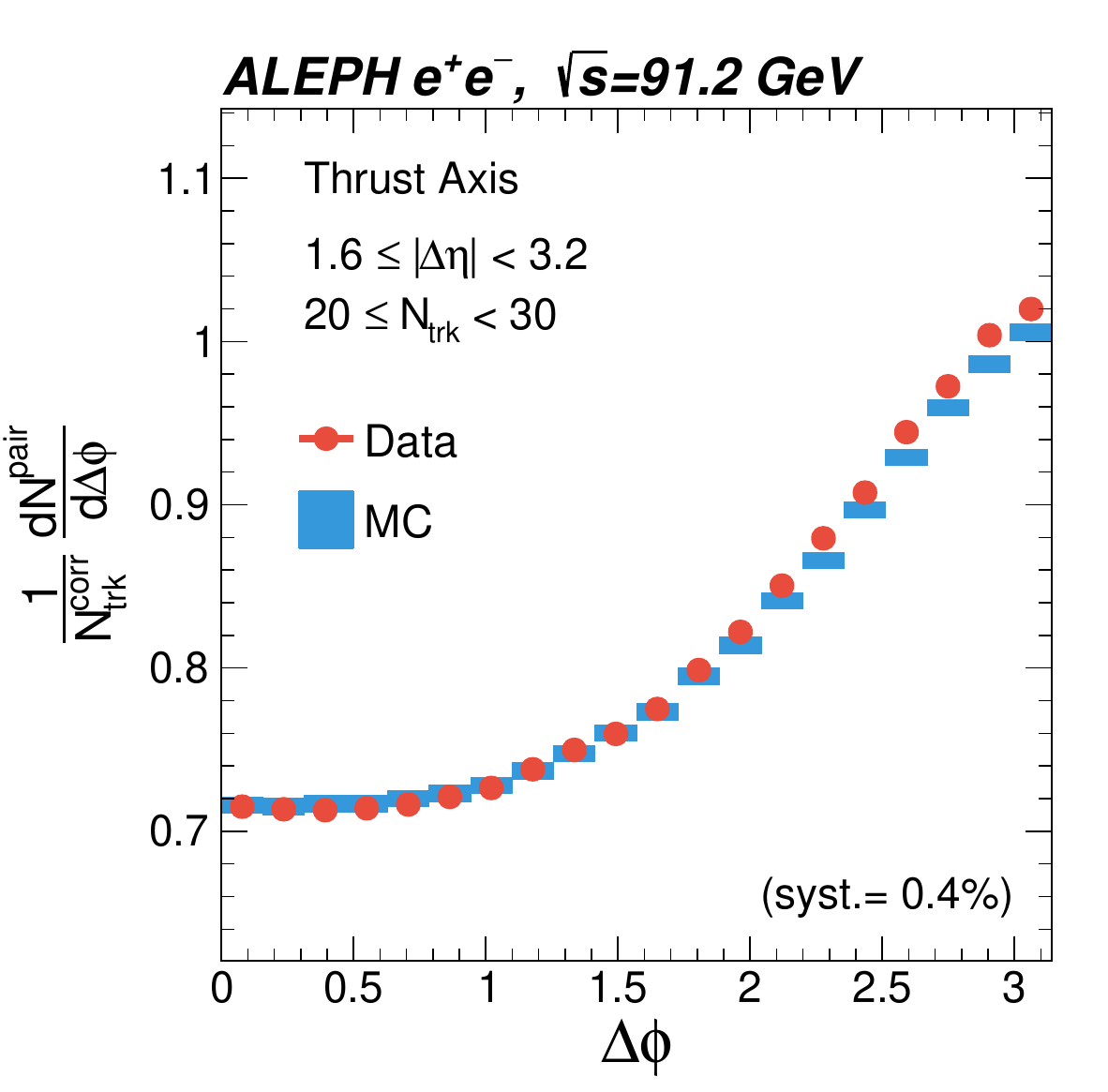}
    \includegraphics[width=0.451\textwidth]{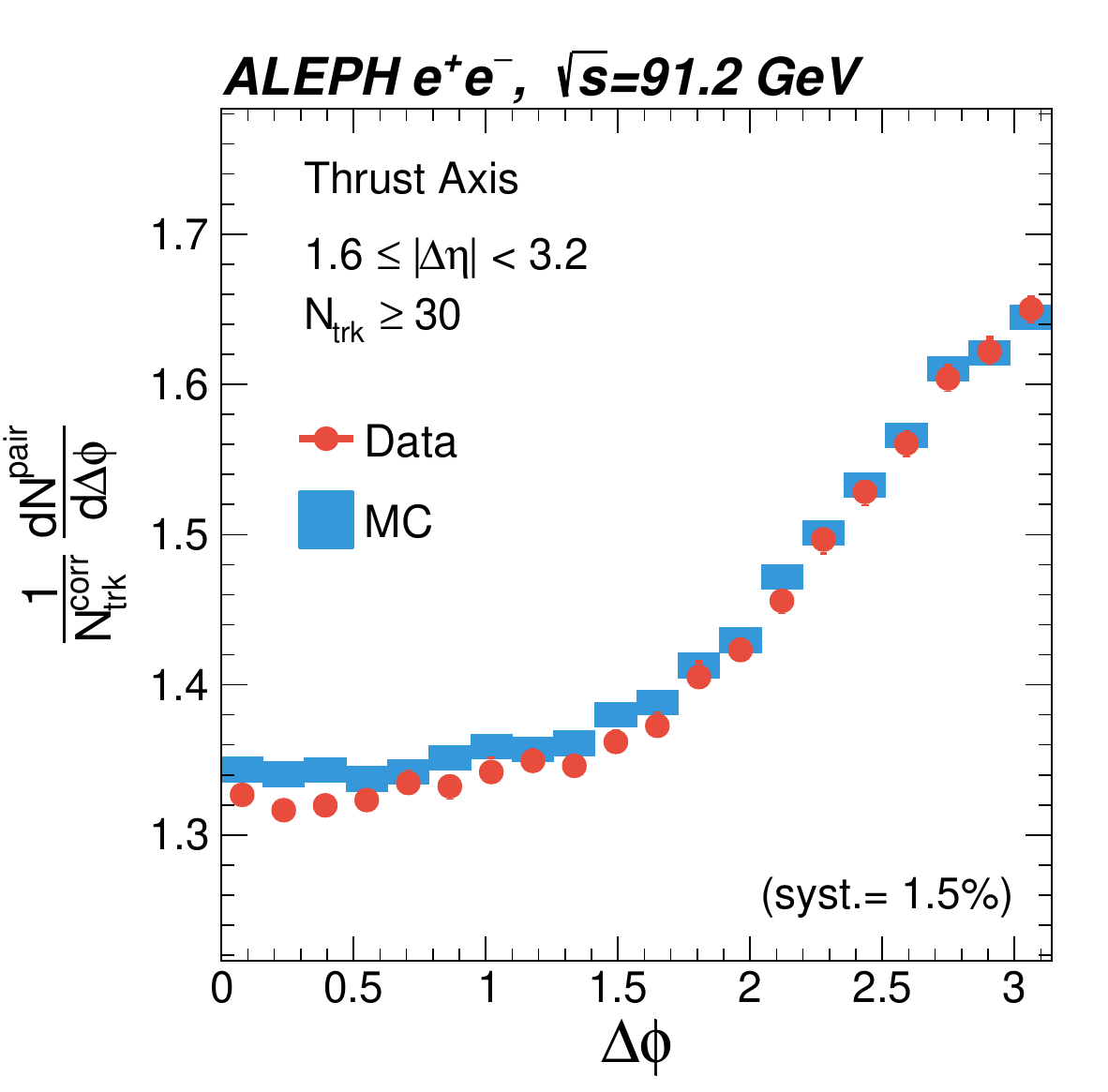}
    \includegraphics[width=0.451\textwidth]{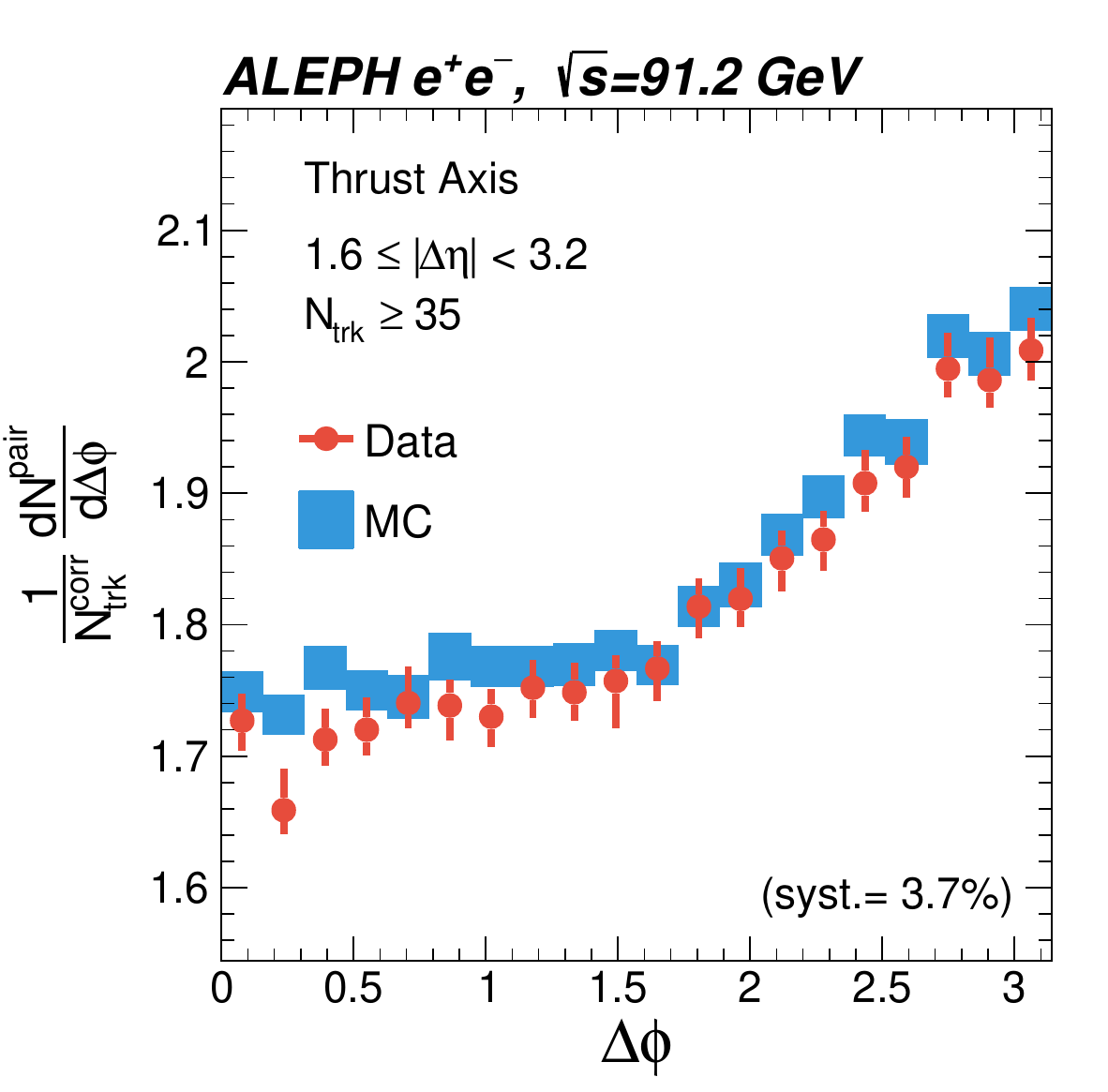}
    \includegraphics[width=0.451\textwidth]{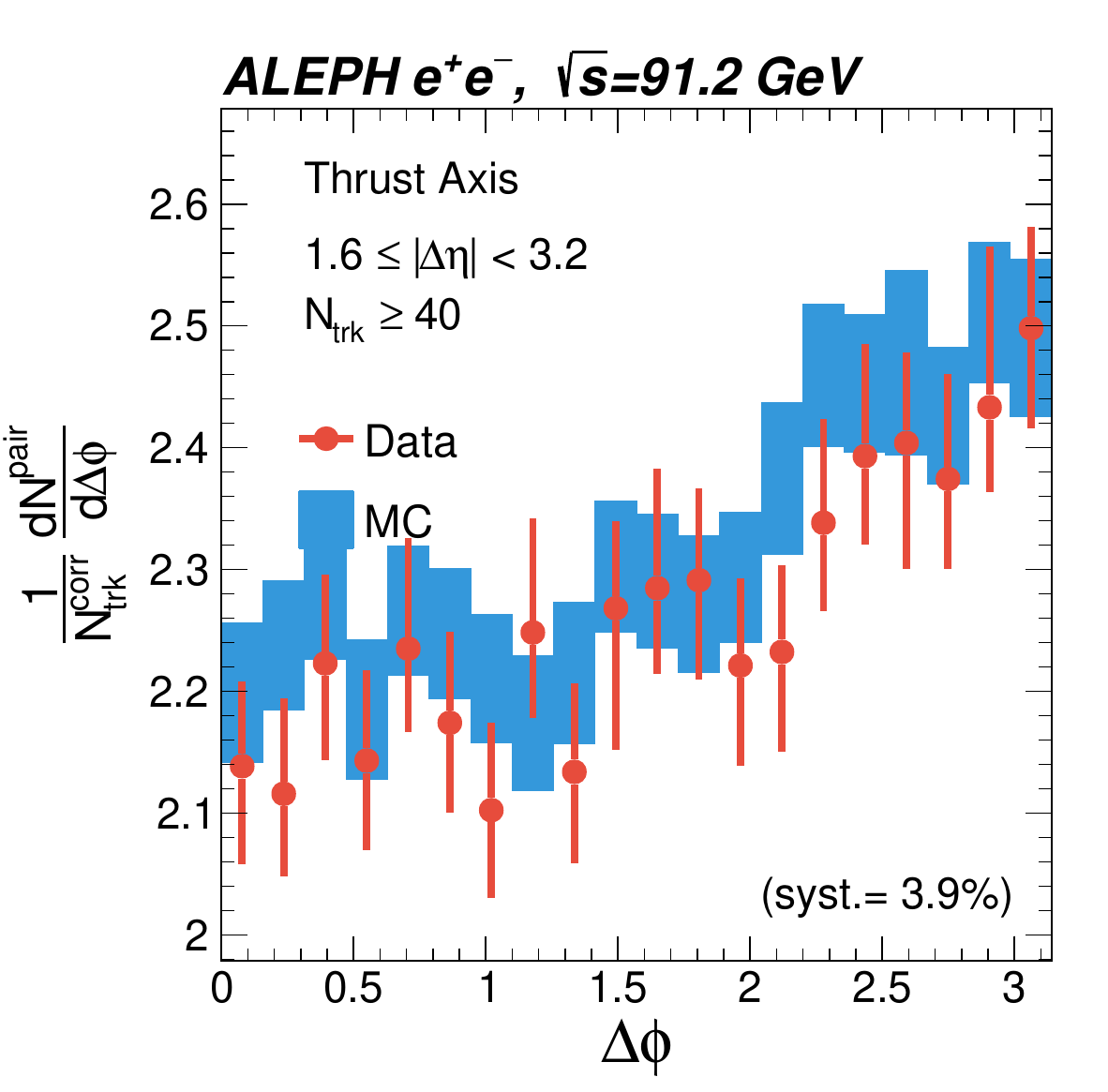}

\caption{Long-range azimuthal differential yields for $Z$-resonance sample in the thrust axis analysis with combined LEP-I and LEP-II statistics. Results are shown as a function of the offline multiplicity in $[5,10), [10,20), [20,30), [30,\infty), [35,\infty), [40,\infty)$ intervals.}
\label{fig:EnergyCut_le_100_Combined_thrust_1d}
\end{figure}

\begin{figure}[ht]
\centering
\includegraphics[width=.45\textwidth]{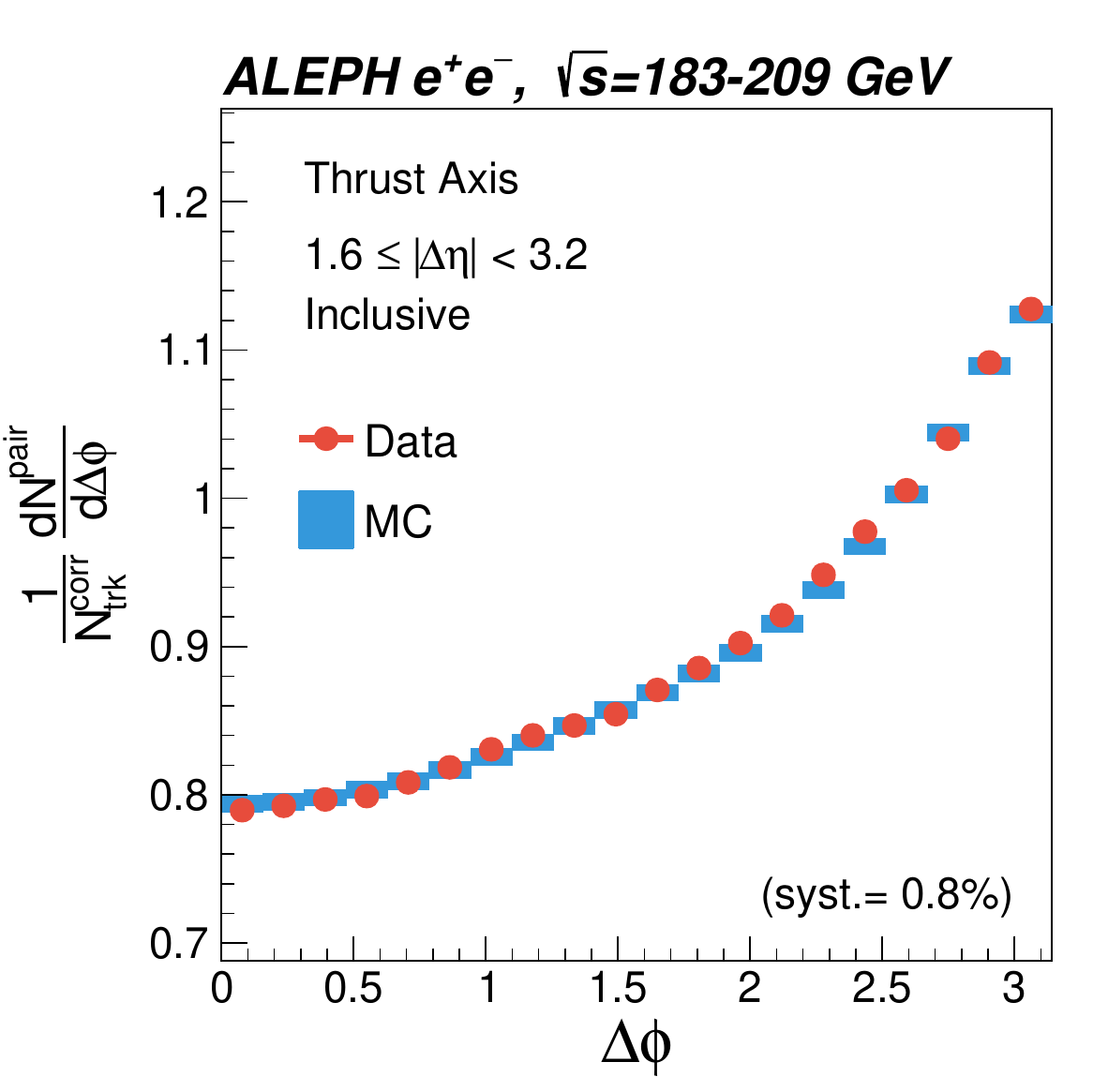}
\includegraphics[width=.45\textwidth]{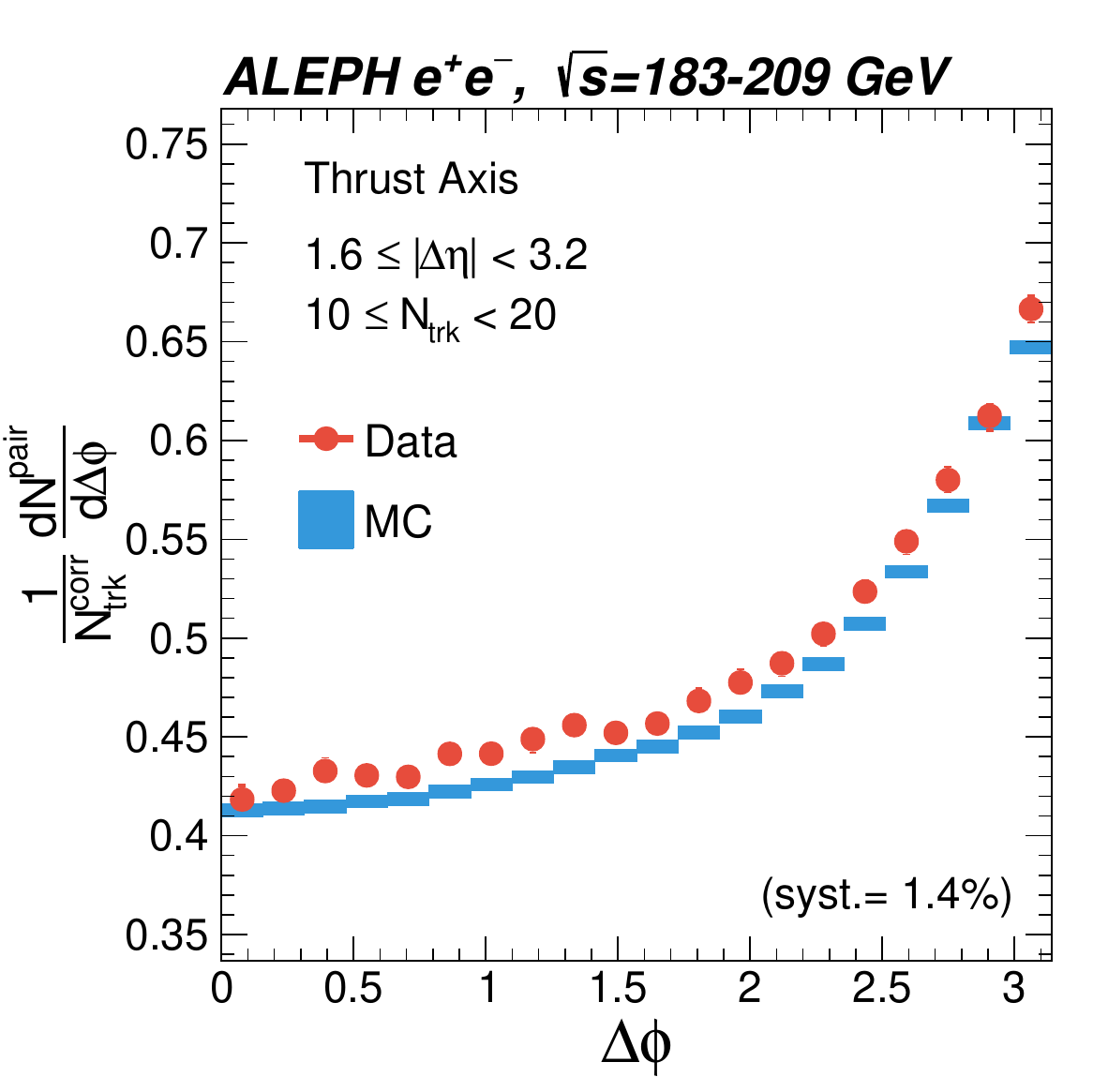}
\includegraphics[width=.45\textwidth]{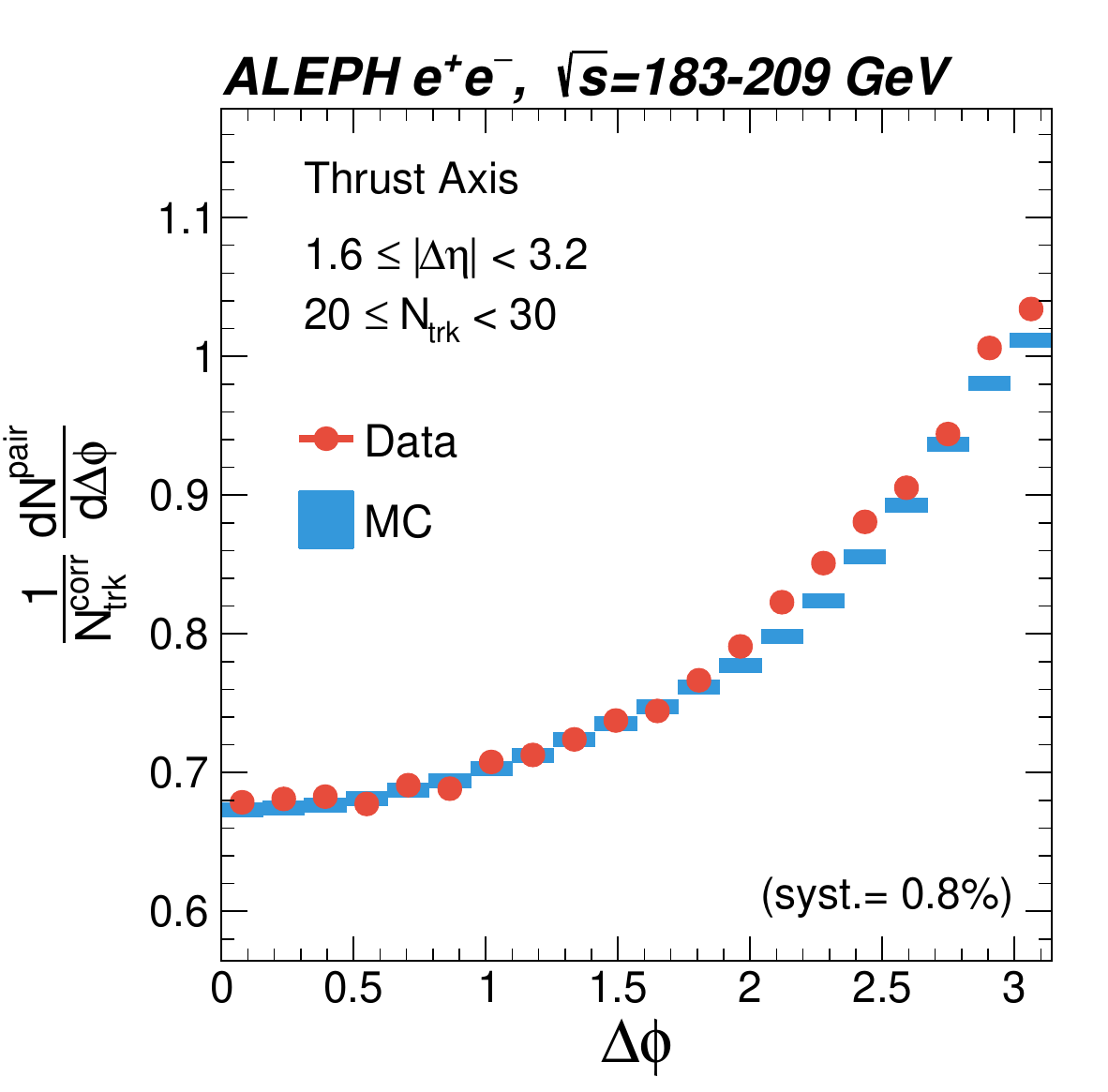}
\includegraphics[width=.45\textwidth]{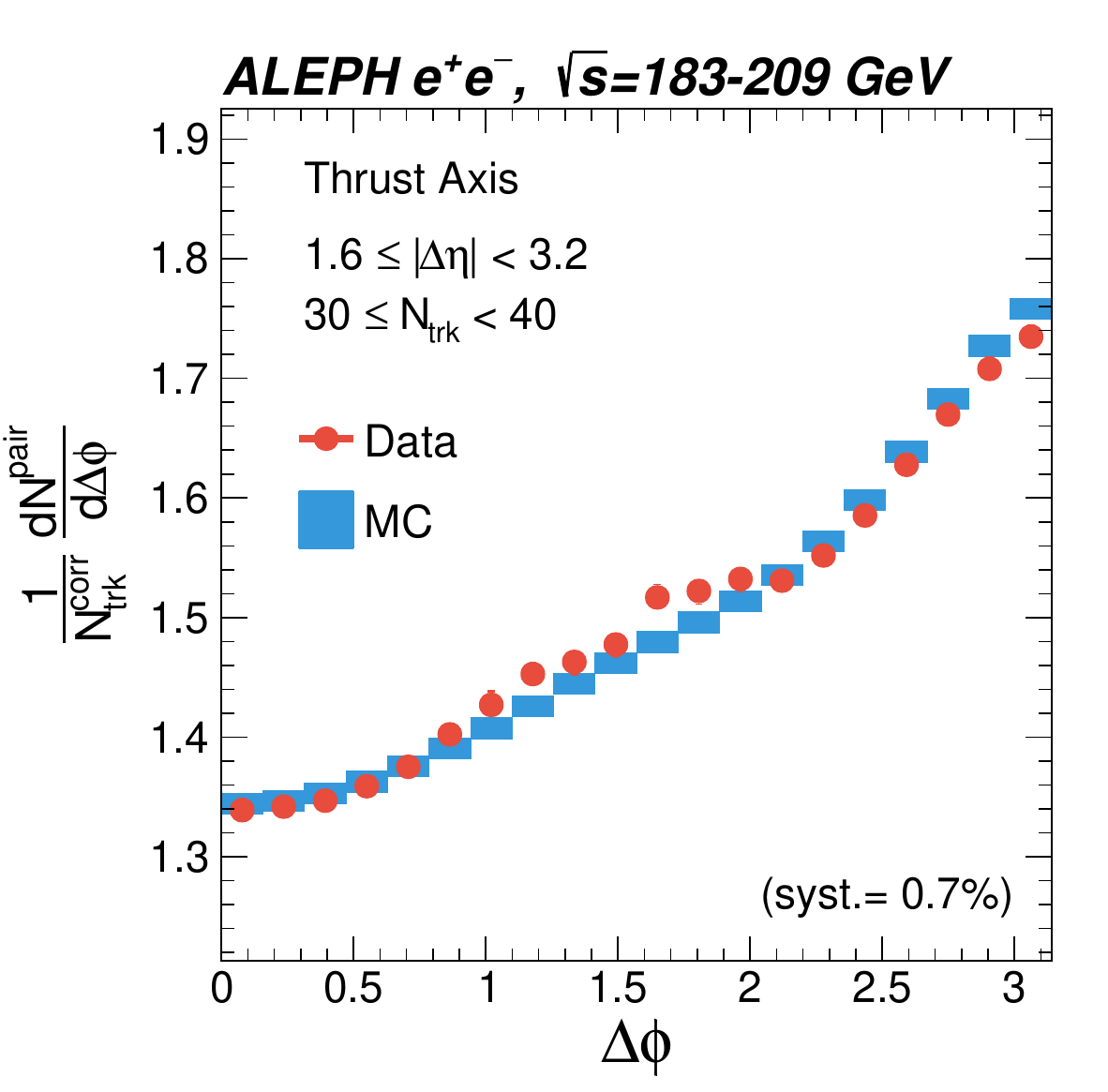}
\includegraphics[width=.45\textwidth]{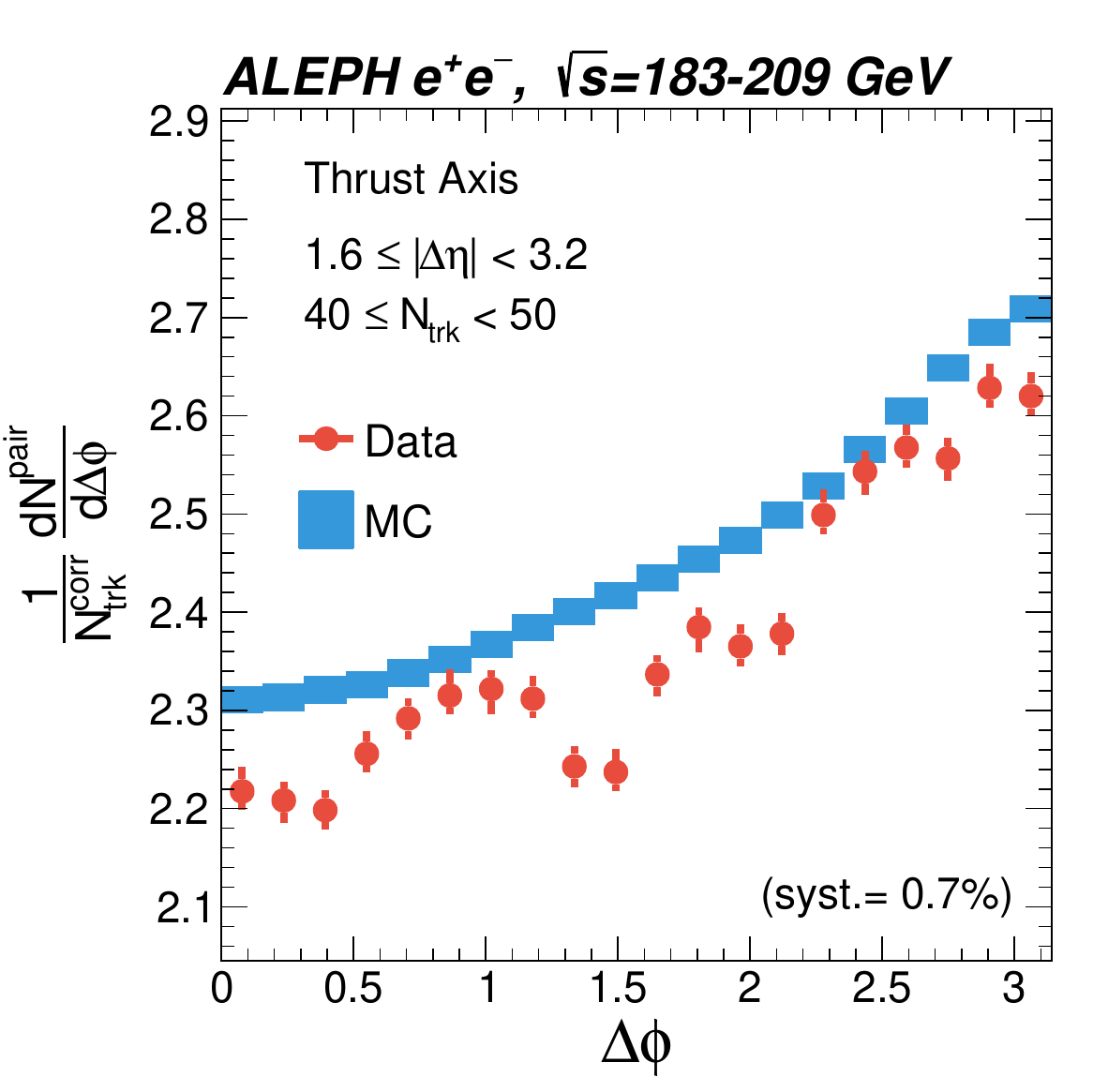}
\includegraphics[width=.45\textwidth]{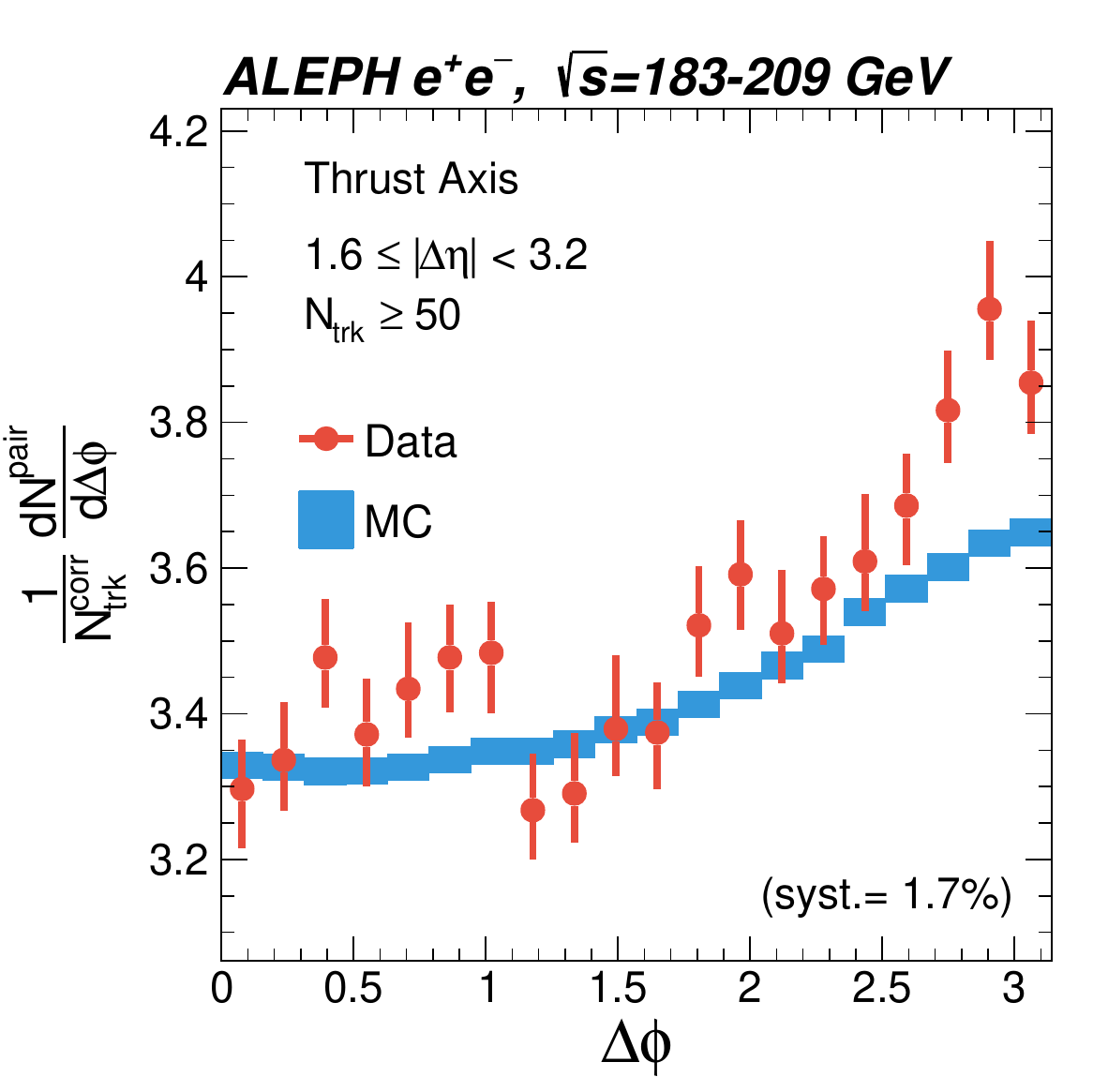}

\caption{Long-range azimuthal differential yields for high-energy sample in the thrust axis analysis. Results are shown multiplicity-inclusively and as a function of the offline multiplicity in $[10,20), [20,30), [30,40), [40,50), [50, \infty)$ intervals.}
\label{fig:EnergyCut_ge_100_thrust_1d}
\end{figure}

Figures~\ref{fig:EnergyCut_le_100_Combined_thrust_1d} and~\ref{fig:EnergyCut_ge_100_thrust_1d} shows the comparisons between data and MC on the long-range azimuthal differential associated yields for the $Z$-resonance and high-energy samples, respectively. For $Z$-resonance sample, the combined LEP-I and LEP-II dataset are descried well by the archived ALEPH MC. Similarly, for the high-energy sample, the MC simulation aligns well with the data for low multiplicity events with $\ntrk<40$. However, in the highest multiplicity class, specifically where $\ntrk>50$, the data reveals a long-range near-side signal that the MC simulation doesn't capture.

\clearpage

\clearpage

\clearpage

\clearpage

\clearpage

\subsection{Systematics evaluation}
\label{sec:Syst}

We consider the systematic uncertainties due to the event and track selections, the $B(0,0)$ normalization factor, and the residual MC correction operations on the long-range associated yield $Y_l(\Delta \phi)$.
We utilized the systematics assessment established in the two-particle correlation analysis work with ALEPH LEP-I archived data~\cite{Badea:2019vey}. We also summarize both the old (LEP-I) values and the new LEP-II results for a comprehensive view in this analysis note.

We consider variations on the ISR and the hadronic event selection criteria for the event selections. 
We vary the ISR selections by relaxing the requirement on the visible two-jet invariant mass $M_{\rm vis}$ from the nominal cut point $0.7\sqrt{s}$ to $0.65\sqrt{s}$, and relaxing the effective center-of-mass energy $\sqrt{s'}$ from $0.9\sqrt{s}$ to $0.87\sqrt{s}$. Two variations give consistent results. The larger uncertainties among the two are quoted as the final uncertainty values.
For the hadronic event selections, the same treatment as the LEP-I work is adopted. We examine their systematic effect by varying the requirements on the number of particles and the reconstructed charged-particle energy. The selection criteria are changed from a $13$ minimal number of particles to $10$ for an event and from $15$~GeV minimal reconstructed charged-particle energy to $10$~GeV. Overall, the hadronic event selections and ISR selections are correlated. The difference from the variation of ISR selections is significantly more prominent. Hence, it is reported as the final event selection uncertainty.
For the track selection, we changed the number of track hits in the time projection chamber from 4 hits to 7 hits.
The inclusion of the $B(0,0)$ factor as the normalization choice (in eq.~\ref{eqn:2PC}) also introduces a systematic uncertainty. We evaluate its impact based on the statistical uncertainty of the $B(0,0)$ normalization factor.
The above systematic uncertainties, in general, impact uniformly across \dphi bins.

The residual MC correction factor (in Section~\ref{sec:residMCCorr}) can lead to an uncorrelated uncertainty across \dphi bins. This systematic uncertainty is estimated by attempting different fits on the residual MC correction factor. Three different functional forms are considered, with half of the maximum deviation between fits considered as the uncertainty from the residual MC correction.

In Table~\ref{tab:Systematics_beam} and~\ref{tab:Systematics_thrust}, the systematic uncertainties due to the aforementioned systematic sources as a function of \ntrk are shown for the beam-axis and thrust-axis analyses, respectively. All numbers are given as percentages. For the systematic evaluation of the Bayesian analysis, we vary the number of toy samples in a pseudo-dataset from 50k to 100k, and vary the granularity of a 1000-step parameter scanning from a 10-sigma interval $[x-10\sigma_x, x+10\sigma_x]$ to a 5-sigma interval $[x-5\sigma_x, x+5\sigma_x]$.

\begin{table}[ht] 
\footnotesize
\caption{Beam axis analysis systematic uncertainties for the LEP-I and LEP-II data as a function of the offline multiplicity \ntrkoff. All values are reported as percentages of the long-range differential associated yield. The systematic uncertainties for LEP-I measurement are from the published two-particle correlation analysis with ALEPH archived data~\cite{Badea:2019vey}. Note that in the residual MC correction column, we report the maximum deviation value between fits. The final considered systematic uncertainty is a half of the maximum deviation.}
\begin{center}
\begin{tabularx}{\textwidth}{C|CCCC}
\hline\hline
\hspace{0.2cm}\ntrkoff \hspace{0.2cm} & TPC hits  & Event energy and track cuts     & $B(0,0)$  & Residual MC correction \\
\hline
\multicolumn{5}{c}{LEP-I}\\
\hline 
$[ 5,10)$ & 0.7 & 0.6 & 0.11 & 10.3 \\
\hline
$[10,20)$ & 0.7 & 0.0 & 0.015 & 2.3 \\
\hline
$[20,30)$ & 0.7 & 0.0 & 0.013 & 0.2 \\
\hline
$[30,\infty)$ & 0.7 & 0.0 & 0.027 & 1.2 \\
\hline
$[35,\infty)$ & 0.7 & 0.0 & 0.057 & 4.4 \\
\hline
\multicolumn{5}{c}{LEP-II}\\
\hline 
$[10,20)$ & 0.28 & 6.84 & 0.10 & 1.52 \\
\hline
$[20,30)$ & 1.99 & 2.97 & 0.06 & 0.61 \\
\hline
$[30,40)$ & 1.13 & 0.64 & 0.06 & 1.10 \\
\hline
$[40,50)$ & 0.45 & 0.10 & 0.09 & 1.50 \\
\hline
$[50,\infty)$ & 2.52 & 0.21 & 0.17 & 1.74 \\
\hline
\hline
\end{tabularx} 
\label{tab:Systematics_beam}
\end{center}
\end{table}

\begin{table}[ht] 
\footnotesize
\caption{Thrust axis analysis systematic uncertainties for the LEP-I and LEP-II data as a function of the offline multiplicity \ntrkoff. All values are reported as percentages of the long-range differential associated yield. The systematic uncertainties for LEP-I measurement are from the published two-particle correlation analysis with ALEPH archived data~\cite{Badea:2019vey}. Note that in the residual MC correction column, we report the maximum deviation value between fits. The final considered systematic uncertainty is a half of the maximum deviation.}
\begin{center}
\begin{tabularx}{\textwidth}{C|CCCC}
\hline\hline
\hspace{0.2cm}\ntrkoff \hspace{0.2cm} & TPC hits  & Event energy and track cuts     & $B(0,0)$  & Residual MC correction \\
\hline
\multicolumn{5}{c}{LEP-I}\\
\hline 
$[ 5,10)$ & 0.3 & 3.4 & 0.88 & 0.50 \\
\hline
$[10,20)$ & 0.3 & 0.0 & 0.09 & 0.21 \\
\hline
$[20,30)$ & 0.3 & 0.0 & 0.05 & 0.06 \\
\hline
$[30,\infty)$ & 0.3 & 0.0 & 0.06 & 0.21 \\
\hline
$[35,\infty)$ & 0.3 & 0.0 & 0.13 & 0.21 \\
\hline
\multicolumn{5}{c}{LEP-II}\\
\hline 
$[10,20)$ & 1.09 & 0.39 & 0.44 & 1.17 \\
\hline
$[20,30)$ & 0.68 & 0.44 & 0.21 & 0.11 \\
\hline
$[30,40)$ & 0.65 & 0.05 & 0.12 & 0.10 \\
\hline
$[40,50)$ & 0.73 & 0.04 & 0.16 & 0.13 \\
\hline
$[50,\infty)$ & 1.60 & 0.50 & 0.27 & 0.02 \\
\hline
\hline
\end{tabularx} 
\label{tab:Systematics_thrust}
\end{center}
\end{table}

\subsection{Quantification of the ridge-like signals}
The structure of the associated yield function is commonly quantified using the "zero yield at minimum" (ZYAM) method~\cite{Ajitanand:2005jj}. In this method, the minimum of the \ydphi associated yield is shifted to zero after subtracting a flat correlation contribution, represented as the constant $C_{\rm ZYAM}$.

Given that correlation functions are symmetric about  $\Delta\phi=0$, we consider the following three even-function fit templates
\begin{equation}
\def\arraystretch{1.}
f(\Delta \phi) = \left\{
\begin{array}{l}
v_0 + 2 \sum\limits_{n=1}^3 v_n \cos(n\Delta \phi),\\
a_0 + a_2 (\Delta \phi)^2 + a_4 \cos(2\Delta \phi),\\
a_0 + a_2 (\Delta \phi)^2 + a_4 (\Delta \phi)^4,
\end{array} \right.
\label{eqn:ZYAMFuncs}
\end{equation}
to determination of the constant $C_{\rm ZYAM}$ correlation and its corresponding $x$ coordinate ($\Delta\phi_{\rm min}$).
Thereafter, the ridge yield is quantified by integrating ZYAM-subtracted azimuthal differential associated yield over the long-range near-side ($0 \le \dphi \le \dphi_{\rm min}$) region,
\begin{equation}
Y_{\rm ridge} = \int\limits_{0}^{\Delta\phi_{\rm min}} [ Y_l(\Delta\phi) - C_{\rm ZYAM} ] d\Delta\phi.
\label{eqn:RidgeYield}
\end{equation}

To quantify the size of the long-range near-side associated yield, the central values or upper limits for each measurement are reported using the bootstrap method~\cite{Efron:1979bst}. In the case of upper limits, the measured azimuthal differential associated yield is varied according to its statistical and systematic uncertainties. It is then re-evaluated with the ZYAM method to obtain an alternative ridge yield.
This procedure is repeated $2 \times 10^5$ times to form the bootstrapped ridge yield datasets using three fit templates (see eq.~(\ref{eqn:ZYAMFuncs})). 

\begin{figure}[ht]
\centering
\includegraphics[width=0.75\textwidth]{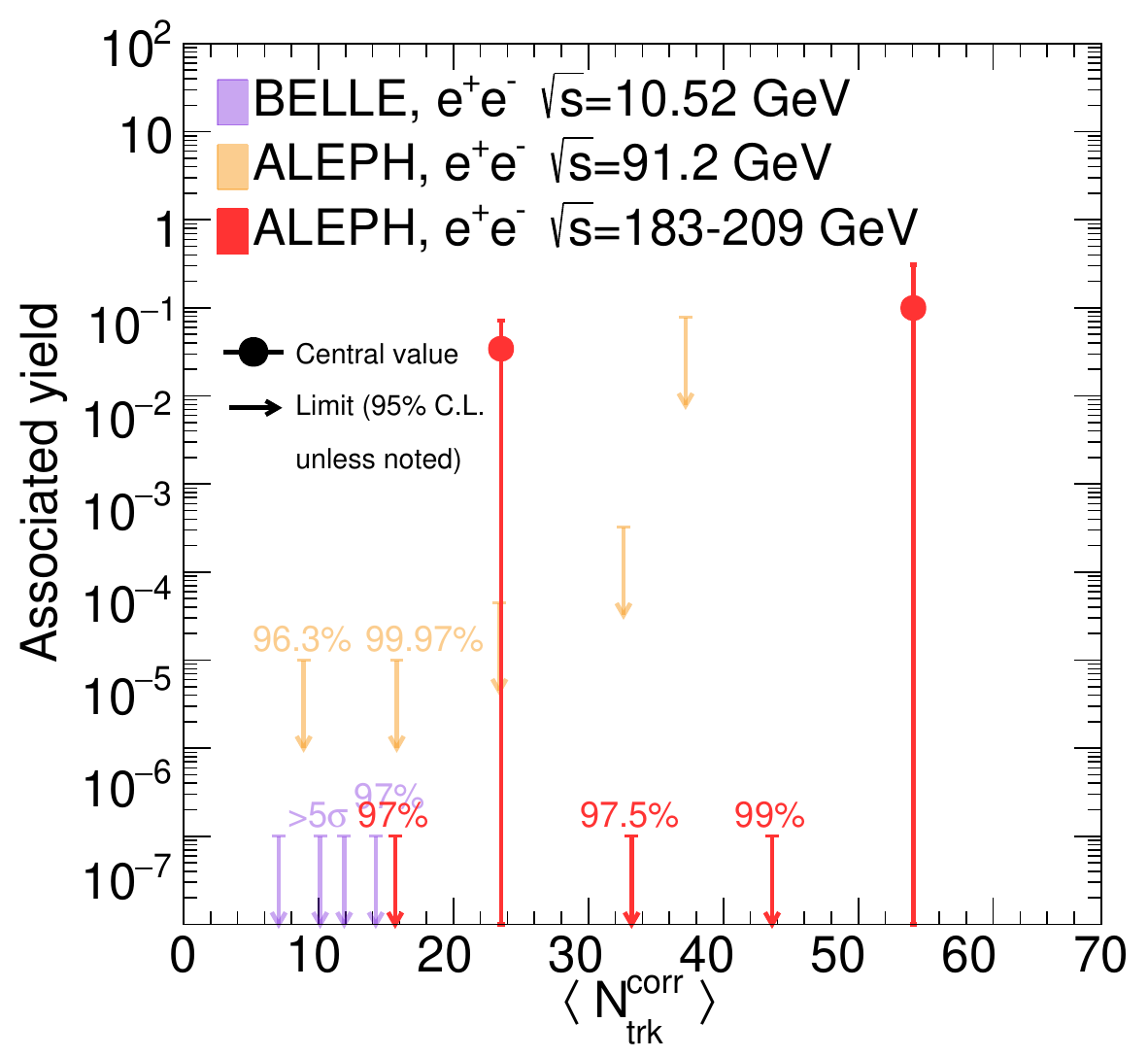}
\includegraphics[width=0.75\textwidth]{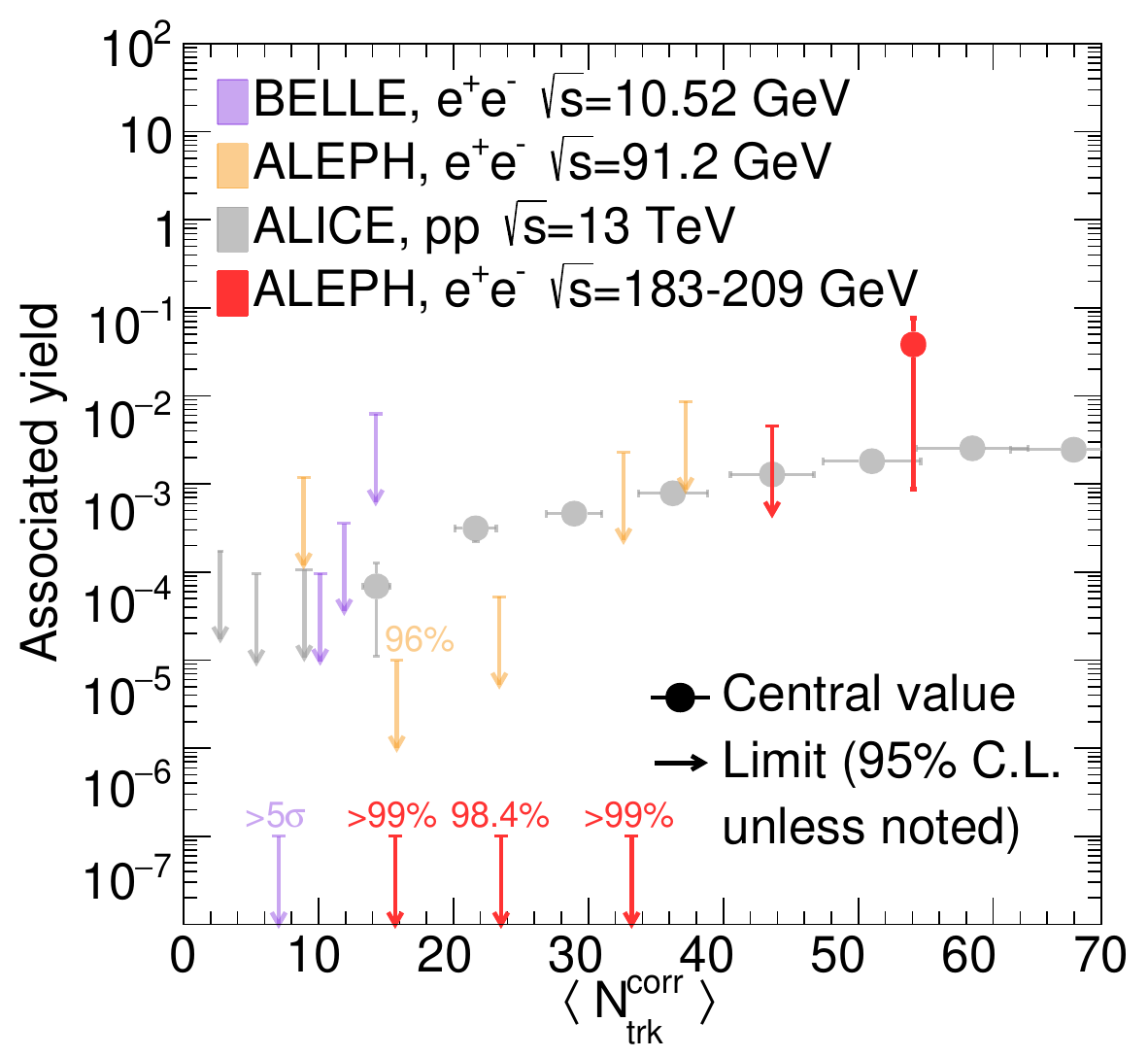}
\caption{
Ridge yield as a function of the $\langle {\rm N}_{\rm trk}^{\rm corr}\rangle$ for the beam (top) and thrust (buttom) axis analysis. This work (LEP-II analysis, $\sqrt{s}=183-209$~GeV) is shown in red, overlapping with results from Belle (pale blue)~\cite{Belle:2022fvl}, LEP-I (pale black)~\cite{Badea:2019vey}, and ALICE (pale orange)~\cite{ALICE:2023ulm}. 
The label ``$>5\sigma$'' indicates the $5 \sigma$ confidence level upper limit.}
\label{fig:CLplots}
\end{figure}

For the measurement with finite long-range near-side yields, we report the central value and its associated 68\% uncertainty band. Otherwise, the 95\% upper limit of the ridge yield or the confidence level of ridge signal exclusions at $10^{-7}$ is reported. In figure~\ref{fig:CLplots}, the ridge-yield evaluations as a function of the averaged corrected multiplicity $\langle {\rm N}_{\rm trk}^{\rm corr}\rangle$ for LEP-II high-energy beam- and thrust-axis analyses are shown in red, overlapping with results from Belle~\cite{Belle:2022fvl}, LEP-I~\cite{Badea:2019vey}, and ALICE~\cite{ALICE:2023ulm} experiments. 
Incorporating the same scaling treatment for $e^{+}e^{-}$ and $pp$ collisions as detailed in ALICE publication~\cite{ALICE:2023ulm}, we scale the $x$ axis of the ALICE data by the acceptance correction coefficients $c_{\rm ee} = 0.78$ and $c_{\rm pp} = 0.57$ for ALEPH and ALICE experiments, respectively. The scaled $\left \langle {\rm N}_{\mathrm{trk}}^{\mathrm{corr}} \right \rangle$ for ALICE data points are displayed with uncertainty ranges from the scaling process. A half of the maximum deviation between the correction coefficients is quoted as the relative uncertainty.

\clearpage

\section{Flow analysis}
\label{sec:flow}
To systematically quantify possible enhanced flow-like signatures, we follow the Fourier decomposition analysis~\cite{CMS:2011cqy,ALICE:2011svq,ATLAS:2012at} to characterize the anisotropy harmonics with the two-particle azimuthal correlations.
The possible non-flow effects are suppressed in the large $|\Delta \eta|$ region. 
Thus, one can describe the long-range azimuthal differential yields  (eq.~\ref{eqn:DeltaPhiAssociatedYield}) using the Fourier series
\begin{equation}
\begin{aligned}
Y_l(\Delta\phi) = \frac{1}{{\rm N}_{\rm trk}^{\rm corr}}\frac{d{\rm N}^{\rm pair}}{d\Delta\phi} = \frac{{\rm N}^{\rm assoc}}{2\pi} \bigg( 1 + \sum_{n=1}^{\infty} 2 V_{n\Delta} \cos(n\Delta\phi) \bigg),
\end{aligned}
\label{eqn:fourier}
\end{equation}
where ${\rm N}^{\rm assoc}$ is the number of associated track pairs in the $|\Delta \eta|$ region of interest ($1.6 \le |\Delta \eta| < 3.2$) and within the full $\Delta \phi$ range for a particular multiplicity and $p_{T}$ class.
The long-range associated yield $Y_l(\Delta \phi)$ is a discrete histogram; one can therefore use the Discrete Fourier Transform (DFT) to obtain the Fourier coefficients ($V_{n\Delta}$) and the normalization (${\rm N}^{\rm assoc}$), which is formulated as follow:
\begin{equation}
\begin{aligned}
Y_l(\Delta \phi)~\text{is defined}~&\text{on the discrete set with}~N=40~\text{elements},\\
\bigg\{\Delta \phi_k &\bigg| \Delta \phi_k = k \frac{2\pi}{N} \quad (k=1,2 \cdots ,N) \bigg\}. \\
{\rm N}^{\rm assoc} &= \frac{2\pi}{N} \sum_{k=1}^{N} Y_l(\Delta \phi_k), \\
V_{n\Delta} &= \frac{1}{ 2 \frac{{\rm N}^{\rm assoc}}{2\pi} } \bigg[ \frac{1}{N/2} \sum_{k=1}^{N} Y_l(\Delta \phi_k) \cdot \cos(n\Delta \phi_k) \bigg]. \\
\end{aligned}
\label{eqn:dft}
\end{equation}
The Fourier coefficients $V_{n\Delta}$ can be further factorized into the single-particle Fourier harmonics product, assuming the azimuthal anisotropy results from the hydrodynamic flow effect. The single-particle harmonic coefficient $v_n$ is related to the Fourier coefficient $V_{n\Delta}$ by
\begin{equation}
\begin{aligned}
V_{n\Delta} (&= v_n^{\rm trig} \times v_n^{\rm assoc}) = v_n^2, \\
v_n &= \frac{V_{n\Delta}}{\sqrt{|V_{n\Delta}|}}.
\end{aligned}
\label{eqn:vn}
\end{equation}
Since in our analysis, we consider the trigger particle and the associated particle within the same $p_T$ bin, we denote $v_n = v_n^{\rm trig} = v_n^{\rm assoc}$.

\subsection{Systematics evaluation}
\label{sec:Syst_flow}
We consider the systematic sources mentioned in Section~\ref{sec:Syst} to evaluate the uncertainty of the single-particle anisotropy coefficients $v_n$ determination. The impacts of event selection and track selection criteria are assessed by comparing the $v_n$'s obtained with the alternative cut points versus the nominal ones. 
The uncertainty on the $B(0,0)$ factor will only affect the overall normalization, but not contribute as a difference in $v_n$.

The residual MC correction systematic  contributes uncorrelatedly across $\Delta \phi$ bins. Hence, we add in independent Gaussian noise to each $\Delta \phi$ bin of the Bayesian toy 1-D correlation function, which is used for the $v_n$ parameter estimation. The standard deviation of the Gaussian noise is assigned as the values reported in Tables~\ref{tab:Systematics_beam} and~\ref{tab:Systematics_thrust} for the beam- and thrust-axis analysis, respectively. The systematic uncertainty on $v_n$ due to the residual MC correction is therefore determined as the difference between the nominal $v_n$ estimation and the one of the alternative configuration with Gaussian noises.

For the systematic uncertainty evaluation of the Bayesian analysis on the $v_n$, we vary the number of toy samples and the granularity of parameter scanning, as described in Section~\ref{sec:Syst}.

\subsection{Results}

\begin{figure}[ht]
\centering
    \begin{subfigure}[b]{0.65\textwidth}
	\includegraphics[width=\textwidth]{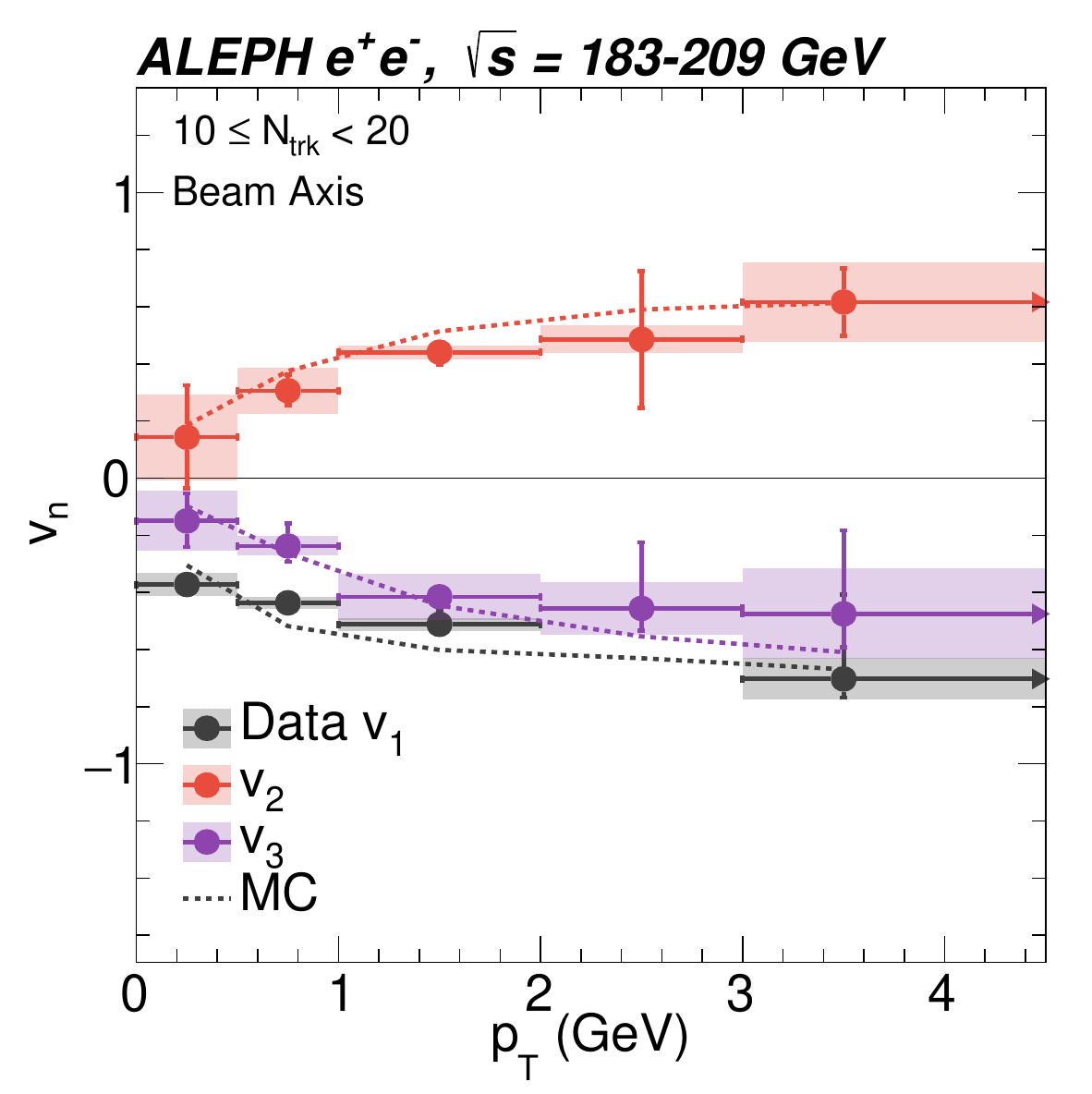}
	\caption{$10 \le {\rm N}_{\rm trk} < 20$}
    \end{subfigure}

    \begin{subfigure}[b]{0.65\textwidth}
    \includegraphics[width=\textwidth]{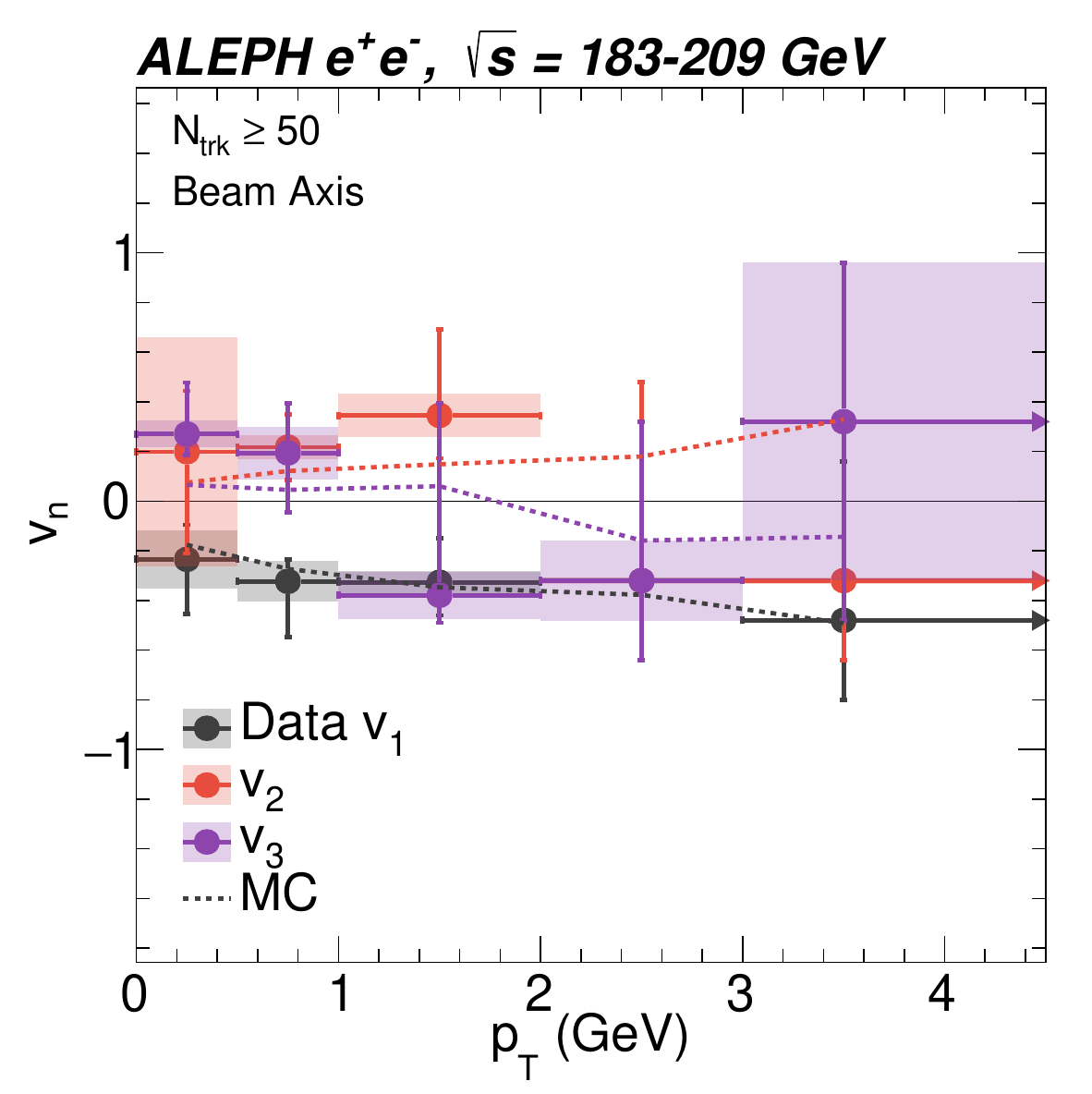}
    \caption{${\rm N}_{\rm trk} \ge 50$}
    \end{subfigure}
\caption{$v_{n}$ as a function of the track pairs' $p_{T}$ requirement in different  multiplicity intervals for the beam axis analysis for LEP-II high-energy sample. Data's $v_1$, $v_2$ and $v_3$ are shown in black, red and purple error bars. MC results are dashed lines with corresponding colors.}
\label{fig:VnVsPt_beam}
\end{figure}

\begin{figure}[ht]
\centering
    \begin{subfigure}[b]{0.65\textwidth}
    \includegraphics[width=\textwidth]{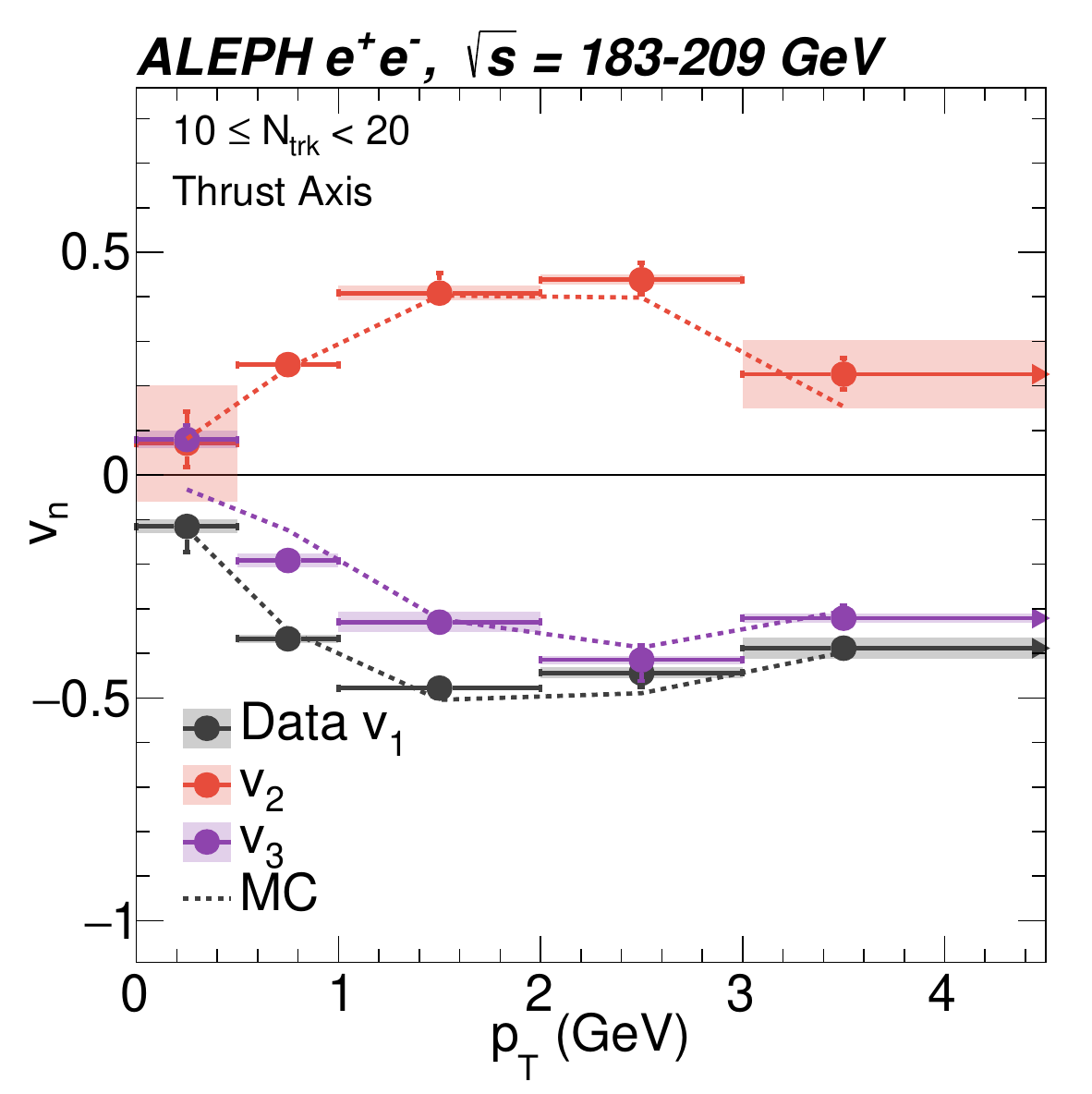}
    \caption{$10 \le {\rm N}_{\rm trk} < 20$}
    \end{subfigure}
    \begin{subfigure}[b]{0.65\textwidth}
    \includegraphics[width=\textwidth]{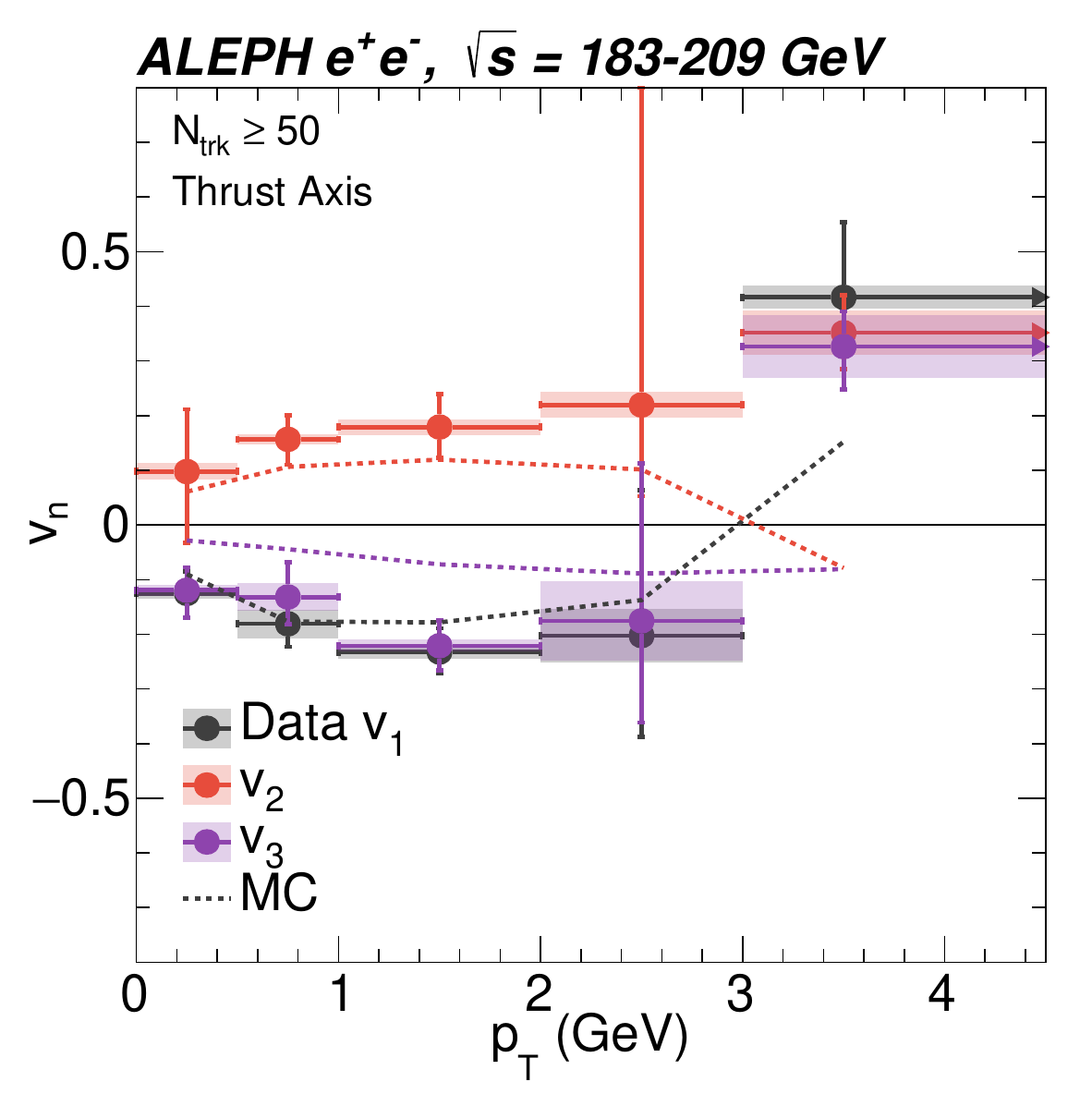}
    \caption{${\rm N}_{\rm trk} \ge 50$}
    \end{subfigure}
\caption{$v_{n}$ as a function of the track pairs' $p_{T}$ requirement in different  multiplicity intervals for the thrust axis analysis for LEP-II high-energy sample. Data's $v_1$, $v_2$ and $v_3$ are shown in black, red and purple error bars. MC results are dashed lines with corresponding colors.}
\label{fig:VnVsPt_thrust}
\end{figure}

The azimuthal anisotropy coefficients for charged particles, namely $v_1$, $v_2$, and $v_3$, are plotted as functions of the $p_{T}$ of the associated particle across various multiplicity intervals. For the beam axis analysis using LEP-II high-energy data and MC samples, these coefficients are depicted in Figure~\ref{fig:VnVsPt_beam}. The results for the thrust axis analysis are presented in Figure~\ref{fig:VnVsPt_thrust}.

In the beam axis analysis, both the data and the MC show a consistent decrease in the values of $v_1$ and $v_3$ with increasing $p_T$. In contrast, $v_2$ demonstrates a rising trend in the low multiplicity class. In high multiplicity scenarios, the $v_n$ values from the MC simulation are smaller than those from the low multiplicity class. It's important to highlight that the statistics in the beam axis analysis are limited, especially in areas of high multiplicity and elevated $p_T$.

For the thrust axis analysis, a notable pattern is observed in the $v_n$ values with respect to the $p_T$ of charged particles. The MC reproduces the $p_T$ dependence of the $v_n$ values. A divergence is particularly evident in the highest multiplicity bin, where ${\rm N}_{\rm trk} \ge 50$. In this bin, the data's $v_2$ value strays from the observed trend, rising continuously with $p_T$ and contrasting sharply with the MC simulation. Specifically, all $v_2$ values in the data are approximately $1 \sigma$ above the central value predicted by the MC. Furthermore, for the highest $p_T$ values exceeding 3 GeV/c, the deduced $v_2$ value is markedly above the MC prediction. Moreover, the $v_3$ value changed sign at the high $p_{T}$.

\begin{figure}[ht]
\centering
    \begin{subfigure}[b]{0.65\textwidth}
    \includegraphics[width=\textwidth]{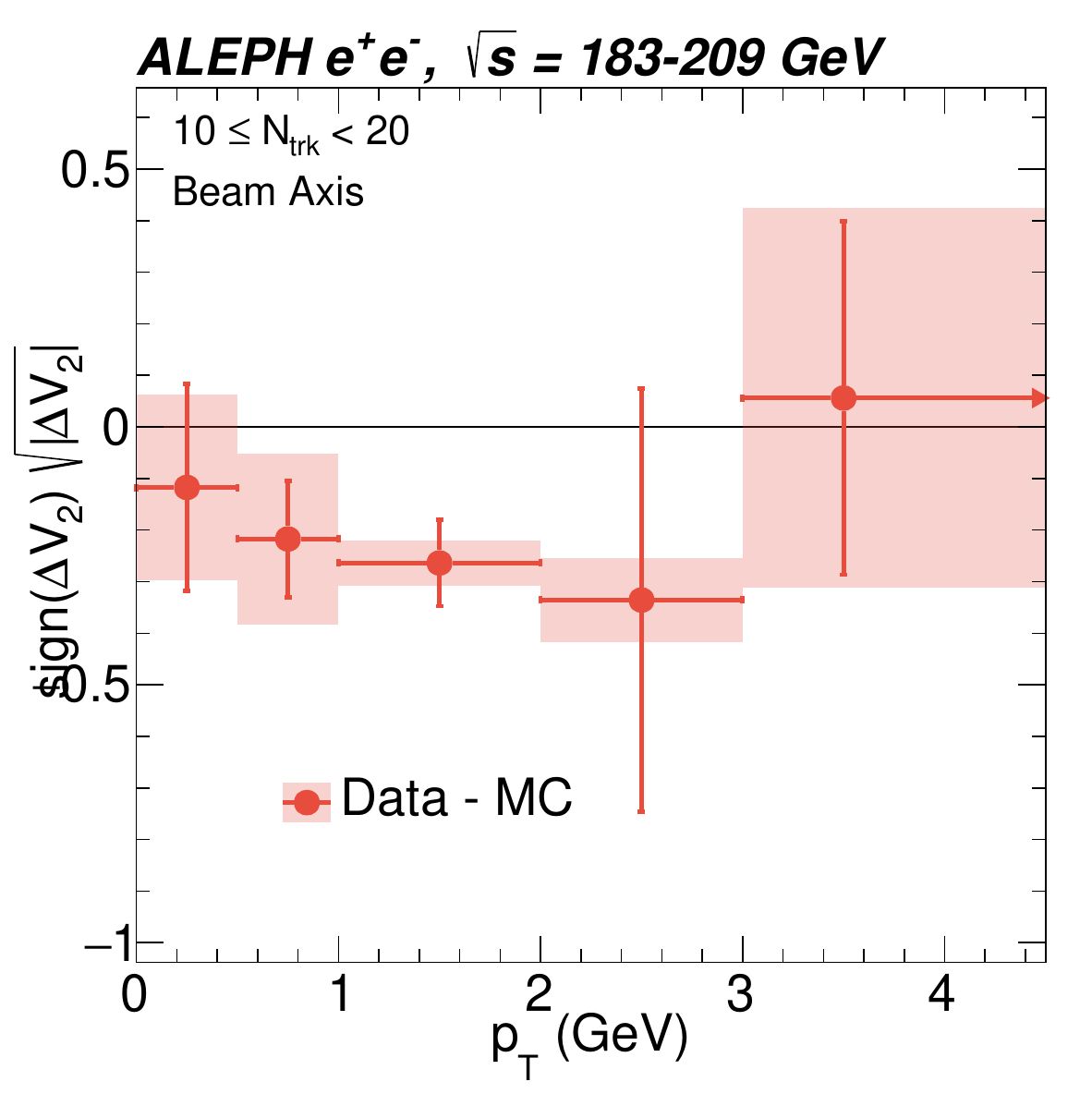}
    \caption{$10 \le {\rm N}_{\rm trk} < 20$}
    \end{subfigure}
    \begin{subfigure}[b]{0.65\textwidth}
    \includegraphics[width=\textwidth]{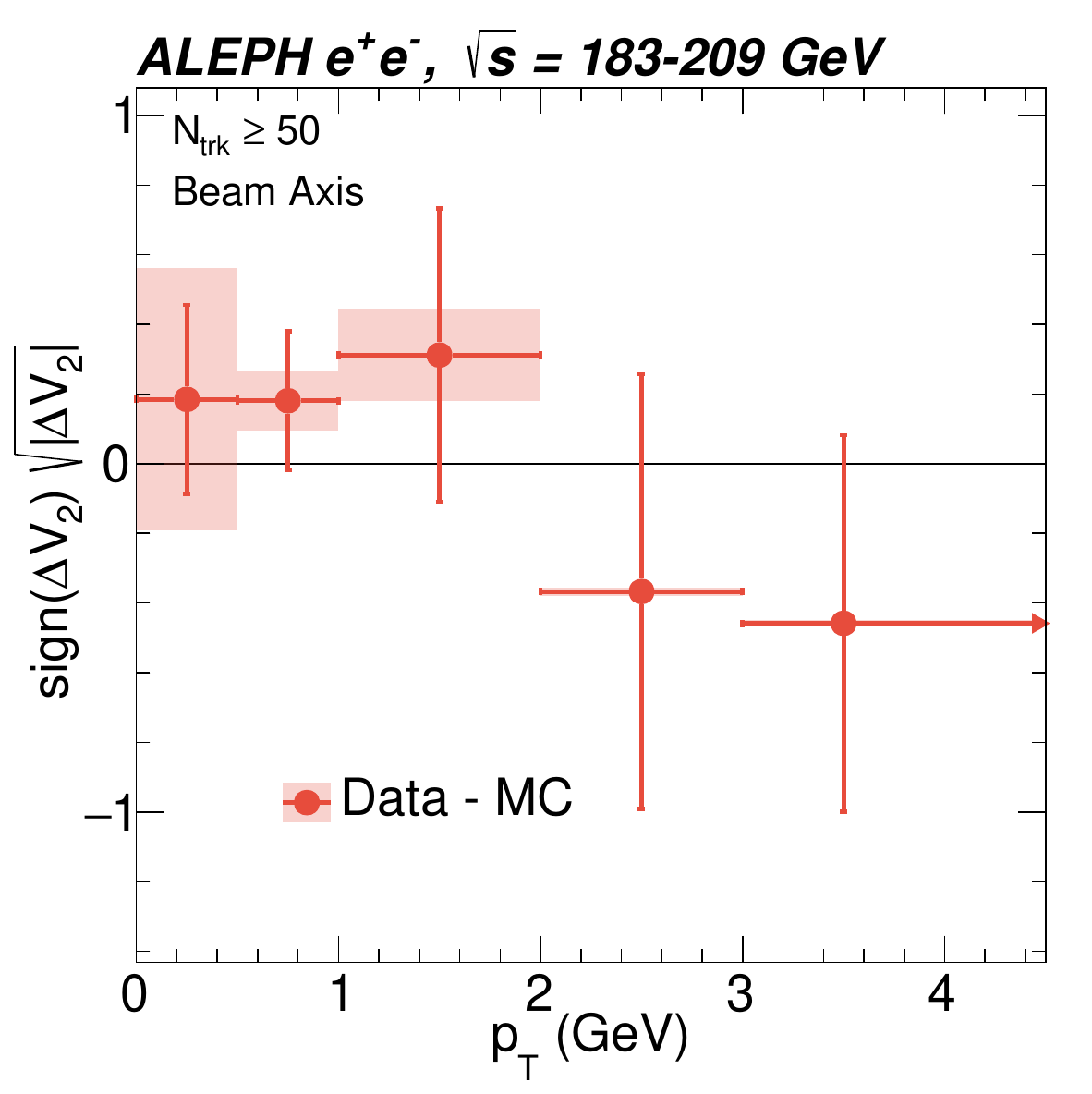}
    \caption{${\rm N}_{\rm trk} \ge 50$}
    \end{subfigure}
\caption{$\text{sign}(\Delta V_{2}) \sqrt{\Delta V_{2}}$ as a function of the track pairs' $p_{T}$ requirement in different  multiplicity intervals for the beam axis analysis for LEP-II high-energy sample.}
\label{fig:Dv2VsPt_beam}
\end{figure}

\begin{figure}[ht]
\centering
    \begin{subfigure}[b]{0.65\textwidth}
    \includegraphics[width=\textwidth]{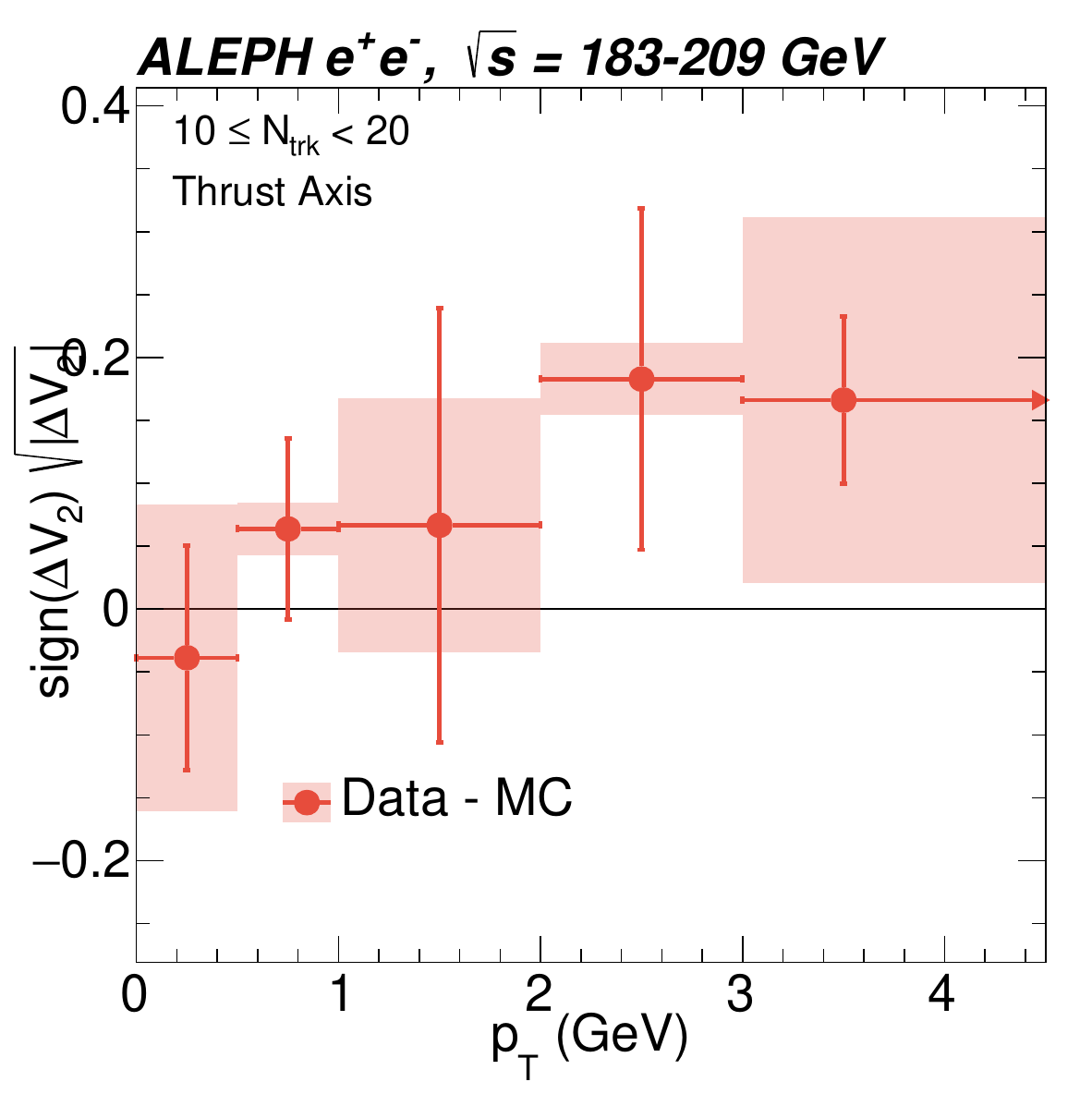}
    \caption{$10 \le {\rm N}_{\rm trk} < 20$}
    \end{subfigure}
    \begin{subfigure}[b]{0.65\textwidth}
    \includegraphics[width=\textwidth]{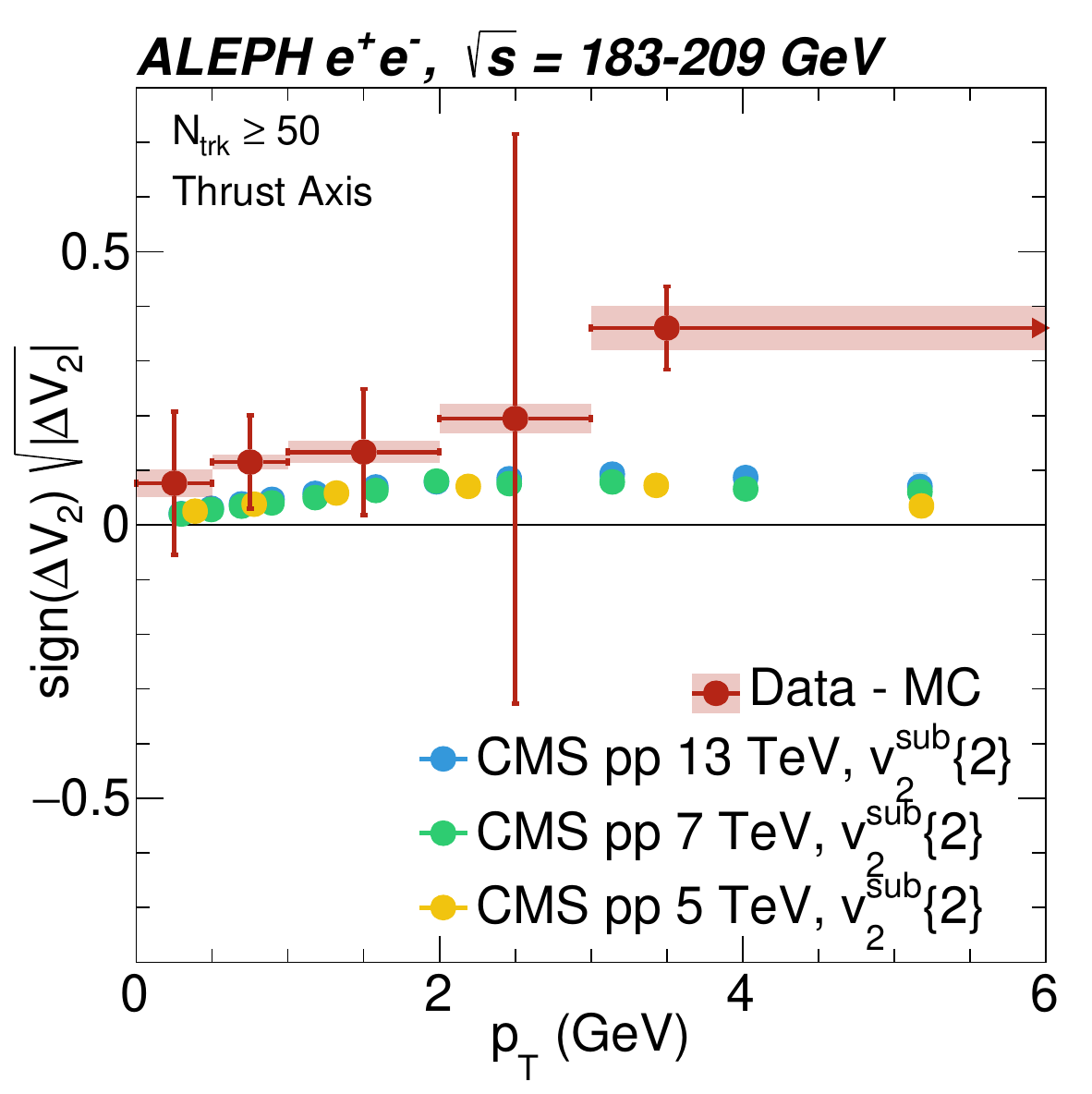}
    \caption{${\rm N}_{\rm trk} \ge 50$}
    \end{subfigure}
\caption{$\text{sign}(\Delta V_{2}) \sqrt{\Delta V_{2}}$ as a function of the track pairs' $p_{T}$ requirement in different  multiplicity intervals for the thrust axis analysis for LEP-II high-energy sample. The result of ${\rm N}_{\rm trk} \ge 50$ is overlaid with CMS subtracted flow coefficient measurements~\cite{CMS:2016fnw}.}
\label{fig:Dv2VsPt_thrust}
\end{figure}

To further elucidate the differences between data and MC, Figure~\ref{fig:Dv2VsPt_beam} and Figure~\ref{fig:Dv2VsPt_thrust} shows the difference of $v_2$ in the data and the MC in the beam and thrust axis analyses, respectively.

\clearpage
\section{Summary}
\label{sec:summary}

We report the first measurement of two-particle angular correlations for charged particles resulting from $e^+e^-$ annihilation, achieving energies up to $\sqrt{s} = 209$ GeV. This study utilized archived ALEPH LEP-II data with center-of-mass energies from 91 to 209 GeV, collected between 1992 and 2000. In the thrust axis analysis of $e^+e^-$ collisions with energies ranging from $\sqrt{s}=183$ to 209 GeV, we reveal a long-range near-side excess in the correlation function.

The decomposition of two-particle correlation functions using a Fourier series was performed for the first time in $e^+e^-$ collisions. The derived Fourier coefficients $v_n$ from LEP-II data offered an insightful contrast to the archived MC. Particularly in high multiplicity events, with particle multiplicity greater than 50, the magnitude of$v_2$ and $v_3$ in data were greater than the corresponding values from the Monte Carlo reference. 

To further visualize these distinctions, we have represented the difference in $v_2$ between data and the MC, focusing on both the beam and thrust axis analyses. The difference between data and MC-derived $v_2$ as a function of associated particle $p_T$ is similar to the $v_2^{\rm sub}$ measured in high multiplicity $pp$ collisions. These intriguing findings not only fortify our understanding of the underlying mechanisms in particle collisions but also shed light on the origins of flow-like signals in smaller collision systems. 



\clearpage

\bibliographystyle{JHEP}
\typeout{}
\bibliography{TPC}
\end{document}

%% file: utility.tex
\def\pp{\ensuremath{pp}\xspace}
\def\ee{\ensuremath{e^+e^-}\xspace}

\def\ntrkoff{\ensuremath{{\rm N}_{\rm Trk}^{\rm Offline}}\xspace}
\def\ntrk{\ensuremath{{\rm N}_{\rm trk}}\xspace}

\def\deta{\ensuremath{\Delta\eta}\xspace}
\def\dphi{\ensuremath{\Delta\phi}\xspace}
\def\ydphi{\ensuremath{Y(\Delta \phi)}\xspace}